\documentclass[journal]{IEEEtran}
\usepackage{amsfonts}
\usepackage{amssymb}
\usepackage{amsmath,amsthm}
\usepackage{graphicx}
\usepackage{cite}
\usepackage{color}

\usepackage{amsopn}
\DeclareMathOperator{\diag}{diag}

\newtheorem{thm}{Theorem}
\newtheorem{lem}{Lemma}
\newtheorem{rem}{Remark}
\newtheorem{prop}{Proposition}
\newtheorem{cor}{Corollary}
\newtheorem{assum}{Assumption}

\usepackage{amsmath,amsthm} % ADDED BY AXEL
\usepackage{amssymb} % Added by Sebin
\usepackage{enumitem}% Added by Sebin
\usepackage[dvipsnames]{xcolor}
\DeclareMathOperator{\cost}{cost}

\usepackage{mathtools} % ADDED BY AXEL

\usepackage{csquotes}

\begin{document}

\title{Networked Multi-Virus Spread with a Shared Resource: Analysis and Mitigation Strategies}

\author{Axel Janson,
        Sebin Gracy,
        Philip E. Par\'{e},
        Henrik Sandberg,
        Karl H. Johansson% <-this % stops a space
%\thanks{Some of the material in this paper was presented at the 3rd IFAC Workshop on Cyber-Physical \& Human Systems \cite{axel2020multi}}
\thanks{Axel Janson, Sebin Gracy,       Henrik Sandberg and
        Karl H. Johansson are with the Division of Decision and Control Systems, School of Electrical Engineering and Computer Science, KTH Royal Institute of Technology, Stockholm, Sweden. (%E-mails: 
        axejan@kth.se, gracy@kth.se, hsan@kth.se, kallej@kth.se).
}
\thanks{Philip E. Par\'{e} is with the School of Electrical and Computer Engineering, Purdue University, IN, USA (%E-mail:
philpare@purdue.edu).}

\thanks{This work was supported in part by the Knut and Alice Wallenberg Foundation, Swedish Research Council under Grants~2016-00861 and~2017-01078, and the National Science Foundation, grants NSF-CNS \#2028738 and NSF-ECCS \#2032258.}
%}
% \thanks{This work is based on research partially sponsored by the National Science Foundation grants ECCS~15-09302, CCF~11-11342, DMS~13-12907, and CNS~15-44953, the Office of Naval Research (ONR) MURI Grant N00014-16-1-2710, US Army Research Office (ARO) Grant W911NF-16-1-0485, and ONR Basic Research grant Navy N00014-12-1-0998.  All material in this paper represents the position of the authors and not necessarily that of the funding agencies.}% <-this % stops a space
% \thanks{Ji Liu is with Stony Brook University (\texttt{ji.liu@stonybrook.edu}).
% Philip E. Par\'{e} is with the Division of Decision and Control Systems at KTH (\texttt{philipar@kth.se}).
% Carolyn L. Beck and Tamer Ba\c sar are with the Coordinated Science Laboratory at the University of Illinois at Urbana-Champaign (\texttt{\{beck3, basar1\}@illinois.edu}).
% Angelia Nedi\'{c} is with the School of ECEE  at Arizona State University
% (\texttt{angelia.nedich@asu.edu}).
% Choon Yik Tang is with the School of ECE  at the University of Oklahoma
% (\texttt{cytang@ou.edu}). }% <-this % stops a space
}

\maketitle
% REQUIRED
\begin{abstract}
The paper studies multi-competitive continuous-time epidemic processes in the presence of a shared resource. %First we derive a model to capture the behavior, called susceptible-infected-water-susceptible (SIWS) model, which is a more general version of the classic susceptible-infected-susceptible (SIS) model. 
We consider the setting where multiple viruses are simultaneously prevalent in the population, and the spread occurs due to not only individual-to-individual interaction but also due to individual-to-resource interaction. %\axel{\textbf{[Inconsistent with later phrasing.]}}.
%could be potentially exacerbated due to the presence of a shared resource.
In such a setting, an individual 
% agent
% (resp. \phil{individual [should this be ``sub-population'' instead?]}) 
is either %\axel{\textbf{[Agent or individual?]}} is either %in the healthy state (in which case it is
not affected by any of the viruses, or infected by one and exactly one of the multiple viruses. 
% \phil{The zero equilibrium corresponds to the the healthy state (i.e., where all the viruses are eradicated) of the population, and we refer to any non-zero equilibrium as an \emph{endemic} equilibrium. [this sentence seems unnecessary... the abstract is a bit long. We could remove it and add what I have in the next sentence]} 
% Furthermore, w
We classify the 
% endemic (nonzero) 
equilibria into three classes: a) the healthy state (all viruses are eradicated),
b) single-virus endemic equilibria %\phil{equilibrium [should this be plural?]}
(all but one viruses are eradicated), and 
c) 
%for the bi-virus case,  
coexisting equilibria %[similarly, is this one always unique?]} 
(%neither of the viruses are eradicated, and,  
multiple viruses simultaneously infect separate fractions of the population). 
%In a broad sense, the system has two equilibria; the healthy state
%In this setting, an agent \axel{\textbf{[Agent or individual?]}} is either in the healthy state (in which case it is not affected by any of the viruses), or it is infected by one and exactly one of the multiple viruses. Similarly, a virus is either in the eradicated state (in which case it does not infect the population), or in the endemic state \axel{\textbf{[Would also be "endemic" during coexistence.]}} (in which case it infects every agent \axel{\textbf{[Different use of agent compared to above.]}} in the network and also the shared resource), or, for the bi-virus setting, in the coexisting state (in which case two viruses simultaneously exist in different fractions of the population).
We provide i) a sufficient condition for exponential (resp. asymptotic) eradication of a virus; %convergence to the eradicated state of a virus; 
ii) %for the single-virus case, 
a sufficient condition for the existence, uniqueness and asymptotic stability of a single-virus endemic equilibrium; iii) a necessary and sufficient condition for the healthy state to be the unique equilibrium; 
and
iv) for the bi-virus setting (i.e., two competing viruses), a sufficient condition and a necessary condition for the existence of a coexisting equilibrium.
%both the viruses to coexist. 
Building on these analytical results, we provide 
% a
two mitigation strategies: a technique that guarantees convergence 
% (both asymptotic and exponential) 
to the healthy state; 
and, in a bi-virus setup, a scheme that employs one virus
%endemic state of one of the viruses, 
to ensure that the other virus is eradicated. The results are illustrated in a numerical study of a spread scenario in Stockholm city.%, corroborating our theoretical results.
% \sebcancel{[This sentence seems a bit off. Maybe rephrase.]}
% convergence to the single-virus endemic equilibrium. 
\end{abstract}

% REQUIRED

\section{Introduction}
%\subsection*{Motivating Example, and general blabla}
%The study of large-scale spread of infectious diseases across structured populations has garnered particular attention with the recent COVID-19 pandemic.
%eradicating one of the viruses by leveraging the notion of competitive exclusion
%Epidemics have been a long-standing feature of human civilization, starting from the Antonine Plague in~165 AD %and  Plague of Justinian in 541 AD to the Spanish flu in $1918$ andto the novel coronavirus (SARS-CoV-2) in the present day. The severity of their impacts have been well-documented. For instance,
In February $1918$ a deadly influenza pandemic (popularly known as the Spanish flu) swept across the globe. It lasted until $1920$, and caused 
approximately
% around 
$50$ million deaths \cite{johnson2002updating}. Influenza viruses have continued to spread across the globe in recurring epidemics \cite{potter2001history}. %Similarly, the Black Death,  an epidemic of bubonic plague during $1347$-$1352$, resulted in $50$ million deaths  (or about $60$ per cent of the entire population of Europe) \cite{benedictow2005black}.} \axel{\textbf{[While I appreciate the need for motivating examples, this is laying it on a bit thick. The Corona-pandemic could be enough motivation on its own.]}}  
%More recently, in December $2019$, a novel coronavirus (SARS-CoV-2), that causes the disease COVID-19, was detected in Wuhan, China. This virus quickly spread throughout China, and before long, cases were reported across Asia. In March~2020, the World Health Organization (WHO) officially declared COVID-19 as a pandemic \cite{who_pandemic}. %In both cases, the ensuing economic damage has also been quite significant \cite{coronaecon}.
Given that the spread of infectious diseases has an enormous impact on society, %and as such has been 
the study of spread %\axel{[spreading processes?]} %\seb{[Response: we need to stick with infectious diseases here, since we started off with the Spanish flu example]} \axel{\textbf{[I want to emphasize that this is a study of infectious disease spread and not the diseases themselves.]}} 
has been an active area of research since Bernoulli's seminal paper \cite{bernoulli1760essai}. The overarching goal %behind
of these research directions is to %gain knowledge of 
find conditions that would cause an epidemic to become eradicated, and leverage the knowledge of these conditions to design spread control strategies. To this end, various infection models have been proposed and studied in the literature;
% susceptible-infected (SI),
susceptible-infected-susceptible (SIS), susceptible-infected-removed %\footnote{\seb{The \enquote{R} in SIR (and SEIR) also stands for Recovered, elsewhere in the literature; see for instance \cite{roy2008effects}}.} %\sebcancel{recovered}
(SIR), susceptible-exposed-infected-removed %\sebcancel{recovered} 
(SEIR), %\axel{\textbf{[Or "removed"?]}}, 
etc. In this paper,
% extended abstract, 
we focus on SIS models.   %While the economic ramifications of COVID-19 have been significant \cite{coronaecon}, the biggest cause of concern remains the growing number of fatalities worldwide. Epidemics are not new %with over $55,000$ deaths being reported \cite{dong2020interactive} as of April 3,~2020, and more fatalities daily

% \par 
More specifically, we consider networked SIS models, in which a population of individuals is divided into subpopulations, %. 
% These subpopulations are subject to internal spread, and also interact via a contact network, allowing the infection to spread within and between subpopulations. 
and the infection can spread both within and between these subpopulations. 
Networked SIS models have been studied extensively using discrete-time \cite{ahn2013global,wang2003epidemic,chakrabarti2008epidemic,BokharaieMTNS10} and continuous-time dynamics %\seb{[SIS models]} 
\cite{van2008virus,khanafer2016stability,van2009virus}.
% \cite{van2008virus,khanafer2016stability,khanafer2014information,van2009virus}.
In the present paper, we will focus on continuous-time dynamics. Although both time-invariant and time-varying continuous-time models have been studied in the literature (see for instance \cite{pare2018epidemic,ogura2016stability} and references therein), we will restrict ourselves to the time-invariant case.
%In an SIS model, an agent is either in the susceptible or infected state. A healthy agent could become infected at some infection rate $\beta$, scaled by the interactions it has with its neighbors. Each agent has its own healing rate $\delta$, that is, the rate at which it recovers from the infection.  %It is assumed that the total number of agents in  the  network remains fixed \cite{yorke2}.

The spread represented by the SIS model has typically been understood as a consequence of %interactions amongst the agents in a network.
human contact. However, 
the spread of infectious diseases
% such a spread 
can significantly worsen 
% also 
due to the presence of a shared resource. For example, waterborne pathogens spread via water distribution systems \cite{kough2015modelling} and droplet-transmitted pathogens spread via surfaces in public transit vehicles \cite{hertzberg2018behaviors}. Such observations have motivated the development of the susceptible-infected-water-susceptible (SIWS) model \cite{liu2019networked,pare2019multi} -- essentially, a networked continuous-time \emph{single-virus} SIS model that incorporates shared resources. For the SIWS model with a \emph{single} shared resource, %\textcolor{orange}{(observe that in such a setting, there is no resource-to-resource virus transmission)},
sufficient conditions for asymptotic convergence to the healthy state (where each subpopulation is infection-free, and the shared resource is contamination-free) have been provided  \cite[Theorem~1]{liu2019networked}. This result has been generalized to also account for \emph{multiple} shared resources %(observe that in such a setting, virus transmission could be due to individual-to-individual contact, individual-to-resource contact and vice-versa, and, in contrast to the single resource case, resource-resource contact) %\axel{\textbf{[Inconsistent with previous phrasing]}}
in \cite[Theorem~1]{pare2019multi}, where %Observe that in such a setting,
virus transmission could be due to individual-to-individual contact, due to individual-to-resource contact% and vice-versa
, and, in contrast to the single resource case, also due to resource-to-resource contact. However, %the SIWS model, as proposed therein, does not account for multiple competing viruses, neither are any
no theoretical guarantees for the endemic behavior (i.e., where the virus persists) of the SIWS model have been provided.% in \cite{liu2019networked,pare2019multi}. 
% \phil{Overcoming this limitation, \cite{axel2020multi} establishes i) a sufficient condition for exponential convergence to the healthy state, ii) a less restrictive sufficient condition for asymptotic convergence to the healthy state, and iii) a sufficient condition for the existence of a single-virus endemic equilibrium in the SIWS model, but no guarantees on the uniqueness and stability of this equilibrium were provided. To the best of our knowledge, this remains an open problem.}

\par The analysis of epidemic spread using SIWS models has been restricted to the single-virus case% (see for instance \cite{liu2019networked,pare2019multi})
. While such analysis provides insights into how to battle %the
an epidemic, %they are limited. More specifically, 
they do not account for settings where multiple strains of a virus could be simultaneously active within a population.
%Depending on the nature of the interaction between strains, this could either lead to \emph{competitive exclusion} (i.e., the infection rate of one virus dominates that of the other viruses, thereby causing those viruses to get eradicated) or \emph{coexistence} (different strains of a virus coexist in  a population by infecting different groups of subpopulation.).
In particular, it is possible that viral strains \emph{compete} with each other to infect the population: 
% in other words, 
% \axel{[In this setting]}, 
each individual %(or subpopulation) 
can be infected by one, and only one, of the multiple viral strains prevalent \cite{castillo1989epidemiological}, i.e., the strains impose \emph{cross-immunity}. Such a phenomenon is not only restricted to epidemics, but may also appear %be exhibited 
% \seb{[we have exhibited here, and exhibits in the next sentence...maybe change this one to manifested]} 
in the context of incompatible  information transmission on social networks \cite{sahneh2014competitive}, or of two competing products within the same market \cite{li2015marketing}. It is well-known that competitive multi-virus propagation exhibits %extremely 
rich behaviour in comparison to single-virus propagation \cite{newman2005threshold}. One possible outcome of competitive multi-virus propagation is \emph{coexistence} (i.e., multiple strains coexist in a population by infecting %different 
separate fractions of each subpopulation), while another is \emph{competitive exclusion} (i.e., the spread parameters of one strain dominate those of the other strains, thereby causing those strains to become eradicated). In particular, %it has been
\cite{bremermann1989competitive} established that cross-immunity %(i.e., being infected with a strain implies immunity from the other strain) 
between strains typically leads to competitive exclusion in single-population SIS models, whereas \cite{li2004coexist,liu2019analysis,pareautomatica,sahneh2014competitive} %\cite{bremermann1989competitive}, whereas %and, hence, merits a detailed investigation \cite{newman2005threshold}.
%SIS models that account for multiple competing strains of viruses have been studied in, among others, \cite{wei2013competing,watkins2016optimal}.
%\seb{\textbf{Depending on the nature of the interaction between strains, this could either lead to \emph{competitive exclusion} (i.e., the infection rate of one virus dominates that of the other viruses, thereby causing those viruses to get eradicated) or \emph{coexistence} (different strains of a virus coexist in  a population by infecting different groups of subpopulation.). 
% \seb{[what about co-infection?]} \axel{\textbf{[Depends on what you mean, but I would put it under coexistence.]}} \seb{[Response: Here is what I mean: same individual being infected with multiple strains at the same time]}.
%Some types of interaction that are known to permit coexistence include super-infection \cite{iannelli2005strain} and co-infection \cite{martcheva2006role}. Cross-immunity (i.e., being infected with a strain implies immunity from the other strain) between the strains typically leads to competitive exclusion in single-population SIS models \cite{bremermann1989competitive}.}}
%\axel{\textbf{[The bolded text is out of order with the unbolded text above it, and the texts are somewhat redundant with respect to each other. Also, the cases of co-infection and super-infection are not compatible with the above definition of coexistence.]}} 
%However, 
%gave 
provided conditions for coexistence in networked SIS models. %it has been shown that coexistence is possible.} %under certain conditions
%\cite{li2004coexist,liu2019analysis,pareautomatica,sahneh2014competitive}.} %In this paper, we consider the case of multiple cross-immune strains of a disease. 
In particular, analysis of the various equilibria of a competing continuous-time time-invariant bi-virus model has been provided in \cite{liu2019analysis}, %\phil{whereas that of a continuous-time time-varying multi-virus model in \cite{pareautomatica}.} \textcolor{purple}{[I cut this previously, since it does not seem reasonable to talk about analysis of the equilibria of a time-varying model. We are also mainly interested in pointing to the time-invariant model equilibria found in pareautomatica. I think a simple fix is to leave the reference to pareautomatica and cut the text in red.]} \sebcancel{[Response by Sebin: I see why you want to cut it. In that case, we should also drop "In particular" from the next sentence.]} %has been provided 
%in \cite{pareautomatica} \axel{\textbf{[Analysis of equilibria doesn't make sense for time-varying case anyway, everything after comma should be cut.]}}. 
%In particular, %for continuous-time time-invariant multi-virus models, 
whereas a necessary and sufficient condition for %the existence of 
a coexisting equilibrium has been established \cite[Theorem~6]{pareautomatica}. %and \cite[Theorems~6 and 7]{liu2019analysis}, respectively.} \axel{\textbf{[While some of the important equilibria are analyzed in \cite{liu2019analysis}, these results are subsumed by \cite{pareautomatica}.]}} %whereas a discrete-time time-invariant multi-virus model has been proposed in \cite{pare2020analysis}.
%Recently, by extending the setup in  \cite{liu2019analysis} to also account for multiple competing viruses and time-varying topologies, a more general  model has been presented in \cite{pareautomatica}.
%\axel{[Stuff about the results in \cite{pareautomatica}.]} 
However, the results obtained in \cite{liu2019analysis,pareautomatica} are restrictive in the following sense:
i) %the conditions for existence of a coexisting equilibrium in
\cite[Theorems~6 and 7]{liu2019analysis} rely on the assumption that the spread parameters with respect to each virus is the \emph{same} for every subpopulation; and ii) %the result in 
\cite[Theorem~6]{pareautomatica} %, while more general than similar results in \cite{liu2019analysis}, 
is reliant on the assumption that the set of spread parameters for each virus
%\seb{[of at least one virus; \phil{I think it's for each/every virus}]} 
is a scaled version of that of other viruses. % i.e. the spread parameters of each virus %\seb{[of at least one virus]} 
%depends on those of the other viruses \axel{\textbf{[Do we need this i.e.?]}}.
%that the conditions that establish coexistence of multiple strains of a virus rely on the assumption that the infection rates and the healing rates with respect to each virus is the \seb{\emph{same} for every subpopulation.}
%\axel{[Stuff about the restrictions of the results in \cite{pareautomatica}.]}
Moreover, none of these works account for the presence of a shared resource in the network. Thus, to the best of our knowledge, for the multi-virus SIWS model, a detailed analysis of the various equilibria and their stability properties remains missing in the literature. The present paper aims to fill this gap and in the process, for bi-virus SIS models, establish conditions for the existence of a coexisting equilibrium under less stringent assumptions.

% \par 
%Observe that the discussion has been centered insofar from the point of view of analysing the spread of an epidemic, whereas 
The overarching goal is to develop 
% the development of 
strategies for the eradication of epidemics.
Several control strategies have been formulated in the context of networked SIS models: % typically with the objective of eradicating the disease. These strategies include 
optimal antidote distribution
\cite{mai2018distributed},
%\cite{torres2015sparse,mai2018distributed}, %and 
contact reduction \cite{theodorakopoulos2012selfish, tomovski2011simple}, etc. In particular, for a directed network with heterogeneous spread, \cite{enyioha2015distributed} provides a fully distributed Alternating Direction Method of Multipliers (ADMM) algorithm that allows for local computation of optimal investment required to boost the healing rate at each node. Note that this ADMM algorithm requires every node to communicate with its neighbors its local estimate of the full network, which %is 
can be undesirable to do in practice.
%considers a directed networkcomprising heterogeneous agents, and %seeks to control the spread of a virus, Towards this end, the authors in \cite{enyioha2015distributed} proposes  a fully distributed  Alternating Direction Method of Multipliers (ADMM) algorithm that allows for local computation of optimal investment required to boost the healing rate at each node. However, the ADMM algorithm in \cite{enyioha2015distributed} involves heavy communication overhead, since every agent needs to share with its neighbors its local estimate of the full network.
% Overcoming the drawbacks with the ADMM algorithm in \cite{enyioha2015distributed},  %\cite{ramirez2018distributed}
% proposes distributed discrete-time nonlinear algorithms for handling a class of distributed resource allocation problems. 
A decentralized algorithm that involves disconnecting nodes and increasing the healing rates, subject to resource constraints, has been proposed in \cite{torres2016sparse}.
%This algorithm also accounts for %the sparse control paradigm; control sparsity, that is, control resources can be allocated only to a subset of nodes, and not necessarily to the whole network.
More recently, a distributed algorithm that, given resource limitations, guarantees the eradication of an epidemic with a \emph{specified} rate has been proposed in~\cite{mai2018distributed}. In the case of multiple cross-immune strains, eradication of all viruses via optimal antidote distribution is featured in \cite{pare2017multi}. In certain multi-strain epidemics, one strain can impose more severe symptoms than other strains \cite{lancet2020coronastrains}. If resource limitations do not permit us to eradicate all strains, we might prioritize eradicating a malignant strain first. The idea of eradicating one strain while sustaining another has been explored in \cite{watkins2016optimal}.
Note that the %distributed control
algorithms provided in %\cite{liu2019analysis},
\cite{liu2019analysis,enyioha2015distributed,torres2016sparse,mai2018distributed}, and the results in \cite{nowzari2016analysis,pareautomatica,watkins2016optimal,gracy2020asymptotic} account only for %\seb{[single-virus]} 
SIS models. %\axel{\textbf{[I think some of these do account for multiple viruses.]}} %and to our knowledge,  distributed control algorithms for eradicating or mitigating virus spread in %\phil{multi-virus SIWS models remain unavailable [We propose this model, so of course this hasn’t been done...]}. The present paper aims to address this gap as well.

\subsection*{Paper Contributions}
%The main contributions of the present paper are analytical results for the multi-virus SIWS model. 
%\sebcancel{[The paper studies the multi-virus SIWS model.]}. 
The paper studies the multi-virus SIWS model. We establish existence, uniqueness and stability of certain equilibria, and develop mitigation strategies based on these results, as follows:
\begin{enumerate}[label=(\roman*)]
    %\item A sufficient condition for exponential eradication of a virus; see Theorem~\ref{thm:expo}.
    %\item %\phil{less strict [??]} %\seb{[Response: "less strict.." is vague. If we add this, then we should also say "less strict in comparison to other results", and how is it less strict etc. Also, in the interest of symmetry, we might also have to have something similar in place for Theorems 5 and 6.]}
    %A sufficient condition, which is less strict than that in \cite[Theorem~1]{liu2019networked} %and Theorem~\ref{thm:expo} \axel{\textbf{[It is not less strict across the board, in the sense that Theorem~\ref{thm:expo} does not require irreducibility. I would prefer to just state that it is a sufficient condition without comparing the conditions here.]}}, %\axel{\textbf{[I don't think that we have to point out the differences between Theorems~\ref{thm:expo} and~\ref{thm:asymp} here, although the items are quite similar.]}} 
    %for asymptotic eradication of a virus; see Theorem~\ref{thm:asymp}.
    \item %For the single-virus case, 
    %\textcolor{purple}{An important problem is determining whether a virus can persist in a population, and the behaviour of such viruses. To this end, we provide a \sebcancel{[Since persistence of a virus has non-trivial impacts on society, it is important to know under what conditions can a virus persist in a population. Furthermore,  assuming that at least one agent is initially infected whether  the entire population (and the shared resource) gets eventually infected]}} %A
    Since the persistence of a virus has an enormous impact on society, it is important to know under what conditions a virus persists and, assuming at least one individual in a subpopulation is infected initially, if the virus can become endemic in the population. To this end, we provide a sufficient condition for the existence, uniqueness, and asymptotic stability of a single-virus endemic equilibrium; see Theorem~\ref{thm:equi}.
    %\item %For the multi-virus case with a shared resource, 
    %A necessary and sufficient condition for the healthy state to be the unique equilibrium; see Theorem~\ref{thm:uniqueness:healthy:state}.
    \item \label{cont:suff} In the bi-virus case with (resp. without) a shared resource, as discussed previously, a natural question is whether both viruses can persist simultaneously. We establish a sufficient condition for the existence of a coexisting equilibrium; %\seb{[COMMENT: choice of terminology: coexisting equilibrium vs coexistence equilibrium? In line with \cite{liu2019analysis}, I am going with coexisting equilibrium.]}
    see Theorem~\ref{thm:joint_eq_exist_shared} (resp. Corollary~\ref{cor:joint_eq_exist}). 
    \item Similarly, in the bi-virus case with (resp. without) a shared resource, 
    % \sebcancel{[conditions that lead to competitive exclusion are of relevance]}
    it is also important to determine conditions that lead to competitive exclusion. Thus, we establish a necessary condition %that rules out 
    for the existence of a coexisting equilibrium; see Theorem~\ref{thm:noequi_single-virus} (resp. Corollary~\ref{cor:noequi_single-virus_noshared}).
    %Observe that the sufficient condition in Theorem~\ref{thm:joint_eq_exist_shared} (resp. Corollary~\ref{cor:joint_eq_exist}) and the necessary condition in Theorem~\ref{thm:noequi_single-virus} (resp. Corollary~\ref{cor:noequi_single-virus_noshared})
      %account for heterogeneous spread, and moreover
      %do not impose any dependency \axel{\textbf{[This can be thought of as inaccurate, since they need to have "compatible" single-virus endemic equilibria, which is a form of dependence.]}} \emph{between} the spread parameters of both the viruses.\\
      %Observe that the sufficient condition in Theorem~\ref{thm:joint_eq_exist_shared} (resp. Corollary~\ref{cor:joint_eq_exist}) and the necessary condition in Theorem~\ref{thm:noequi_single-virus} (resp. Corollary~\ref{cor:noequi_single-virus_noshared}), in contrast to \cite[Theorem~6]{pareautomatica}, do not require the spread parameters of each virus to be a scaled version of those of the other viruses.
      %\axel{\textbf{[If the point is to emphasize the novelty of Theorem~\ref{thm:joint_eq_exist_shared} here, the problem is that \cite[Theorem~5]{pareautomatica} does allow heterogeneous spread.]}} \seb{[Response: Yes, that's correct.]}
    %\item A distributed control strategy for the stabilization (exponential or asymptotic) of the healthy state; see Theorem~\ref{thm:all_deltas_chosen_to_heal}.
    %\item A distributed control strategy for the eradication of a malignant virus in the bi-virus case; see Theorem~\ref{thm:virus-as-vaccine}.
    \item Utilizing our analytical results, we have developed two mitigation strategies: one
    %, one 
    for the stabilization (exponential or asymptotic) of the healthy state, and the other for the eradication of a malignant virus in the bi-virus case; see Theorems~\ref{thm:all_deltas_chosen_to_heal} and~\ref{thm:virus-as-vaccine}, respectively.
\end{enumerate}

%\subsection*{Differences with \cite{axel2020multi}}
%Corollary~\ref{cor:all_deltas_chosen_to_heal} and Proposition~\ref{prop:deltas_and_shared_chosen_to_heal}.
%\seb{TBD:} \axel{\textbf{[Bullet list of result extensions. Relaxed assumptions, proof improvements etc. More in-depth simulations. The new results are "novel".]}}
%
Additionally, we present results on the stability and uniqueness of the healthy state, see Theorems~\ref{thm:expo},~\ref{thm:asymp} and~\ref{thm:uniqueness:healthy:state}. A preliminary version of some of the results in this paper %  \footnote{\sebcancel{COMMENT: This paragraph could be dropped, and added in the cover letter. Thoughts?}} 
% of some of the results in this paper has been 
% accepted
% % presented 
% for  publication
is scheduled for presentation
at the $3^{\rm{rd}}$ IFAC Workshop on Cyber-Physical \& Human Systems; see \cite{axel2020multi}. 
% \\

The aforementioned findings have several consequences. In an epidemiological context, 
% firstly, 
the results on the healthy state 
%and that of \sebcancel{[the]} endemic equilibrium 
shed light on the conditions that ensure the eradication of a virus in the population and the shared resource, and, therefore, aid in developing a mitigation strategy (see Section~\ref{sect:dist:control}). The results on the single-virus endemic equilibrium are useful %in the context of 
for learning the spread parameters (i.e., healing rate and infection rate), as discussed for the special case of SIS models in~\cite{pare2018analysis}.
%and conditions that lead to the virus becoming endemic in the entire population along with the shared resource,
% to an entire population along with the shared resource
% becoming endemic,
% getting infected, 
%respectively.
One of the results concerning the coexisting equilibrium inspires a novel virus eradication strategy, where the goal is to push the dynamics towards the single-virus endemic equilibrium of the benign virus, as opposed to the healthy state (see Section~\ref{sect:dist:control}). %\sebcancel{[The mitigation strategy developed herein presents a novel approach]}
% Secondly, 
The results on the existence of a coexisting equilibrium
%\sebcancel{[Furthermore, these find applications 
%can be applied not just in epidemiology but also in opinion dynamics in social networks]}
find applications 
%can be applied
not just in epidemiology, but also in opinion dynamics in social networks,
where it provides conditions under which
%(for example, economic %\phil{\textbf{[don't understand what `compting economic ideas' means]})
competing ideas (for instance, pro-Covid-19 restrictions versus anti-Covid-19 restrictions) coexist in different groups of the same population.

%The present paper builds on our contributions in \cite{axel2020multi}, and presents a more comprehensive treatment. \textcolor{orange}{In particular, the differences are the following:
%
%\begin{enumerate}[label=(\roman*)]
%    \item The proof of Theorem~\ref{thm:expo} improves upon \cite[Theorem~ 5]{axel2020multi}. 
%    \item Theorem~\ref{thm:asymp} relaxes the conditions of \cite[Theorem~6]{axel2020multi}, owing to a new proof design. As a consequence, Theorem~\ref{thm:uniqueness:healthy:state}, particularized for the single-virus case, relaxes the conditions in \cite[Theorem~9]{axel2020multi}.
%    \item Theorem~\ref{thm:equi} strengthens \cite[Theorem~8]{axel2020multi} 
    % with additional results of 
%    by establishing uniqueness and stability of the endemic equilibrium.
%    \item Several analytical results in this paper, including the control strategies, are novel; see Theorems~\ref{thm:joint_eq_exist_shared}-\ref{thm:virus-as-vaccine}.
%    \item An in-depth set of simulations to illustrate our %theoretical 
%    findings.
%\end{enumerate}}
%
\subsection*{Paper Outline}
The paper is organized as follows: %We conclude the present section by listing all the notation used in the sequel. 
We introduce the model %that captures the setting discussed insofar, 
and formally present the main problems being investigated % in the present paper,
in Section~\ref{sect:Model}. We collect the necessary technical background needed for establishing our main results in Section~\ref{sect:prelims}. The main results %of the present paper 
are split up across Sections~\ref{sect:stab:analysis:healthystate}%,\ref{sec:persist},\ref{sect:coexist} and~
--\ref{sect:dist:control}. More specifically, Sections~\ref{sect:stab:analysis:healthystate}%,\ref{sec:persist},
--\ref{sect:coexist} concern the analysis of equilibria, whereas mitigation strategies for eradicating or mitigating the spread of viruses are provided in Section~\ref{sect:dist:control}. We illustrate our %theoretical 
findings via simulations in Section~\ref{sect:simulations}. Finally, we summarize the main findings of our paper, and highlight relevant problems of future interest in Section~\ref{sect:conclusions}.

\subsection*{Notations}
%\seb{[TBD: Say what $\mathbb{R}_+$ stands for..]}
We denote the set of real numbers by $\mathbb{R}$, and the set of nonnegative real numbers by $\mathbb{R}_+$. For any positive integer $n$, we use $[n]$ to denote the set $\{1,2,...,n\}$. The $i^{\rm{th}}$ entry of a vector $x$ is denoted by $x_i$. The element in the $i^{\rm{th}}$ row and $j^{\rm{th}}$ column of a matrix $M$ is denoted by $M_{ij}$. We use $\textbf{0}$ and $\textbf{1}$ to denote the vectors whose entries all equal $0$ and $1$, respectively, and use $I$ to denote the identity matrix, while the sizes of the vectors and matrices are to be understood from the context. For a vector $x$ we denote the square matrix with $x$ along the diagonal by $\diag(x)$. For any two real vectors $a, b \in \mathbb{R}^n$ we write $a \geq b$ if $a_i \geq b_i$ for all $i \in [n]$, $a>b$ if $a \geq b$ and $a \neq b$, and $a \gg b$ if $a_i > b_i$ for all $i \in [n]$. Likewise, for any two real matrices $A, B \in \mathbb{R}^{n \times m}$, we write $A \geq B$ if $A_{ij} \geq B_{ij}$ for all $i \in [n]$, $j \in [m]$, and $A>B$ if $A \geq B$ and $A \neq B$. %, and $A \gg B$ if $A_{ij} > B_{ij}$ for all $i \in [n]$, $j \in [m]$. 
For a square matrix $M$, we use $\sigma(M)$ to denote the spectrum of $M$, $\rho(M)$ to denote the spectral radius of $M$, and $s(M)$ to denote the largest real part among the eigenvalues of $M$, i.e., $s(M) = \max\{\rm{Re}(\lambda) : \lambda \in \sigma(M)\}$. Given a matrix $A$, $A \prec 0$ (resp. $ A\preccurlyeq 0 $) indicates that $A$ is negative definite (resp. negative semidefinite), whereas $A \succ 0$ (resp. $ A\succcurlyeq 0 $) indicates that $A$ is positive definite (resp. positive semidefinite). %Further, the definiteness of a real square matrix $M$ is denoted as follows: Positive definite is $M \succ 0$, positive semidefinite is $M \succeq 0$, negative definite is $M \prec 0$ and negative semidefinite is $M \preccurlyeq 0$. %\seb{If not done already, then add notations for positive (semi)definiteness, negative (semi)definiteness etc.}
We denote a subset by $P \subseteq Q$, a proper subset by $P \subset Q$, and set difference by $P \setminus Q$.%The cardinality of a set $P$ is denoted by $|P|$.

\begin{figure}
\centering
    \scalebox{0.6}[0.6]{\includegraphics[width=\columnwidth]{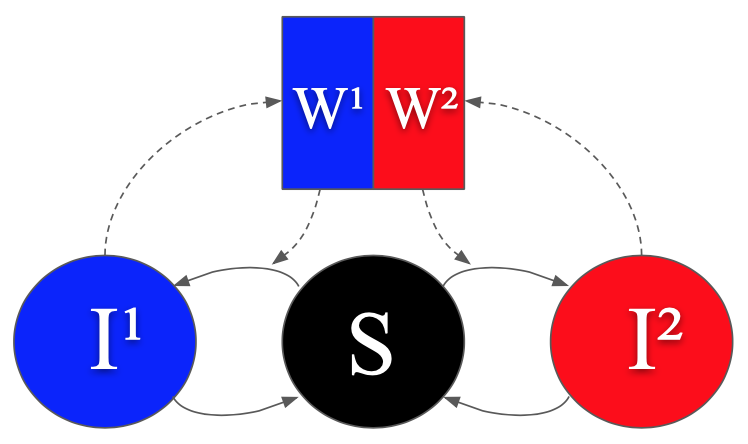}}
    \caption{Visualization of the model for the case when $m=2$. An individual is either susceptible (S), infected with virus~$1$ (I$^1$), or infected with virus~$2$ (I$^2$).
    The shared resource (W) is contaminated by individuals infected with either virus, and in turn augments the corresponding infection rate.} %\axel{[gonna fix image]}}
    \label{fig:model_visualized}
\end{figure}

\section{Problem Formulation}\label{sect:Model}
%
%\seb{[Explain in brief (4-5 lines) as to what this section is all about.]}
%
In this section, we detail a model of multi-viral spread across a population network with a shared resource. %First we establish the premise of the model, from which model variables and dynamics are derived. 
We then formally state the problems being investigated. Finally, we introduce pertinent assumptions and definitions for use in later sections. %\seb{[Try rephrasing the last sentence]} \axel{\textbf{[How about now?]}}.

\subsection{The multi-virus SIWS model} \label{subsec:model}

%\axel{\textbf{[This section will be reworded.]}}

Consider a population of individuals, subdivided into $n$ %subpopulations 
\emph{population nodes} in a network, with a resource $W$ being shared among some or all of the population nodes. Suppose that $m$ viruses are active in the population. An individual can become infected by a virus, either by coming into contact with an infected individual, or due to interaction with the (possibly) contaminated shared resource. We make the assumption that viral infection causes cross-immunity, meaning that an individual can be infected by no more than one virus at the same time. An infected individual can then recover, returning
% back 
to the susceptible state. The model is visualized from an individual's perspective in Figure~\ref{fig:model_visualized}. The spread of the $m$ viruses across the population can be represented by a multi-layer network $G$ with $m$ layers, where the vertices correspond to population nodes and the shared resource, and each layer contains a set of directed edges, $E^k$, specific to each virus~$k$.
%\axel{The spread of each virus~$k$ across the network can be represented by a directed graph $\mathcal{G}$, 
%specific to the virus~$k$, 
%with a directed edge from node $j$ to node $i$ if individuals 
%infected by virus~$k$ 
%in node $j$ can cause infection in node $i$.}
There exists a directed edge from node $j$ to node $i$ in $E^k$ if an individual, infected by virus~$k$ in node $j$, %can directly cause infection in node $i$ %commented by Sebin
can directly infect individuals in node $i$.
Furthermore, the existence of a directed edge in $E^k$ from node $j$ (resp. from the shared resource $W$) to the shared resource $W$ (resp.  to the node $j$) signifies that the shared resource $W$ (resp. node $j$) can be contaminated with virus~$k$ by infected individuals in node $j$ (resp. by the shared resource $W$). %In turn, a directed edge in $E^k$ from the shared resource $W$ to a node $i$ signifies that individuals in node $i$ can be infected by the (possible) contamination of virus~$k$ in the shared resource.
%We say that $\mathcal{G}$ is strongly connected 
We say that the $k^{\rm{th}}$ layer of $G$ is strongly connected if there is a path via the directed edges in $E^k$ from each node, and from the shared resource, to every node, and to the shared resource. In real-world scenarios, it is often the case that viral epidemics can spread from each subpopulation to every other subpopulation, %and so it can be reasonable to 
in which case we assume that each layer is strongly connected. %\seb{[Note: I commented out the last line, since we talk about infection and healing rates in the next para]}%Once infected, an individual recovers from infection at some fixed rate, and we assume that once recovered, the individual is once again susceptible to infection by any virus.
%Consider a network of $n$ nodes, where each node represents an individual, or a population subgroup, and a common resource $W$ being shared among the $n$ nodes. Suppose that there are $m$  viruses active in the network.  A node becomes infected as a consequence of either coming in contact with other infected nodes, or due to contact with the (possibly) contaminated shared resource. We assume that the viruses are competing with each other to infect each node in the network, which implies that, at a given time instant, an agent may get infected by no more than one virus. The spread of the competing $m$ viruses among the $n$ nodes can be represented by a directed graph $\mathcal{G}$, with existence of a directed edge from node $j$ to node $i$ if individuals in node $j$ can infect those in node $i$. We also assume that not only does a node get (possibly) infected due to contact with $W$ but also that $W$ could be contaminated whenever an infected node comes in contact with it. Thus, each node in $\mathcal{G}$ possibly has bidirectional connections with $W$. We say that a graph is strongly connected if, for each node, there is a path to every other node in the graph.
%In the model, time is a continuous variable denoted by~$t$. There are $m$ viruses active in the population and $n$ population nodes. The viruses are identified by $k \in [m]$ and the population nodes are identified by $i \in [n]$. 
%
\par Each population node $i$ contains $N_i$ individuals, with a birth rate $\mu_i$ equal to its death rate $\Bar{\mu}_i$. At any time $t \geq 0$, $S_i(t)$ is the number of susceptible individuals in node $i$, while $I_i^k(t)$ is the number of individuals infected by virus~$k$ in node $i$, while $S_i(t) + \textstyle \sum_{k=1}^m I_i^k(t) = N_i$. The rate at which individuals infected by virus~$k$ in node $j$ infect susceptible individuals in node $i$ is denoted by $\alpha_{ij}^k$, where $\alpha_{ij}^k = 0$ corresponds to the absence of a directed edge from node $j$ to node $i$ in $E^k$. In node $i$, individuals infected by virus~$k$ recover to the susceptible state at a rate $\gamma_i^k$. The shared resource contains a viral mass with respect to each virus~$k$, denoted by $W^k(t)$, representing the level of contamination at time $t \geq 0$. The viral mass of virus~$k$ grows at a rate proportional to all $I_i^k(t)$ scaled by their corresponding rates $\zeta_i^k$, and decays at a rate $\delta_w^k$. The resource-to-node infection rate to node $i$ with respect to virus~$k$ is denoted by $\alpha_{iw}^k$.
The time evolution of the number of susceptible and infected individuals (with respect to each virus~$k \in [m]$) in population node $i \in [n]$ is given by
\begin{equation} \label{eq:orig}
 \begin{split}
  \dot{S}_i(t) = & \, \mu_i N_i - \Bar{\mu}_i S_i(t) + \textstyle \sum_{k=1}^{m} \gamma_i^k I_i^k(t) \\
  &
  - \textstyle\sum_{k=1}^{m} \big{(} \alpha_{iw}^k W^k(t) - \textstyle \sum_{j=1}^{n} \alpha_{ij}^k \frac{I_j^k(t)}{N_i} \big{)} S_i(t), \\
  \dot{I}_i^k(t) = &- \, (\Bar{\mu}_i + \gamma_i^k) I_i^k(t) \\
  &+ \big{(} \alpha_{iw}^k W^k(t) + \textstyle \sum_{j=1}^{n} \alpha_{ij}^k \frac{I_j^k(t)}{N_i} \big{)} S_i(t), \\
  \dot{W}^k(t) = &- \delta_w^k W^k(t) +\textstyle \sum_{j=1}^{n} \zeta_j^k I_j^k(t).
 \end{split}
\end{equation}
%Simplifying~\eqref{eq:orig} with equal death and birth rates yields:
%
%\axel{This step should be skipped, and the equal death and birth rates mentioned in passing while defining~\eqref{eq:split}.}
%
% \begin{equation} \label{eq:simp}
%  \begin{split}
%   \dot{S}_i(t) = &\sum_{k=1}^{m} (\mu_i + \gamma_i^k) I_i^k(t) \\
%   &- \big{(} \alpha_{iw}^k W^k(t) + \sum_{j=1}^{n} \alpha_{ij}^k \frac{I_j^k(t)}{N_i} \big{)} S_i(t), \\
%   \dot{I}_i^k(t) = &- (\mu_i + \gamma_i^k) I_i^k(t) \\
%   &+ \big{(} \alpha_{iw}^k W^k(t) + \sum_{j=1}^{n} \alpha_{ij}^k \frac{I_j^k(t)}{N_i} \big{)} S_i(t), \\
%   \dot{W}^k(t) = &-\delta_w^k W^k(t) + \sum_{j=1}^{n} \zeta_j^k I_j^k(t).
%  \end{split}
% \end{equation}
%
% \noindent 
We define new variables to simplify the system. Let %\seb{COMMENT: Do we need all these varaibles to be explained? Does it not suffice to just define them? }
\begin{align*}
&p_i^k(t) \coloneqq \frac{I_i^k(t)}{N_i}, \, z^k(t) \coloneqq \frac{\delta_w^k W^k(t)}{\sum_{j=1}^{n} \zeta_j^k N_j}, \, \delta_i^k \coloneqq \gamma_i^k + \mu_i, \\
&\beta_{ij}^k \coloneqq \alpha_{ij}^k \frac{N_j}{N_i}, \, \beta_{iw}^k \coloneqq \frac{\alpha_{iw}^k}{\delta_w^k} \textstyle \sum_{j=1}^{n} \zeta_j^k N_j, c_i^k \coloneqq \frac{\zeta_i N_i}{\sum_{j=1}^{n} \zeta_j^k N_j},
\end{align*}
where the variables can be interpreted as follows: With respect to virus~$k$, $p_i^k(t)$ is the fraction of currently infected individuals in node $i$, $z^k(t)$ is a scaled contamination level in the shared resource, $\delta_i^k$ is the healing rate in node $i$, $\beta_{ij}^k$ is the node-to-node infection rate from node $j$ to $i$, scaled with respect to population ratios, $\beta_{iw}^k$ is a scaled resource-to-node infection rate to node $i$, and $c_i^k$ is a scaled node-to-resource contamination rate from node $i$.
%
% \begin{itemize}
% \item $p_i^k(t) = \frac{I_i^k(t)}{N_i}$ be the proportion of population $i$ infected by virus~$k$.
% \item $z^k(t) = \frac{\delta_w^k W^k(t)}{\sum_{j=1}^{n} \zeta_j^k N_j}$ be a normalized measure of the viral mass of type $k$ in the reservoir.
% \item $\delta_i^k = \gamma_i^k + \mu_i$ be the total rate of decay (either by recovery or death) for $p_i^k(t)$.
% \item $\beta_{ij}^k = \alpha_{ij}^k \frac{N_j}{N_i}$ be a normalized coefficient of exposure for virus~$k$ from node $j$ to node $i$.
% \item $\beta_{iw}^k = \frac{\alpha_{iw}^k}{\delta_w^k} \sum_{j=1}^{n} \zeta_j^k N_j$ be a normalized coefficient of exposure from the viral mass of type $k$ to the node $i$.
% \item $c_i^k = \frac{\zeta_i N_i}{\sum_{j=1}^{n} \zeta_j^k N_j}$ be a normalized growth coefficient of the viral mass of type $k$ due to the proportion of infected individuals in node $i$.
% \end{itemize}
% \noindent 
Then, assuming that the birth rates and death rates are equal for each node,~\eqref{eq:orig} can be rewritten as
\begin{align} 
%  \begin{split}
  \dot{p}_i^k(t) = - \delta_i^k p_i^k(t) + \big{(} &1 - \textstyle \sum_{l=1}^m p_i^l(t) \big{)} \nonumber \\ 
  &\times \big{(} \beta_{iw}^k z^k(t) + \textstyle \sum_{j=1}^{n} \beta_{ij}^k p_j^k(t) \big{)}, \nonumber\\
  \dot{z}^k(t) = \delta_w^k \big{(} - z^k(t) +& \textstyle \sum_{i=1}^{n} c_i^k p_i^k(t) \big{)}.  \label{eq:split}
%  \end{split}
\end{align}
%We gather the variables into vectors and matrices using  $p^k(t)$ as our state vector. % with $p_i^k(t)$ as element $i$..
Using vector notation,~\eqref{eq:split} can be rewritten as
\begin{equation} \label{eq:vec}
 \begin{split}
  \dot{p}^k(t) = & \, \Big{(} \big{(} I - \textstyle \sum_{l=1}^m \diag(p^l(t)) \big{)} B^k - D^k \Big{)} p^k(t) \\
  &\,\,+ \big{(} I - \textstyle \sum_{l=1}^m \diag(p^l(t)) \big{)} b^k z^k(t), \\
  \dot{z}^k(t) = & \, \delta_w^k \big{(} - z^k(t) + c^{k} p^k(t) \big{)}, \\
 \end{split}
\end{equation}
%\axel{\textbf{[The following paragraph is still clunky and dissonant with respect to the notations.]} 
where 
%$P^k(t)$ is the diagonal matrix of $p^k(t)$, 
$B^k$ is an $n \times n$-matrix with 
%matrix of 
$\beta_{ij}^k$ as the $(i,j)^{\rm{th}}$ element, %corresponding to the $i^{\rm{th}}$ row and $j^{\rm{th}}$ column, % of $B^k$.
$D^k$ is a diagonal $n \times n$-matrix with $D_{ii}^k = \delta_i^k$ for all $i \in [n]$, $b^k$ is a column vector with $\beta_{iw}^k$ as the $i^{\rm{th}}$ element, and $c^k$ is a row vector with $c_i^k$ as the $i^{\rm{th}}$ element.
To simplify notation further, we define
\begin{gather*}
 y^k(t)
 \coloneqq
 \begin{bmatrix}
 p^k(t)\\
 z^k(t)
 \end{bmatrix}
 ,
 y(t)
 \coloneqq
 \begin{bmatrix}
 y^1(t) \\
 \vdots \\
 y^m(t) \\
 \end{bmatrix}
 ,
 \normalsize B_w^k
 \coloneqq
 \begin{bmatrix}
 B^k & b^k \\
 \delta_w^k c^k & 0 
 \end{bmatrix}
 ,
\end{gather*}

\begin{gather*}
 X(y^k(t))
 \coloneqq
 \begin{bmatrix}
 \diag(p^k(t))  & 0 \\
 0 & 0 
 \end{bmatrix}
 ,
 \,\,
 D_w^k
 \coloneqq
 \begin{bmatrix}
 D^k & 0 \\
 0 & \delta_w^k 
 \end{bmatrix}
 .
\end{gather*}
With these variables in place, 
% system~
we can rewrite~\eqref{eq:vec} as
%This yields system~\eqref{eq:vec} of the form:
\begin{equation}\label{eq:yk}
 %\dot{y}^k(t) = \underbrace{\Big{(} - D_w^k + B_w^k - X\big{(}y(t)\big{)}B_w^k \Big{)}}_{= A^k(y(t))} y^k(t).
 \dot{y}^k(t) =\big{(} - D_w^k + (I - \textstyle \sum_{l=1}^m X(y^l(t)))B_w^k \big{)}y^k(t).
\end{equation}
%\seb{TBD: put problem statement here}
Defining $A_w^k(y(t)) \coloneqq \big{(} - D_w^k + (I - \textstyle \sum_{l=1}^m X(y^l(t)))B_w^k \big{)}$, the dynamics of the system of all $m$ viruses are given by
\begin{gather} \label{eq:full}
 \dot{y}(t)
 =
 \begin{bmatrix}
 A_w^1 \big{(} y(t) \big{)} & 0 & \dots & 0 \\
 0 & A_w^2 \big{(} y(t) \big{)} & \dots & 0 \\
 \vdots & \vdots & \ddots & \vdots \\
 0 & 0 & \dots & A_w^m \big{(} y(t) \big{)}
 \end{bmatrix}
 y(t).
\end{gather}
%
%Throughout this paper, we will primarily be dealing with the following types of equilibria: %of system~\eqref{eq:full}: 
%With respect to a single virus~$k \in [m]$ 
We say that virus~$k \in [m]$ is \emph{eradicated}, or in its \emph{eradicated state}, if $y^k = \textbf{0}$, which is clearly an equilibrium of~\eqref{eq:yk}. %We say that a virus~$k \in [m]$ is active if $y^k(t) > \textbf{0}$. 
When considering the system of $m$ viruses~\eqref{eq:full}, we say that the system is in the \emph{healthy state} if %and only if 
all viruses are eradicated, i.e., $y = \textbf{0}$. %The healthy state is clearly an equilibrium of~\eqref{eq:full}. 
If~\eqref{eq:full} has an endemic (non-zero) equilibrium, it can belong to one of two types: \emph{single-virus endemic equilibrium}, where $y^k>\textbf{0}$ for some $k \in [m]$ and all other viruses in $[m]$ are eradicated; or \emph{coexisting equilibrium}, where $y^k>\textbf{0}$ for multiple $k \in [m]$.
\subsection{Problem Statements}\label{subsec:pb:statement}
%With the setup as given in~\eqref{eq:yk} and
For the model~\eqref{eq:full}, we formally state the problems being investigated in this paper. %\sebcancel{[I think we had almost the same material in the version that we had previously sent. Kalle wants these things to be more mathematical.]} \textcolor{purple}{[We had this statement list before we defined the equilibria previously \sebcancel{[check again with the version that Kalle commented on]}, so when we changed the order, the contrast in language was a bit jarring. For instance, talking about coexistence without the defined term "coexisting equilibrium". I think Kalle just wanted us to apply our terminology formally here.]}
\begin{enumerate}[label=(\roman*)]
    \item \label{q1} Under what conditions does $y^k(t)$ converge exponentially, or asymptotically, to its eradicated state, i.e., $y^k=\textbf{0}$, for some $k \in [m]$?
    \item \label{q2} For a single-virus setup, i.e., $m=1$, under what conditions does the system have a unique single-virus endemic equilibrium, $y^* > \textbf{0}$, and under such conditions, does the system converge asymptotically to $y^*$ from any non-zero initial condition?
    %\sebcancel{[Assuming that, for all $k \in [m]\setminus\{\ell\}$, $y^k =0$, find conditions such that  $\lim_{t\rightarrow \infty}y(t)^\ell = y^{\star}$, where $0 \ll y^{\star} \ll 1$]}
    \item \label{q3} What is a necessary and sufficient condition for the healthy state, i.e., $y=\textbf{0}$, to be the unique equilibrium? %\sebcancel{[Find a necessary and sufficient condition such that $y=0$ is the unique equilibrium.]}
    \item \label{q4} For a bi-virus setup, i.e., $m=2$, under what conditions does the system have a coexisting equilibrium, i.e., $(\hat{y}^1, \hat{y}^2)$ such that $\hat{y}^1>\textbf{0}$ and $\hat{y}^2>\textbf{0}$?
     %\sebcancel{[For a bi-virus setup, i.e. $m=2$, under what conditions does the system have an equilibrium $(\hat{y}^1, \hat{y}^2) \gg \textbf{0}$ in an appropriate domain such that $\hat{y}^1 + \hat{y}^2 \leq \textbf{1}$]}
    \item \label{q5} For a bi-virus setup, under what conditions does the system \emph{not} have a coexisting equilibrium, i.e., $(\hat{y}^1, \hat{y}^2)$ such that $\hat{y}^1>\textbf{0}$ and $\hat{y}^2>\textbf{0}$?
     %\sebcancel{[For a bi-virus setup, i.e. $m=2$, under what conditions does the system \emph {not} have an equilibrium $(\hat{y}^1, \hat{y}^2) \gg \textbf{0}$ in an appropriate domain such that $\hat{y}^1 + \hat{y}^2 \leq \textbf{1}$]}
    \item \label{q6} How can the healing rates, i.e., $\delta_i^k$, be chosen to ensure that the system converges exponentially, or asymptotically, to the healthy state, i.e., $y=\textbf{0}$?
    \item \label{q7} For a bi-virus setup, how can the healing rates of virus~$2$, i.e., $\delta_i^2$, be chosen to ensure that the system converges to the single-virus endemic equilibrium of virus~$1$?
\end{enumerate}

\noindent Before we address these questions, we point out two connections between the considered setup and the existing literature.
\begin{rem} \label{rem:m=1}
If $m=1$, 
system~\eqref{eq:yk} coincides with the single-virus model proposed in \cite{liu2019networked}%; see %the single-virus system from~\eqref{eq:yk} in line with the model in
%\cite[Equation~(12)]{liu2019networked}.
~$\blacksquare$ %\axel{\textbf{[The equation reference could be cut.]}} \seb{[Response: I suggest we keep it.]}.~$\blacksquare$
\end{rem}
%In the rest of this paper,
% extended abstract, 
Hereafter, when a single-virus system is considered, we drop the superscripts identifying the virus from all variables:

\noindent $B_w\coloneqq B_w^1$, $D_w\coloneqq D_w^1$, etc.

% \axel{\textbf{[I think the definition of system~\eqref{eq:full} below should be moved to subsection~\ref{subsec:model}, seeing as how it is referred to as being the actual model system in many places.]}}

% Defining $A_w^k(y(t)) = \big{(} - D_w^k + (I - \textstyle \sum_{l=1}^m X(y^l(t)))B_w^k \big{)}$, and by gathering all viruses into the same system, we can rewrite~\eqref{eq:yk} as follows: 
% \begin{gather} \label{eq:full}
%  \dot{y}(t)
%  = \tiny
%  \begin{bmatrix}
%  A_w^1 \big{(} y(t) \big{)} & 0 & \dots & 0 \\
%  0 & A_w^2 \big{(} y(t) \big{)} & \dots & 0 \\
%  \vdots & \vdots & \ddots & \vdots \\
%  0 & 0 & \dots & A_w^m \big{(} y(t) \big{)}
%  \end{bmatrix}
%  y(t).
% \end{gather}

%\seb{TBD: Put a transitory text here; telling why the remarks are in place.}

\begin{rem} \label{rem:no_W}
Note that if $b^k = \textbf{0}$ for all $k \in [m],$ the influence of the shared resource on the population is nullified. Then, the multi-virus dynamics of the population nodes, $p^k(t)$, in~\eqref{eq:full} are equivalent to the time-invariant multi-virus setup of \cite{pare2017multi}.~$\blacksquare$
\end{rem}
%
% \subsection{Single-virus system}
%
% Note that if $m=1$ we recover the single-virus model from~\eqref{eq:split}:
%
% \begin{equation} \label{eq:single}
%  \begin{split}
%   \dot{p}_i(t) = & - \delta_i p_i(t) + \big{(}~1 - p_i \big{)} \big{(} \beta_{iw} z(t) + \sum_{j=1}^{n} \beta_{ij} p_j(t) \big{)}, \\
%   \dot{z}(t) = & \,\delta_w \big{(} - z(t) + \sum_{i=1}^{n} c_i p_i(t) \big{)}.
%  \end{split}
% \end{equation}
%
% Defining $B, D, b, c, B_w$ and $D_w$ analogously, as well as:
%
% \begin{gather*}
%  x(t)
%  =
%  \begin{bmatrix}
%  p_1(t) \dots z(t)
%  \end{bmatrix}^T.
% \end{gather*}
%
% We can write~\eqref{eq:single} as:
%
% \begin{equation}\label{eq:singlemat}
%  \dot{x}(t) = \Big{(} - D_w + B_w - X\big{(}x(t)\big{)}B_w \Big{)} x(t).
% \end{equation}
%
%\subsection{Assumptions}
%
% \phil{
\subsection{Positivity assumptions}
% [the title of this subsection is a bit too long]
% }

%
In order for~\eqref{eq:full} to be well-defined and realistic, we make %make use of 
the following assumption.
\begin{assum} \label{assum:base}
Suppose that $\delta_i^k>0, \delta_w^k > 0, \beta_{ij}^k \geq 0, \beta_{iw}^k \geq 0$ and $c_i^k \geq 0$ for all $i, j \in [n]$ and $k \in [m]$, with $c_l^k > 0$ for at least one $l \in [n]$.~$\blacksquare$ 
\end{assum}
Note that if Assumption~\ref{assum:base} holds, then for all $k \in [m]$, $B_w^k$ is a nonnegative matrix and $D_w^k$ is a positive diagonal matrix %(and is therefore invertible). %Thus, $(B_w^k - D_w^k)$ is a Metzler matrix, for all $k \in [m]$. 
Moreover, recall that a square matrix $M$ is said to be irreducible if, replacing the non-zero elements of $M$ with ones and interpreting it as an adjacency matrix, the corresponding graph is strongly connected. Then, noting that non-zero elements in $B_w^k$ represent directed edges in the set $E^k$, we see that $B_w^k$ is irreducible whenever the $k^{\rm{th}}$ layer of the multi-layer network $G$ is strongly connected. %when $B_w^k$ is interpreted as an adjacency matrix representing the set of directed edges $E^k$, $B_w^k$ is said to be irreducible if and only if the $k^{\rm{th}}$ layer of the multi-layer network $G$ is strongly connected.

Thanks to %these assumptions
Assumption~\ref{assum:base}, we can
% will be able to 
restrict our analysis to the sets $\mathcal{D} \coloneqq \{y(t): p^k(t) \in [0,1]^n, z^k(t) \in [0, \infty) \, \forall k \in [m]\}$ and $\mathcal{D}^k \coloneqq \{y^k(t): p^k(t) \in [0,1]^n, \, z^k(t) \in [0, \infty)\}$. %for system~\eqref{eq:full} and~\eqref{eq:yk}, respectively. 
%\axel{\textbf{[It would be prudent to restrict S further, so that the sum of all the $p^k(t)$ has be less than one in every component.]}} 
Since $p_i^k(t)$ is to be interpreted as a fraction of a population, and $z^k(t)$ is a nonnegative quantity, these sets represent the %realistic
sensible domain of the %model 
system. 
That is, %, for any $k \in [m]$, 
if $y^k(t)$ takes values outside of $\mathcal{D}^k$, then those values would lack physical meaning. The following lemma shows that once $y(t)$ enters $\mathcal{D}$, %$[0,1]^n$ %\axel{COMMENT: [Should be the set $\mathcal{D}$.]}, 
it never leaves $\mathcal{D}$.
%
%In keeping with this, we can establish the following result for these sets:
%
\begin{lem}\label{lem:pos}
Let Assumption~\ref{assum:base} hold. Then $\mathcal{D}$ is positively invariant with respect to~\eqref{eq:full}%, and if $y(t) \in \mathcal{D}$ for $t \geq 0$, $\mathcal{D}^k$ is positively invariant with respect to~\eqref{eq:yk} for all $k \in [m]$.
~$\blacksquare$
\end{lem}
\textit{Proof:}
Consider $y(t) \in \mathcal{D}$. If $p_i^k(t) = 1$, then $\dot{p}_i^k(t) < 0$, so if $p_i^k(0) \leq 1$ then $p_i^k(t) \leq 1$, for all $t \geq 0$, $k \in [m], i \in [n]$. Further, if $p_i^k(t) = 0$ and $y(t) \geq \textbf{0}$, then $\dot{p}_i^k(t) \geq 0$, for all $t \geq 0$. Similarly, if $z^k(t) = 0$ and $y(t) \geq \textbf{0}$, then $\dot{z}^k(t) \geq 0$. Thus, $y(t) \geq \textbf{0}$ if $y(0) \geq \textbf{0}$, for all $t \geq 0$.~$\square$
\section{Preliminaries} \label{sect:prelims} %stability notions, results from Khalil etc.}
%\seb{Comment: Add relevant definitions and theorems on stability analysis from \cite{khalil2002nonlinear}.}
%
In this section, we recall some preliminary results, pertinent to the analysis of system~\eqref{eq:full}. A real square matrix is said to be Metzler if all elements outside the diagonal are nonnegative. We require the following results for Metzler matrices.
%
% \begin{lem} \label{lem:metz}
% \cite[Proposition~2]{rantzer2011distributed} Suppose that $M$ is a Metzler matrix such that $s(M) < 0$. Then, there exists a positive diagonal matrix $P$ such that $M^T P + P M \prec 0$.~$\blacksquare$ %\hfill  
% \end{lem}
%
\begin{lem} \label{lem:metz_irreduc}
\cite[Lemma A.1]{khanafer2016stability} Suppose that $M$ is an irreducible Metzler matrix such that $s(M) = 0$. Then there exists a positive diagonal matrix $P$ such that $M^T P + P M \preccurlyeq 0$.~$\blacksquare$ \end{lem}
\begin{lem} \label{lem:eigspec}
\cite[Proposition~1]{liu2019analysis} Suppose that $\Lambda$ is a negative diagonal matrix and $N$ is an irreducible nonnegative matrix. 
%Let $M = \Lambda+N$.
Let $M$ be the irreducible Metzler matrix $M = \Lambda+N$. 
Then, $s(M) < 0$ if and only if $\rho(-\Lambda^{-1} N) < 1, s(M)=0$ if and only if $\rho(-\Lambda^{-1} N) = 1$, and $s(M)>0$ if and only if, $\rho(-\Lambda^{-1} N) > 1$.~$\blacksquare$
\end{lem}
We will also be making use of the following variants of the Perron-Frobenius theorem for irreducible matrices.
\begin{lem} \label{lem:perron_frob}
\cite[Chapter 8.3]{meyer2000matrix} \cite[Theorem~2.7]{varga1999matrix} %(Perron-Frobenius Theorem) 
Suppose that $N$ is an irreducible nonnegative matrix. Then,
\begin{enumerate}[label=(\roman*)]
    \item $r = \rho(N)$ is a simple eigenvalue of $N$. \label{item:perfrob_simpleeig}
    \item There is an eigenvector $\zeta \gg \textbf{0}$ corresponding to the eigenvalue $r$. \label{item:perfrob_pos_exists}
    \item $x > \textbf{0}$ is an eigenvector of $N$ only if $Nx = rx$ and $x \gg \textbf{0}$. %\axel{\textbf{[Not sure this needs to be mentioned]}} 
    \label{item:perfrob_pos_necess}
    \item If $A$ is a nonnegative matrix such that $A < N$, then $\rho(A) < \rho(N)$. \label{item:perfrob_matrix_ineq}~$\blacksquare$
\end{enumerate}
\end{lem}
%
%\seb{[Comment: Double-check whether we are indeed using all 4 items in Lemma~\ref{lem:perron_frob}]} 
\begin{lem} \label{lem:perron_frob_metz}
\cite[Lemma~2.3]{varga1999matrix} Suppose that $M$ is an irreducible Metzler matrix. Then $r = s(M)$ is a simple eigenvalue of $M$, and a corresponding eigenvector is $\zeta \gg \textbf{0}$.~$\blacksquare$
\end{lem}
%\seb{[Lemma~\ref{lem:perron_frob_metz} concerns Metzler matrices, so maybe we could have it immediately after Lemma~\ref{lem:eigspec}]} 
%
%In the sequel, we will be interested in the stability and domains of attraction of equilibria of system~\eqref{eq:full}. The following results will be relevant for this analysis.\\
%
\noindent Let $\mathbb{G} \subseteq \mathbb{R}^n$. Consider 
\vspace{-1ex}
\begin{equation} \label{eq:genericsys}
    \dot{x}(t) = f(x),
\end{equation}

\vspace{-1ex}

\noindent where $f : \mathbb{G} \rightarrow \mathbb{R}^n$ is a locally Lipschitz map.
%
% \axel{\textbf{[Should this be a corollary?]}}
\begin{prop} \label{prop:asymp_domatt}
Let $\textbf{0}$ be an equilibrium of~\eqref{eq:genericsys} and $\mathbb{E} \subseteq \mathbb{G}$ be a positively invariant and connected set with respect to~\eqref{eq:genericsys}, containing $\textbf{0}$. Let $V: \mathbb{E} \rightarrow \mathbb{R}$ be a continuously differentiable and positive definite function, such that $\dot{V}(x(t))$ is negative definite. %, for all $x \in \mathbb{E}$. 
Further, let it hold that
\begin{equation} \label{eq:bounded_contours}
\{ x : V(x) \leq c\} \cap \mathbb{E}
\end{equation}
is a bounded set for any constant $c > 0$, and is equal to $\mathbb{E}$ as $c \rightarrow \infty$. Then the equilibrium $\textbf{0}$ is asymptotically stable, with domain of attraction containing $\mathbb{E}$.~$\blacksquare$
\end{prop}
This proposition can be proven using \cite[Theorem~4.1]{khalil2002nonlinear}. % and the discussion on pages~122-123,~316-322 in \cite{khalil2002nonlinear}.
% %
% \begin{prop} \label{prop:expo_domatt}
% Let $\textbf{0}$ be an equilibrium of~\eqref{eq:genericsys} and $\mathbb{E} \subset \mathbb{G}$ be a positively invariant and connected set with respect to~\eqref{eq:genericsys}, containing $\textbf{0}$. Let $V: \mathbb{E} \rightarrow \mathbb{R}$ be a continuously differentiable function such that %\seb{[the spacing is weird, use align environment.]} %\axel{\textbf{[Ok, in align now]}}:
% %
% \begin{align*}
% k_1\|x\|^a \leq & \, V(x) \leq k_2 \|x\|^a, \\
% \dot{V}(x) &\leq -k_3 \|x\|^a,
% \end{align*}
% % 
% for all $x \in \mathbb{E}$, where $k_1, k_2, k_3$ and $a$ are positive constants. Then the equilibrium $\textbf{0}$ is exponentially stable, with domain of attraction containing $\mathbb{E}$. \hfill $\blacksquare$
% \end{prop}
% %
% This proposition can be proven using \cite[Theorem~4.10]{khalil2002nonlinear} and the discussion on pages~122, 317-320 in \cite{khalil2002nonlinear}.
%
%\axel{\textbf{[Proposition~\ref{prop:asymp_domatt} has been formulated by me and it is proven with a handwave. Here I am less certain that the statement is correct as written. That is, the way it is used in this paper should be legit, but it might be necessary to add more qualifying conditions to the statement.]}} \seb{[Response: Could you check whether (or not) Prop.~\ref{prop:asymp_domatt} as stated currently can indeed be proven? If not, please see how it could be modified. Then check whether the statement (and proof) of Theorem~\ref{thm:asymp} is correct.]}\\
%
The following lemma pertains to system~\eqref{eq:full}, providing a constraint on any endemic equilibrium. %of system~\eqref{eq:full}. 
%
%\seb{COMMENT: I think moving Lemmas~\ref{lem:equi_non-zero_nonone} and~\ref{lem:max_not_shared} to the Preliminaries section could improve readability. Thoughts?} \axel{\textbf{Would be more balanced.}}
%
\begin{lem} \label{lem:equi_non-zero_nonone}
Consider system~\eqref{eq:full} under Assumption~\ref{assum:base}. Suppose, for all $k \in [m]$, that $B_w^k$ is irreducible. If $y = (y^1, \dots, y^m) \in \mathcal{D}$ is an equilibrium of~\eqref{eq:full}, then, for each $k \in [m]$, either $y^k = \textbf{0}$, or $\textbf{0} \ll y^k \ll \textbf{1}$. Moreover, %we have that 
$\textstyle \sum_{k=1}^m y^k \ll \textbf{1}$.~$\blacksquare$
\end{lem}
 \textit{Proof:} See Appendix.~$\square$ %Consider an equilibrium $y = (y^1, y^2, \dots, y^m) \in \mathcal{D}$ of system~\eqref{eq:full}. 
Lemma~\ref{lem:equi_non-zero_nonone} states that when the underlying network is strongly connected, any endemic equilibrium involves each active virus infecting a separate fraction of each population node, and contaminating the shared resource to some degree. 

\section{Analysis of the eradicated state of a virus}\label{sect:stab:analysis:healthystate}
%\seb{[Explain in brief (4-5 lines) as to what this section is all about. Also, this section is just way too big, may have to break it down into smaller sections so as to improve readability. ]}
%
In this section, we present
sufficient conditions for the exponential (resp. asymptotic)
%two results related to the
stability of the eradicated state of a virus. %\axel{\textbf{[Could use a lead-in here. I have the following suggestion.]}}
The key condition is found through eigenvalue analysis of $(B_w^k - D_w^k)$, as seen in the following theorem.
%In this section, we present a number of analytical results on the existence and stability of equilibria of system~\eqref{eq:full}. First, we establish conditions under which a virus will converge to its eradicated state. Next we show that violating the above-mentioned conditions implies existence of a unique single-virus endemic equilibrium, which is also asymptotically stable with an appropriately defined domain of attraction. Finally, we provide novel conditions for the existence of coexisting equilibria in a bi-virus system, as well as conditions under which one virus dominates the other, so that no coexistence is possible.
%as well as conditions precluding the existence of coexisting equilibria in a system.
%
%
%
%\label{sect:stab:analysis:healthystate}
% \subsection*{Eradication of a virus} \label{subsec:erad}
%
%\axel{\textbf{In this subsection, some transitions are still plagiarized from our previous submission to CPHS.}} \seb{[Response: We should try to rephrase as much as we can]}
%
%\seb{Comment: Already done in \cite{axel2020multi}; see corresponding results.\\}
%The first result is a sufficient condition for exponential stability of the %\seb{[a virus is either active in the population, or it has been eradicated. That is, it cannot have \emph{multiple} eradicated states. Hence, it should be \enquote{the} instead of \enquote{an}.]} 
%eradicated state of a virus, as seen in the following theorem, while the second result is a sufficient condition for asymptotic stability of the eradicated state of a virus. %\seb{[what about the second result?]}.
%
\begin{thm} \label{thm:expo}
Consider the SIWS model~\eqref{eq:yk} under Assumption~\ref{assum:base} 
with
% and assume that 
$y(0) \in \mathcal{D}$. Suppose that for some virus~$k \in [m]$ we have $s(B_w^k - D_w^k) < 0$. Then the eradicated state of virus~$k$ is exponentially stable, with domain of attraction containing~$\mathcal{D}^k$.~$\blacksquare$
% \hfill $\blacksquare$
\end{thm}
%
%\textcolor{purple}{The proof is straightforward, and follows by observing that \eqref{eq:yk} is bounded from above by a linear system, which is exponentially stable whenever $s(B_w^k - D_w^k) < 0$. Then, applying Grönwall-Bellman's Inequality \cite[pg 651]{khalil2002nonlinear}, the proof is concluded.}
%\textit{Proof:} \seb{See proof of Theorem~1 in REF.}
%See Appendix.~$\square$
%
%\axel{\textbf{[This is a new proof of Theorem~\ref{thm:expo}, the previous proof is hidden. This proof removes the need for Proposition~\ref{prop:expo_domatt} and Lemma~\ref{lem:metz}. Let me know if the proof is unclear.]}}
%
\textit{Proof:} See Appendix.~$\square$ 
Theorem~\ref{thm:expo} states that if %the largest real part of the eigenvalue of 
the linearized state matrix of virus~$k$ %(linearized around the eradicated state of virus~$k$) 
is Hurwitz, then, for all initial conditions in the sensible domain, virus~$k$ is eradicated exponentially fast. %\axel{\textbf{[I prefer the Hurwitz term for brevity. Though I guess that would clash with the explanation of Theorem~\ref{thm:asymp}.]}}
Theorem~\ref{thm:expo} answers the first part of question~\ref{q1} in Section~\ref{subsec:pb:statement}. 
% \phil{Observe that while the statements of Theorem~\ref{thm:expo} and \cite[Theorem~5]{axel2020multi} are the same; the proofs are different. [Is this necessary to point out since we already state that \cite{axel2020multi} is a preliminary version of this paper? I think we should delete it]} %of Theorem~\ref{thm:expo}
% %is different}
% %from that of \cite[Theorem~5]{axel2020multi}} 
With respect to \cite[Proposition~2]{liu2019analysis}, Theorem~\ref{thm:expo} is an improvement in the sense that it %following senses: i) Theorem~\ref{thm:expo} 
holds globally (on the sensible domain), and accounts for the multi-virus case, whereas %, for the same condition as in Theorem~\ref{thm:expo} but particularized for $m=1$, 
\cite[Proposition~2]{liu2019analysis} established \emph{local} exponential stability for the \emph{single-virus} case.%; \textcolor{orange}{and ii) in terms of local exponential stability of the eradicated state of a virus,  Theorem~\ref{thm:expo} accounts for the multi-virus case, while \cite[Proposition~2]{liu2019analysis} only accounts for the single-virus case.} % only locally,whereas Theorem~\ref{thm:expo} holds globally, on the sensible domain. Hence, Theorem~\ref{thm:expo} is a stronger version of \cite[Proposition~2]{liu2019analysis} when $m=1$, and more general since it applies to the competing virus case.
%In particular, Theorem~\ref{thm:expo} says that insofar that the matrix $(B_w^k - D_w^k)$ is Hurwitz, then, irrespective of the initial condition of the network,
% with respect to virus~$k$, virus~$k$ is eradicated exponentially fast.

%Observe that while
Theorem~\ref{thm:expo} indeed guarantees exponential eradication of virus~$k$, however, the condition is quite strict. For certain viruses it suffices to know whether or not the virus will be eradicated, but the \emph{speed} with which this eradication takes place is of less importance.
Indeed, it turns out that a relaxation of the strict inequality of the eigenvalue condition in Theorem~\ref{thm:expo} guarantees asymptotic eradication of a virus, as stated in the following theorem.

%This motivates %It is, therefore, pertinent
%the question: whether (or not) eradication of virus~$k$ can be achieved even if the condition in Theorem~\ref{thm:expo} were to be relaxed? \axel{\textbf{[This question is now redundant with question~\ref{q1} in Section~\ref{subsec:pb:statement}.]}} In such a case, one would naturally expect the speed of eradication to decrease. The following theorem provides a sufficient condition for asymptotic eradication of virus~$k$. 
%Next we present a result on sufficient conditions for asymptotic stability of an eradicated state.
% 
\begin{thm} \label{thm:asymp}
Consider the SIWS model~\eqref{eq:yk} under Assumption~\ref{assum:base} with $y(0) \in \mathcal{D}$. Suppose that for some virus~$k \in [m]$ we have $s(B_w^k - D_w^k) \leq 0$, and that the matrix $B_w^k$ is irreducible. Then the eradicated state of virus~$k$ is asymptotically stable, with domain of attraction containing $\mathcal{D}^k$.~$\blacksquare$
\end{thm}
\textit{Proof:} See Appendix.~$\square$\\
Assuming that the $k^{\rm{th}}$ layer of the graph is strongly connected, Theorem~\ref{thm:asymp} states that if the largest real part of any eigenvalue of the linearized state matrix of virus~$k$ %\phil{[Comment: I would still include "linearized around the eradicated state of virus~$k$)". since it is not that obvious that here we are linearizing around the eradicated state of a virus]} \textcolor{purple}{[I think it is sufficient information for this simplified explanation, the location of the linearization is apparent from the theorem statement.]} %(linearized around the eradicated state of virus~$k$) 
is non-positive, then, for all initial conditions in the sensible domain, virus~$k$ is eradicated asymptotically. Theorem~\ref{thm:asymp} %provides and
answers the second part of question~\ref{q1} in Section~\ref{subsec:pb:statement}. 
Observe that  
% it improves upon \cite[Theorem~6]{axel2020multi} \axel{\textbf{[Should we be talking about the CPHS paper in the main body like this?]}} \seb{[Response: if we keep the part on differences wrt \cite{axel2020multi}, then yes; otherwise, no.]} by relaxing the requirement that $\beta_{iw}^k > 0$, $c_i^k > 0$ for all $i \in [n], k \in [m]$. Similarly, 
for the single-virus case ($m=1$), Theorem~\ref{thm:asymp} 
% also 
improves \cite[Theorem~1]{liu2019networked} by relaxing the requirements $\beta_{iw} > 0$ and $c_i > 0$ for all $i \in [n]$.

\begin{rem}[Epidemiological Interpretation]
\label{rem:asymp:stab}
Observe that, due to Lemma~\ref{lem:eigspec}, the conditions in Theorem~\ref{thm:expo} (Theorem~\ref{thm:asymp})
%Note that by applying Lemma~\ref{lem:eigspec} to the conditions in Theorem~\ref{thm:asymp}, we see that $s(B_w^k - D_w^k) \leq 0$
are equivalent to $\rho((D_w^k)^{-1} B_w^k) < 1$ ($\rho((D_w^k)^{-1} B_w^k) \leq 1$).
The fact that Theorem~\ref{thm:expo} (resp. Theorem~\ref{thm:asymp}) guarantees exponential (asymptotic) eradication of a virus~$k$ whenever $\rho((D_w^k)^{-1} B_w^k) < 1$ ($\rho((D_w^k)^{-1} B_w^k) \leq 1$) is consistent with an interpretation of $\rho((D_w^k)^{-1} B_w^k)$ as the basic reproduction number of the virus in the network, typically denoted by $\mathcal{R}_0$. In epidemiology, this is the average number of secondary infections caused by an infected individual, before recovering. %can be expected to pass on the infection to, before they recover.
%we can apply Lemma~\ref{lem:eigspec} to the statement of Theorem~\ref{thm:asymp}. Then the condition $s(B_w - D_w) \leq 0$ is equivalent to $\rho(D_w^{-1} B_w) \leq~1$. 
Thus, Theorem~\ref{thm:expo} (Theorem~\ref{thm:asymp}) states that whenever the basic reproduction number of a virus %in this network
is strictly less than (less than or equal to) one, the virus will exponentially (asymptotically) converge to its eradicated state.~$\blacksquare$
\end{rem}%\axel{\textbf{[Something about "this is expected from epidemiology".]}} \seb{[Response: not sure what you mean..]} %\seb{[Note: We should add a similar interpretation for Theorem~\ref{thm:expo} as well.]} \axel{\textbf{[The only problem is that Lemma~\ref{lem:eigspec} can not be applied right off the bat to Theorem~\ref{thm:expo}.]}}\seb{[Response: Is it because of the irreducibility assumption? ]} \axel{\textbf{[Yes.]}}
%is said to be irreducible if, when it is interpreted as an adjacency matrix of a graph, the graph is strongly connected, and vice-versa.
% \subsection{Persistence of a virus} \label{subsec:persist}
\section{Persistence of a virus}\label{sec:persist}
In this section, we study the possibility of viruses persisting in the network, corresponding to non-zero equilibria of~\eqref{eq:full}. Naturally, the persistence of a virus must follow from the violation of
%\phil{[this what?]} 
the conditions of Theorem~\ref{thm:asymp}. %The following proposition states an important consequence of %violating the eigenvalue condition in Theorem~\ref{thm:asymp}
The resulting behavior is detailed in the rest of this section. %\axel{the sequel} \seb{Response: I use the word "sequel" to indicate that the rest of the paper does one and only one thing -- not the case here. }.
Before we proceed, we state the following result:

\begin{lem} \label{lem:pos_never_zero}
Let Assumption~\ref{assum:base} hold. Then $\mathcal{D} \setminus \{ \textbf{0} \}$ is positively invariant with respect to system~\eqref{eq:full}.~$\blacksquare$
\end{lem}
 \textit{Proof:} See Appendix.~$\square$
 %Lemma~\ref{lem:pos} states that $\mathcal{D}$ is positively invariant with respect to system~\eqref{eq:full}, so it remains to be shown that if $y(0) \in \mathcal{D}$ and $y(0) > \textbf{0}$, then $y(t) > \textbf{0}$ for all $t > 0$. We will prove this by bounding $y(t)$ from below. Since $y(0) > \textbf{0}$, $y^k(0) > \textbf{0}$ for some virus~$k \in [m]$, and by Lemma~\ref{lem:pos}, $y^k(t) \geq \textbf{0}$ for all $t > 0$. Considering such a virus~$k$, note that~\eqref{eq:yk} gives us
% %
% \begin{align}
% \dot{y}^k(t) &= \big{(} - D_w^k + (I - \textstyle \sum_{l=1}^m X(y^l(t)))B_w^k \big{)} y^k(t) \nonumber \\
% &\geq -D_w^k y^k(t), \label{eq:lower_bound_derivative}
% \end{align}
% %
% where~\eqref{eq:lower_bound_derivative} follows from the fact that $y^k(t) \geq \textbf{0}$ for all $t > 0$, and that $(I - \textstyle \sum_{l=1}^m X(y^l(t)))B_w^k$ is a nonnegative matrix. Note that~\eqref{eq:lower_bound_derivative} can be integrated directly, yielding:
% %
% \begin{equation} \label{eq:lower_bound}
% y^k(t) \geq e^{- t D_w^k } y^k(0) > \textbf{0}, \text{ for all } t>0. 
% \end{equation}
% %
% The final inequality in~\eqref{eq:lower_bound} holds due to the properties of exponential functions, and $D_w^k$ being a positive diagonal matrix. Hence, if $y(0) > \textbf{0}$, $y(t) > \textbf{0}$ for all $t>0$, so $\mathcal{D} \setminus \{ \textbf{0} \}$ is positively invariant with respect to system~\eqref{eq:full}. \hfill $\square$

With Lemma~\ref{lem:pos_never_zero} in place, we have the following theorem for a single-virus system, guaranteeing existence of a unique, asymptotically stable, single-virus endemic equilibrium when the eigenvalue condition in Theorem~\ref{thm:asymp} is violated.% in \seb{for} the single-virus case.
\begin{thm} \label{thm:equi}
Consider the SIWS model~\eqref{eq:yk} under Assumption~\ref{assum:base} with $m=1$. Suppose that $B_w$ is irreducible and $s(B_w - D_w) > 0$. Then there exists a unique single-virus endemic equilibrium $\Tilde{y} \in \mathcal{D}$, with $\textbf{0} \ll \Tilde{y} \ll \textbf{1}$, and it is asymptotically stable with domain of attraction containing $\mathcal{D} \setminus \{ \textbf{0} \}$.~$\blacksquare$
% Let $m=1$ and consider~\eqref{eq:yk}. Let Assumption~\ref{assum:base} hold and assume that $y^1(0) \in \mathcal{D}^1$. Suppose that $B_w^1$ is irreducible and $s(B_w^1 - D_w^1) > 0$. Then there exists at least one non-zero equilibrium in $\mathcal{D}^1$.
\end{thm}
%
%\noindent This proof is inspired by the proof of \cite[Theorem~2..4.]{fall2007epidemiological}. 
%\phil{[this structure is a bit weird because we start with a proof environment in which we say we're going to split the proof into three parts but never close the initial proof environment. So it appears that Lemma 8 is inside the proof environmental while it's not really. I think this can be remedied by removing `\textit{Proof:}' from this paragraph or changing it to `\textit{Proof Outline:}']}. 
\textit{Proof:} See Appendix.~$\square$
For a single-virus system, Theorem~\ref{thm:equi} states that when the eigenvalue condition in Theorem~\ref{thm:asymp} is violated, then as long as some viral infection is present initially, the viral spreading process will converge to a unique infection ratio in each population node and a unique contamination level in the shared resource. This answers question~\ref{q2} in Section~\ref{subsec:pb:statement}. %for the single-virus setting.
\begin{rem}[Epidemiological Interpretation] \label{rem:endemic:equi}
Applying Lemma~\ref{lem:eigspec}, we see that the conditions of Theorem~\ref{thm:equi} are equivalent to $\rho((D_w)^{-1}B_w) > 1$. This is again consistent with the interpretation of $\rho((D_w)^{-1}B_w)$ as the basic reproduction number of the virus, since a persisting virus should have a basic reproduction number greater than one.~$\blacksquare$
\end{rem}
Theorem~\ref{thm:equi} establishes the existence, uniqueness and asymptotic stability of a single-virus endemic equilibrium, extending \cite[Theorem~2.4.]{fall2007epidemiological} %\seb{[check for possible typo - ''2..4'']} \axel{\textbf{[This is how the notation looks in \cite{fall2007epidemiological}.]}}
to the setting with a shared resource, whereas \cite{liu2019networked} 
%did not provide any theoretical guarantees regarding the non-zero equilibrium\phil{, 
illustrated this extension in simulations, without providing theoretical guarantees. %\seb{[Response: the phrasing could be improved]}
%[I wanted this to be a bit more precise... please clean it up]} 
%Moreover, Theorem~\ref{thm:equi}, in contrast to \cite[Theorem~8]{axel2020multi}, establishes uniqueness and asymptotic stability of the single-virus endemic equilibrium. 
%\seb{[TBD: compare and contrast Theorem~\ref{thm:equi} with \cite[Theorem~2]{liu2019analysis}]} 
We can partially extend Theorem~\ref{thm:equi} to the multi-virus case, specifically the existence and uniqueness of a single-virus endemic equilibrium for a virus, resulting in the following proposition.
%
%\axel{\textbf{[This proposition should state the sets in which each equilibrium is stable as well. As an extension, if only one virus is viable, there should be a result stating that the single-virus endemic equilibrium of the viable virus is "globally" asymptotically stable.]}}
%
\begin{prop} \label{prop:necessity}
Consider the SIWS model~\eqref{eq:full} under Assumption~\ref{assum:base}. For each $k \in [m]$ such that $B_w^k$ is irreducible and $s(B_w^k - D_w^k) > 0$, there is a unique single-virus endemic equilibrium $(\textbf{0}, \dots, \Tilde{y}^k, \dots, \textbf{0})$ in $\mathcal{D}$, with $\textbf{0} \ll \Tilde{y}^k \ll \textbf{1}$.~$\blacksquare$
\end{prop}
\textit{Proof:} See Appendix.~$\square$
% \textcolor{purple}{The proof is straightforward, following immediately by noting that the multi-virus system \eqref{eq:yk} is equivalent to a single-virus system whenever all other viruses are nullified. \sebcancel{[Comment: this explanation might not be easily understood. I am adding the proof in the Appendix.]}}
%\textit{Proof:} \seb{See proof of Proposition~2 in REF.} 
%\seb{Follows along simlar lines as that of, for instance, \cite{pareautomatica}; see arxiv for details.}%See Appendix.~$\square$
%Suppose that, for some $k \in [m]$, $B_w^k$ is irreducible and $s(B_w^k - D_w^k) > 0$, and that $y^l = \textbf{0}$ for all $l \in [m]$, $l \neq k$. Then the dynamics of virus~$k$ can be written as:
% %
% \begin{align} \label{eq:yk_new}
%  \dot{y}^k(t) =\big{(} - D_w^k + (I - X(y^k(t)))B_w^k \big{)}y^k(t).
% \end{align}
% %
% Note that~\eqref{eq:yk_new} corresponds to the dynamics of the single-virus case. Therefore, since $B_w^k$ is irreducible and $s(B_w^k - D_w^k) > 0$, it follows from the first and second parts of the proof of Theorem~\ref{thm:equi} that there exists a unique single-virus endemic equilibrium of the form $(\textbf{0}, \dots, \Tilde{y}^k, \dots, \textbf{0})$, with $\textbf{0} \ll \Tilde{y}^k \ll \textbf{1}$ in $\mathcal{D}$. This holds for each $k \in [m]$ such that $B_w^k$ is irreducible and $s(B_w^k - D_w^k) > 0$, by repeating the arguments above. \hfill $\square$

Proposition~\ref{prop:necessity} states that each virus violating the eigenvalue condition in Theorem~\ref{thm:asymp} has a unique single-virus endemic equilibrium. % answering question~\ref{q2} in Section~\ref{subsec:pb:statement} for the multi-virus setting.Commented by Sebin
This result is unsurprising, since nullifying the other viruses in the system reduces it to a single-virus system. However, note that Proposition~\ref{prop:necessity} does not say anything about the stability of these equilibria. Nor does the proposition say anything about the case of more than one virus persisting in the population simultaneously. %\seb{[I suggest moving the comment(s) regarding stability of these equilibria to the simulation section.]}
Due to Theorem~\ref{thm:asymp} and Proposition~\ref{prop:necessity}, we have the following necessary and sufficient condition for the healthy state to be the unique equilibrium.%system~\eqref{eq:full}.
%
%\axel{\textbf{[Might add asymptotic stability to this theorem as well. Though I am uncertain how to phrase it.]}}
%
\begin{thm}\label{thm:uniqueness:healthy:state}
Consider the SIWS model~\eqref{eq:full} under Assumption~\ref{assum:base}. Suppose that, for all $k \in [m]$, $B_w^k$ is irreducible. Then the healthy state is the unique equilibrium in $\mathcal{D}$ if, and only if, for all $k \in [m]$, $s(B_w^k - D_w^k) \leq 0$.~$\blacksquare$
\end{thm}
%
%
%\axel{\textit{Proof:} Suppose, for all $k \in [m]$, that $B_w^k$ is irreducible. Note that if $s(B_w^k - D_w^k) \leq 0$ for all $k \in [m]$, the conditions for Theorem~\ref{thm:asymp} are met for all viruses. Then, any equilibrium has to have $y^k = \textbf{0}$ for all $k \in [m]$. Hence, the healthy state $y=\textbf{0}$ is the unique equilibrium in $\mathcal{D}$. Should $s(B_w^k - D_w^k) > 0$ for any $k \in [m]$, then Proposition~\ref{prop:necessity} guarantees the existence of a nonzero equilibrium in $\mathcal{D}$. Hence, the healthy state $y=\textbf{0}$ is the unique equilibrium in $\mathcal{D}$ only if $s(B_w^k - D_w^k) \leq 0$}
% \seb{
\noindent
Theorem~\ref{thm:uniqueness:healthy:state} states that as long as the largest real part of the eigenvalue of the linearized state matrix of \emph{each} virus is non-positive, the healthy state is the only equilibrium of~\eqref{eq:full}.
% }
Theorem~\ref{thm:uniqueness:healthy:state} answers question~\ref{q3} in Section~\ref{subsec:pb:statement}. 
% We see \seb{[Notice]} 
Note that Theorem~\ref{thm:uniqueness:healthy:state} extends \cite[Theorem~1]{liu2019analysis} to the setting with more than two viruses and a shared resource, 
% \seb{[
albeit under the assumption that the healing rate of each agent with respect to each virus is strictly positive.
% ]}
% \phil{except for the case when [this wording is a bit awkward...]} $\delta_i^k = 0$ for some $i \in [n]$ and $k \in [m]$.
%\begin{rem}[Epidemiological Interpretation]\label{rem:uniqueness:healthy:equi}
%In an epidemiological context, %the condition in 
%Theorem~\ref{thm:uniqueness:healthy:state} may be understood as follows: Insofar the basic reproduction number of each virus is no greater than one, all viruses become eradicated asymptotically, thereby leading  %\seb{all the agents} \axel{[the system]} 
%the population to the healthy state.~$\blacksquare$

%\end{rem}

% \seb{[Comment: I don't follow this paragraph.]}
%\seb{[Comment: Theorem~\ref{thm:uniqueness:healthy:state} is immediate from the preceding results, so no need for proof. Am I missing something?]} \axel{no, it is immediate}
%We now explore the case of more than one virus persisting in the population.

% \subsection{Coexistence of viruses} \label{subsec:coexist}
\section{Coexistence of viruses}\label{sect:coexist}

Beyond the single-virus endemic equilibria from 
% subsection 
Section~\ref{sec:persist}, we would like to know when multiple viruses can persist in a population simultaneously, corresponding to a coexisting equilibrium. %of system~\eqref{eq:full}. 
The following theorem makes use of a particular %\axel{novel} \seb{[Response: I am not sure about the novelty here, since I am guessing for the two-node case this condition has already been established in [28], is it not?]}
eigenvalue condition to show the existence of a coexisting %coexistence
equilibrium in a bi-virus system. Before proceeding to the statement of the theorem, recall from Proposition~\ref{prop:necessity} that, if i) $B_w^1$, $B_w^2$ are irreducible, ii) $s(B_w^1-D_w^1) > 0$, and iii) $s(B_w^2-D_w^2) > 0$, then there are exactly two single-virus endemic equilibria, namely
\begin{equation}\label{eq:endemicm2}
    (\Tilde{y}^1, \textbf{0}) \text{ and } (\textbf{0}, \Tilde{y}^2). 
\end{equation}
Moreover, we know that $\textbf{0} \ll \Tilde{y}^1 \ll \textbf{1}$ and $\textbf{0} \ll \Tilde{y}^2 \ll \textbf{1}$.

\begin{thm} \label{thm:joint_eq_exist_shared}
Consider the SIWS model~\eqref{eq:full} under Assumption~\ref{assum:base} with $m=2$. Suppose that $B_w^1$ and $B_w^2$ are irreducible matrices,
%\seb{[since we have Assumption~1 in the statement, and Assumption~1 implies that $D_w^k$ are positive diagonal matrices for $k=1,2$, we do not need to \enquote{suppose} this part ]}
and that $s(B_w^1 - D_w^1) > 0$ and $s(B_w^2 - D_w^2) > 0$. 
% \phil{
% By Proposition~\ref{prop:necessity} there exist two non-zero equilibria, namely $(\Tilde{y}^1, \textbf{0})$ and $(\textbf{0}, \Tilde{y}^2)$, such that $\textbf{0} \ll \Tilde{y}^1 \ll \textbf{1}$ and $\textbf{0} \ll \Tilde{y}^2 \ll \textbf{1}$
% [this is a bit weird to have inside a theorem statement. It would be better to state this right before and reference it here; I've taken a stab it this. If you like it feel free to comment out this sentence].} 
If $s(- D_w^1 + (I - X(\Tilde{y}^2))B_w^1) > 0$ and $s(- D_w^2 + (I - X(\Tilde{y}^1))B_w^2) > 0$, with $\Tilde{y}^1,\Tilde{y}^2$ defined in~\eqref{eq:endemicm2}, %\phil{[. Then]} \textcolor{purple}{[]Doesn't make sense while the sentence starts with "If".]}
then there exists at least one coexisting equilibrium $(\hat{y}^1, \hat{y}^2) \gg \textbf{0}$ in $\mathcal{D}$ such that $\hat{y}^1 + \hat{y}^2 \leq \textbf{1}$.~$\blacksquare$
\end{thm}

\textit{Proof:} See Appendix.~$\square$
\noindent
With each virus satisfying the condition for the existence of its single-virus endemic equilibrium, %being at its single-virus endemic equilibrium, %\axel{\textbf{[This is a bit unclear in terms of whether we are at both equilibria at once.]}} 
% and the state matrix of each virus being linearized around the single-virus endemic equilibrium of the other virus, 
Theorem~\ref{thm:joint_eq_exist_shared} states that if, for each virus, the largest real part of any eigenvalue of the matrix of the dynamics linearized around the single-virus endemic equilibrium of the other virus
% \axel{\textbf{[There are a lot of "of" here.]}} \seb{Response: I agree, but I don't have a better solution}
% linearized state matrix of each virus 
is positive, then both the viruses can \emph{simultaneously} infect separate fractions of each population node. %subpopulations of a given population. 
Theorem~\ref{thm:joint_eq_exist_shared} answers question~\ref{q4} in Section~\ref{subsec:pb:statement}.% for the bi-virus setting.
% }
%We would like to note that the proof is novel and non-trivial, but has been moved to the appendix to improve readability of this section. %Theorem~\ref{thm:joint_eq_exist_shared} answers question~\ref{q4} in Section~\ref{subsec:pb:statement} for the bi-virus setting. 
%In an epidemiological context, Theorem~\ref{thm:joint_eq_exist_shared} may be interpreted as follows: Supposing that the conditions in Theorem~\ref{thm:joint_eq_exist_shared} are satisfied, then neither virus is able to overwhelm the other, and, therefore, must have some balanced state of coexistence where both viruses infect different groups of all population nodes, and contaminate the shared resource, to some degree.

\begin{rem} [Epidemiological Interpretation] \label{rem:coexist:equi}
% In an epidemiological context, Theorem~\ref{thm:joint_eq_exist_shared} may be interpreted as follows: 
% \textcolor{orange}{Supposing that the conditions in Theorem~\ref{thm:joint_eq_exist_shared} are satisfied, then neither virus is able to overwhelm the other, and, therefore, must have some balanced state of coexistence where both viruses infect separate fractions of each population node, and contaminate the shared resource, to some degree.
% Furthermore,} 
% a
Applying Lemma~\ref{lem:eigspec} to the conditions in Theorem~\ref{thm:joint_eq_exist_shared}, we see that they are equivalent to having $\rho((I-X(\Tilde{y}^2))(D_w^1)^{-1}B_w^1)>1$ and $\rho((I-X(\Tilde{y}^1))(D_w^2)^{-1}B_w^2)>1$. This is consistent with an interpretation of $\rho((I-X(\Tilde{y}^2))(D_w^1)^{-1}B_w^1)$ and $\rho((I-X(\Tilde{y}^1))(D_w^2)^{-1}B_w^2)$ as the invasion reproduction numbers of virus~1 invading virus~2 and virus~2 invading virus~1, respectively. The invasion reproduction number is defined for an invading pathogen, introduced into a setting with another, endemic pathogen at equilibrium. It is defined as the average number of secondary 
% \seb{[this term is potentially confusing; the reader might think that there is a possibility of co-infection here, which contradicts the competitive behavior.. thoughts?]} \axel{\textbf{[I am not sure, lets see what Phil says.]}} 
% \phil{[In the context of Theorem~\ref{thm:joint_eq_exist_shared}, I think it's okay]}
infections caused by an individual infected by the invading pathogen, at the time of introduction \cite{porco1998designing}. In line with this interpretation, Theorem~\ref{thm:joint_eq_exist_shared} shows that coexistence is possible whenever both invasion reproduction numbers are greater than one.~$\blacksquare$
\end{rem}
%An interpretation of this result is that two competing viruses, where neither virus is able to overwhelm the other, must have some balanced state of coexistence where both viruses infect all population nodes, and contaminate the shared resource, to some degree. 
%
While conditions that guarantee existence of coexisting equilibria may be found in \cite{li2004coexist} %\cite{liu2019analysis}
and \cite{pareautomatica}, these references do not account for the presence of a shared resource.
% Similar results on the existence of coexisting equilibria %can
% \seb{may} be found in \cite{li2004coexist}, \cite{liu2019analysis} and \cite{pareautomatica}. However, the \seb{relevant} results therein do not account for a shared resource.
% \textcolor{orange}{and furthermore suffer from the following limitations: \cite[Theorem~5.2]{li2004coexist} accounts only for a two-node network; \cite[Theorem~7]{liu2019analysis} considers only
% homogeneous spread; and \cite[Theorem~6]{pareautomatica} requires that the spread parameters of each virus be a scaled version of those of the remaining viruses.} %\phil{other limitations. [this is vague]}]} only for the case without a shared resource.
In order to compare %these 
the results in \cite{li2004coexist,pareautomatica} with Theorem~\ref{thm:joint_eq_exist_shared}, we particularize our model to the setting without a shared resource. %, in line with Remark~\ref{rem:no_W}. 
Specifically, % for multi-competitive SIS models, %with no shared resource, 
\eqref{eq:yk} reduces to
\vspace{-1ex}
\begin{equation} \label{eq:nosharedsys}
\dot{p}^k(t) = %& \,
\Big{(} - D^k + \big{(} I - \textstyle \sum_{l=1}^m \diag(p^l(t)) \big{)} B^k \Big{)} p^k(t).
\end{equation} 

\vspace{-1ex}

\noindent Furthermore, we can let $p(t) \coloneqq [p^1(t),  \dots , p^m(t)]^T$. Then, with $A^k(p(t)) \coloneqq \big{(}- D^k + (I - \textstyle \sum_{l=1}^m \diag(p^l(t)))B^k \big{)}$, the dynamics of $p(t)$ are given by
\begin{gather} \label{eq:fullnosharedsys}
 \dot{p}(t)
 = 
 \begin{bmatrix}
 A^1 \big{(} p(t) \big{)} & 0 & \dots & 0 \\
 0 & A^2 \big{(} p(t) \big{)} & \dots & 0 \\
 \vdots & \vdots & \ddots & \vdots \\
 0 & 0 & \dots & A^m \big{(} p(t) \big{)}
 \end{bmatrix}
 p(t).
\end{gather}
%
%In order for~\eqref{eq:fullnosharedsys} to be well-defined and realistic, we can also particularize 
Assumption~\ref{assum:base}, particularized for the setting without a shared resource, is given as follows:
\begin{assum} \label{assum:noshared}
Suppose that $\delta_i^k>0$ and $\beta_{ij}^k \geq 0$ for all $i, j \in [n]$ and $k \in [m]$.~$\blacksquare$ %\hfill  
\end{assum}
%
%We will also need the following corollary of Proposition~\ref{prop:necessity}, particularized for the setting without a shared resource.
%
%\begin{cor} \label{cor:necessity_noshared}
%Consider system~\eqref{eq:fullnosharedsys} under Assumption~\ref{assum:noshared}. For each $k \in [m]$ such that $B^k$ is irreducible and $s(B^k - D^k) > 0$, there is a unique single-virus endemic equilibrium of the form $(\textbf{0}, \dots, \Tilde{p}^k, \dots, \textbf{0})$ in $[0,~1]^{mn}$, with $\textbf{0} \ll \Tilde{p}^k \ll \textbf{1}$.~$\blacksquare$
%\end{cor}
%
%\textit{Proof:} The proof is analogous to the proof of Proposition~\ref{prop:necessity} and is therefore omitted.~$\square$ \\
%\axel{\textbf{[Might add a separate set of assumptions for this particularized model, since Assumption~\ref{assum:base} pertains to the shared resource as well.]}} \seb{[Response: I think \cite[Assumption~2]{liu2019analysis} could help.]}
%
Applying Proposition~\ref{prop:necessity} %Corollary~\ref{cor:necessity_noshared} 
to the bi-virus setting without a shared resource, we see that if i) $B^1$ and $B^2$ are irreducible, ii) $s(B^1-D^1) > 0$, and iii) $s(B^2-D^2) > 0$, then there exist exactly two single-virus endemic equilibria, namely
\begin{equation}\label{eq:endemicm2_noshared}
    (\Tilde{p}^1, \textbf{0}) \text{ and } (\textbf{0}, \Tilde{p}^2), 
\end{equation}
such that $\textbf{0} \ll \Tilde{p}^1 \ll \textbf{1}$ and $\textbf{0} \ll \Tilde{p}^2 \ll \textbf{1}$. Hence, we have the following corollary to Theorem~\ref{thm:joint_eq_exist_shared}, for the setting without a shared resource. %\axel{\textbf{[Actually, this is not a corollary in a strict sense. It is a trivial extension, but the irreducibility condition is not "the same". Could add comment on this and say that proof is analogous.]}} of Theorem~\ref{thm:joint_eq_exist_shared}, for the case without a shared resource:
\begin{cor} \label{cor:joint_eq_exist}
Consider the SIS model~\eqref{eq:fullnosharedsys} under Assumption~\ref{assum:noshared} with $m=2$. 
%Suppose that $\beta_{ij}^k \geq 0$ and $\delta_i^k > 0$ for all $i, j \in [n]$ and $k \in [2]$. 
Suppose that $B^1$ and $B^2$ are irreducible matrices, and that $s(B^1 - D^1) > 0$ and $s(B^2 - D^2) > 0$. 
%\phil{By Proposition~\ref{prop:necessity} there exist two non-zero equilibria, namely $(\Tilde{p}^1, \textbf{0})$ and $(\textbf{0}, \Tilde{p}^2)$, such that $\textbf{0} \ll \Tilde{p}^1 \ll \textbf{1}$ and $\textbf{0} \ll \Tilde{p}^2 \ll \textbf{1}$.}
If $s(- D^1 + (I - diag(\Tilde{p}^2))B^1) > 0$ and $s(- D^2 + (I - diag(\Tilde{p}^1))B^2) > 0$, with $\Tilde{p}^1,\Tilde{p}^2$ defined in~\eqref{eq:endemicm2_noshared}, then there exists at least one coexisting equilibrium $(\hat{p}^1, \hat{p}^2) \gg \textbf{0}$ in $[0,1]^{2n}$ such that $\hat{p}^1 + \hat{p}^2 \leq \textbf{1}$.~$\blacksquare$
\end{cor}
%\textit{Proof:} The proof is analogous to that of Theorem~\ref{thm:joint_eq_exist_shared} and is therefore omitted.~$\square$
%
%\noindent \axel{\textbf{[Do you want no indent here?]}}
Corollary~\ref{cor:joint_eq_exist} guarantees existence of a coexisting equilibrium in the bi-virus setup. It is an improvement of a similar result in \cite[Theorem~5.2]{li2004coexist}, wherein the same is established for $n=2$. However, \cite[Theorem~5.2]{li2004coexist} %under certain conditions 
establishes uniqueness of the coexisting equilibrium, whereas Corollary~\ref{cor:joint_eq_exist} provides no such guarantees.
%\axel{The same result was shown in \cite{li2004coexist} in a two-node setup ($n=2$), where it was also shown that under certain conditions, the coexisting equilibrium was unique and asymptotically stable. Note that Corollary~\ref{cor:joint_eq_exist} does not address the uniqueness of the coexisting equilibrium.}
%\textcolor{orange}{Conditions, different from the ones in Corollary~\ref{cor:joint_eq_exist}, that guarantee existence of a coexisting equilibrium in a multi-virus setup have been provided in %\cite[Theorem~7]{liu2019analysis} and 
%\cite[Theorem~6]{pareautomatica}. In fact,} 
Note that \cite[Theorem~6]{pareautomatica}, particularized for $m=2$, establishes the existence of not just one but infinitely many coexisting equilibria, thus implying that a coexisting equilibrium in a bi-virus setup %is \axel{[are not necessarily]} \emph{not}
is not necessarily unique.  However, it turns out that the conditions in Corollary~\ref{cor:joint_eq_exist} do \emph{not} coincide with the conditions in \cite[Theorem~6]{pareautomatica}, as discussed in the following remark.

\begin{rem} \label{rem:not_compatible_conditions}
%Consider system~\eqref{eq:fullnosharedsys} under Assumption~\ref{assum:noshared}.
%Notice that one of the conditions in Corollary~1,is as follows:
%
Suppose that, for $m=2$, the conditions in \cite[Theorem~6]{pareautomatica} are satisfied. That is, for some $v > 0$, $B^1$ is an irreducible nonnegative matrix with $B^1 = v B^2$, and $D^1$ is a positive diagonal matrix with $D^1 = v D^2$, $s(B^1 - D^1) > 0$ and $s(B^2 - D^2) > 0$. Let $\Tilde{p}^1$ and $\Tilde{p}^2$ be the unique single-virus endemic equilibria defined in~\eqref{eq:endemicm2_noshared}. %associated with virus~$1$ and $2$, respectively. 
It follows from~\eqref{eq:nosharedsys}, and the fact that $D^1, D^2$ are invertible, 
%$B^1 = \nu B^2$, $D^1 = \nu D^2$ gives $(D^1)^{-1} B^1 = (D^2)^{-1} B^2$, 
that %$\Tilde{p}^1 = \Tilde{p}^2 = \Tilde{p}$ where $\Tilde{p}$ fulfills 
$\Tilde{p}^1, \Tilde{p}^2$ fulfill
\begin{equation} \label{eq:rem_not_compat_equiequa}
\begin{split}
    (I - \diag(\Tilde{p}^1))(D^1)^{-1} B^1 \Tilde{p}^1 &= \Tilde{p}^1, \\
    (I - \diag(\Tilde{p}^2))(D^2)^{-1} B^2 \Tilde{p}^2 &= \Tilde{p}^2. 
\end{split}
\end{equation}
Since $B^1 = \nu B^2$, $D^1 = \nu D^2$ gives $(D^1)^{-1} B^1 = (D^2)^{-1} B^2$, it 
%is immediate
follows from~\eqref{eq:rem_not_compat_equiequa} that $\Tilde{p}^1 = \Tilde{p}^2 = \Tilde{p}$, and therefore
\begin{equation} \label{eq:rem_not_compat_equiequa_equal}
\begin{split}
    (I - \diag(\Tilde{p}))(D^1)^{-1} B^1 \Tilde{p} &= \Tilde{p}, \\
    (I - \diag(\Tilde{p}))(D^2)^{-1} B^2 \Tilde{p} &= \Tilde{p}.  
\end{split}
\end{equation}
By Proposition~\ref{prop:necessity} we have $\Tilde{p} \ll \textbf{1}$, implying that $(I-\diag(\Tilde{p}))$ is a positive diagonal matrix. Then, for $k \in [2]$, it follows that $(I - \diag(\Tilde{p}))(D^k)^{-1} B^k$ is an irreducible nonnegative matrix. Therefore, item~\ref{item:perfrob_pos_necess} in Lemma~\ref{lem:perron_frob} can be applied to~\eqref{eq:rem_not_compat_equiequa_equal}, from which it follows that $\rho ((I - \diag(\Tilde{p}))(D^1)^{-1} B^1) = 1$ and $\rho ((I - \diag(\Tilde{p}))(D^2)^{-1} B^2) = 1$. %\seb{[not obvious the way it is currently written...perhaps number the equations above, and then argue]}
Applying Lemma~\ref{lem:eigspec} we obtain
\begin{equation} \label{eq:rem_cond_zero}  
\begin{split}
    s(-D^1 + (I - \diag(\Tilde{p}^2)) B^1) &= 0, \\
    s(-D^2 + (I - \diag(\Tilde{p}^1)) B^2) &= 0. 
\end{split}
\end{equation}
Observe ~\eqref{eq:rem_cond_zero} is incompatible with the following  condition in Corollary~\ref{cor:joint_eq_exist}, %is as follows: %specifically the condition: 
\vspace{-3ex}
\begin{equation} \label{eq:rem_cond_pos} 
\begin{split}
    s(-D^1 + (I - \diag(\Tilde{p}^2)) B^1) &> 0, \\
    s(-D^2 + (I - \diag(\Tilde{p}^1)) B^2) &> 0.
\end{split}
\end{equation}
Hence, it follows that the conditions in Corollary~\ref{cor:joint_eq_exist} and \cite[Theorem~6]{pareautomatica} are mutually exclusive.~$\blacksquare$
%
%Hence, it follows that Corollary~\ref{cor:joint_eq_exist} does not imply \cite[Theorem~6]{pareautomatica}, \phil{nor the other way around [I think you actually showed the latter: \cite[Theorem~6]{pareautomatica} does not imply  Corollary~\ref{cor:joint_eq_exist}... perhaps we should just conclude that the conditions are `mutually exclusive']}.\hfill ~$\square$
%it is clear that \cite[Theorem~6]{pareautomatica} and Corollary~\ref{cor:joint_eq_exist} cover different conditions.
\end{rem}

Since the conditions in Corollary~\ref{cor:joint_eq_exist} are not compatible
%\phil{do not coincide [`are not compatible with'?]}
with the conditions in \cite[Theorem~6]{pareautomatica}, the %following 
question: \enquote{do the conditions in Corollary~\ref{cor:joint_eq_exist} guarantee uniqueness of the coexisting equilibrium?} is worth investigating. Our simulations show that such a coexisting equilibrium may indeed be unique, and further, asymptotically stable; see Section~\ref{sect:simulations} and Figure~\ref{fig:coexistence}. %However, a rigorous proof establishing uniqueness and asymptotic stability of the coexisting equilibrium remains elusive, and is left for future work.} % we provide simulation results indicating that the coexisting equilibrium may indeed be unique, and further, asymptotically stable.}
%\phil{[do we think this is true? Can we say that, and refer to the Simulations Section?]}

While Theorem~\ref{thm:joint_eq_exist_shared} provides conditions for the existence of coexisting equilibria, a related problem is finding conditions under which no coexisting equilibria can exist. The following theorem makes use of a nontrivial condition to eliminate the possibility of coexisting equilibria in a bi-virus setting, and establishes one virus as being dominant.
\begin{thm} \label{thm:noequi_single-virus}
Consider the SIWS model~\eqref{eq:full} under Assumption~\ref{assum:base} with $m=2$. Suppose that $B_w^1$ and $B_w^2$ are irreducible matrices, and that $s(B_w^1 - D_w^1) > 0$ and $s(B_w^2 - D_w^2) > 0$. 
%By Proposition~\ref{prop:necessity}, there exist two non-zero equilibria, namely $(\Tilde{y}^1, \textbf{0})$ and $(\textbf{0}, \Tilde{y}^2)$, such that $\textbf{0} \ll \Tilde{y}^1 \ll \textbf{1}$ and $\textbf{0} \ll \Tilde{y}^2 \ll \textbf{1}$. 
If $(D_w^1)^{-1}B_w^1 > (D_w^2)^{-1}B_w^2$ %or $(D_w^1)^{-1}B_w^1 < (D_w^2)^{-1}B_w^2$, 
then there are exactly three equilibria in $\mathcal{D}$, namely the healthy state, which is unstable, $(\textbf{0}, \Tilde{y}^2)$ with $\textbf{0} \ll \Tilde{y}^2 \ll \textbf{1}$, which is unstable, and $(\Tilde{y}^1, \textbf{0})$ with $\textbf{0} \ll \Tilde{y}^1 \ll \textbf{1}$, which is locally exponentially stable.~$\blacksquare$
%then there can exist no equilibrium $(\hat{y}^1, \hat{y}^2)$ such that $\hat{y}^1 > \textbf{0}$, $\hat{y}^2 > \textbf{0}$, $\hat{y}^1 + \hat{y}^2 \leq \textbf{1}$.~$\blacksquare$ %\hfill 
\end{thm}
\textit{Proof:} See Appendix.~$\square$
\noindent
Theorem~\ref{thm:noequi_single-virus} states that if one virus has a stronger set of spread and healing parameters than the other virus, then these two viruses cannot coexist in the population.
% }
Theorem~\ref{thm:noequi_single-virus} answers question~\ref{q5} in Section~\ref{subsec:pb:statement}. The following remark aids in understanding the result in Theorem~\ref{thm:noequi_single-virus}.
\begin{rem} \label{rem:no-coexist}
%\seb{[Comment: We need to talk about competitive exclusion here]}
%Theorem~\ref{thm:noequi_single-virus} states that in a bi-virus setting, if one virus has a stronger set of spread and healing parameters than the other virus, then these two viruses cannot coexist in the population. 
Underlying Theorem~\ref{thm:noequi_single-virus} is the so-called competitive exclusion principle, which states that complete competitors cannot coexist \cite{hardin1960competitive}. For instance, if two strains of viruses (say virus~$1$ and virus~$2$) compete with each other to infect the same population, and if virus~$1$ %(resp. virus~$2$) 
has a slight advantage over virus~$2$ %(resp. virus~$1$) 
(for example higher infection rates or lower healing rates), then virus~$1$ %(resp. virus~$2$) 
will eventually displace virus~$2$, %(resp. virus~$1$)
which will get eradicated.%\textcolor{orange}{This interpretation will play an important role in devising a distributed control strategy for mitigation of epidemics, as we will see in Section~\ref{subsect:competive:exclusion}.}
~$\blacksquare$ 
\end{rem}
Similar results to Theorem~\ref{thm:noequi_single-virus} can be found in \cite[Theorem~5]{liu2019analysis}, for the setting without a shared resource. In order to make a comparison, we employ the SIS model~\eqref{eq:fullnosharedsys} to state the following corollary of Theorem~\ref{thm:noequi_single-virus}.
\begin{cor} \label{cor:noequi_single-virus_noshared}
Consider the SIS model~\eqref{eq:fullnosharedsys} under Assumption~\ref{assum:noshared} with $m=2$. Suppose that $B^1$ and $B^2$ are irreducible matrices, and that $s(B^1 - D^1) > 0$ and $s(B^2 - D^2) > 0$. 
%By Proposition~\ref{prop:necessity} there exist two non-zero equilibria, namely $(\Tilde{p}^1, \textbf{0})$ and $(\textbf{0}, \Tilde{p}^2)$, such that $\textbf{0} \ll \Tilde{p}^1 \ll \textbf{1}$ and $\textbf{0} \ll \Tilde{p}^2 \ll \textbf{1}$. 
If $(D^1)^{-1}B^1 > (D^2)^{-1}B^2$ 
%or $(D^1)^{-1}B^1 < (D^2)^{-1}B^2$, 
then there are exactly three equilibria in $[0,1]^{2n}$, namely the healthy state, which is unstable, $(\textbf{0}, \Tilde{p}^2)$ with $\textbf{0} \ll \Tilde{p}^2 \ll \textbf{1}$, which is unstable, and $(\Tilde{p}^1, \textbf{0})$ with $\textbf{0} \ll \Tilde{p}^1 \ll \textbf{1}$, which is locally exponentially stable.~$\blacksquare$
%then there can exist no equilibrium $(\hat{p}^1, \hat{p}^2)$ such that $\hat{p}^1 > \textbf{0}$, $\hat{p}^2 > \textbf{0}$, $\hat{p}^1 + \hat{p}^2 \leq~\textbf{1}$.~$\blacksquare$
\end{cor}
%
%\textit{Proof:} The proof is analogous to the proof of Theorem~\ref{thm:noequi_single-virus} and is therefore omitted.~$\square$\\
%
Corollary~\ref{cor:noequi_single-virus_noshared} is directly comparable to \cite[Theorem~5]{liu2019analysis}. The main difference is that \cite[Theorem~5]{liu2019analysis} requires that $\beta_{ij}^k = \beta^k$ or $\beta_{ij}^k = 0$, and $\delta_i^k = \delta^k$, for all $i,j \in [n]$, $k \in [2]$, and some $\beta^k > 0$, $\delta^k > 0$, whereas Corollary~\ref{cor:noequi_single-virus_noshared} has no such restrictions. Hence, Corollary~\ref{cor:noequi_single-virus_noshared} subsumes and improves upon \cite[Theorem~5]{liu2019analysis}, while Theorem~\ref{thm:noequi_single-virus} extends the result to the setting with a shared resource. %\axel{\textbf{[Might add something here about Corollary 3 in "Sufficient condition for survival of the fittest in a bi-virus epidemics".]}}

\section{Mitigation strategies} \label{sect:dist:control}
%
%\axel{\textbf{[Another promising possibility is to emulate \cite[Problem~1]{pareautomatica}.]}}
%
In this section, we present two strategies for ensuring that some (or all) viruses are eradicated. First, we %demonstrate
establish that it is possible to cause convergence to the eradicated state of a virus~$k$ by boosting the corresponding healing rates~$\delta_i^k$. Applying this technique to all $m$ viruses, the system converges to the healthy state. Second, we show that it is possible to leverage a benign virus in order to eradicate a malignant virus, in a bi-virus setting.
%consider the possibility of leveraging a benign virus to eradicate a malignant one in a bi-virus system. 
%\phil{[we probably need to discuss this in the Introduction with some justification via citing the medical literature...]}
%
\subsection{Boosting the healing rates} \label{subsec:boost_healing}
When managing the epidemic spread of a virus, a natural strategy is to boost the healing rates in the population nodes. The following result shows that the healing rates can always be chosen to ensure asymptotic or exponential eradication of a virus, using an algorithm inspired by \cite{liu2019analysis,gracy2020asymptotic}. %\seb{[we need asymptotic convergence to the healthy state!]}.
%
%\seb{[COMMENT: I suggest calling the next result as a Proposition.]}
%\phil{[we need to point out the similarities to our TCNS paper and/or the inspiration from \cite{liu2019analysis}]}
\begin{prop} \label{prop:deltas_chosen_to_heal}
Consider the SIWS model~\eqref{eq:yk} under Assumption~\ref{assum:base} with
% and assume that 
$y(0) \in~\mathcal{D}$. Suppose, for some~$k \in [m]$, that $B_w^k$ is irreducible, %\axel{Then, suppose that $\beta_{iw}^k$ is set to zero for some $i \in [n]$, while maintaining the irreducibility of $B_w^k$, and subsequently the healing rates have been chosen to fulfill
%For
and that the healing rates are of the form
\begin{equation} \label{eq:deltas_chosen_to_heal}
\delta_i^k = \beta_{iw}^k + \textstyle \sum_{j=1}^n \beta_{ij}^k + \epsilon_i^k,
\end{equation}
for each $i \in [n]$, where $\epsilon_i^k \geq 0$. Then, if $\epsilon_i^k > 0$ for some $i \in [n]$, the eradicated state of virus~$k$ is exponentially stable, with domain of attraction containing $\mathcal{D}^k$. Otherwise, %, if $\epsilon_i^k = 0$ for all $i \in [n]$, 
the eradicated state of virus~$k$ is asymptotically stable, with domain of attraction containing $\mathcal{D}^k$.~$\blacksquare$
\end{prop}
%
%\axel{\textbf{[I would prefer to contextualize this result a bit more. Something like, $\delta_i^k$ is the base healing rate and $u_i^k$ is the "boost", so that $\delta_i^k + u_i^k$ as the boosted healing rate. Although this would run into problems whenever $\delta_i^k$ exceeds the condition in~\eqref{eq:deltas_chosen_to_heal} to begin with.]}}
%
The proof is straightforward, following from Theorem~\ref{thm:asymp} and similar arguments as in \cite[Section~V]{liu2019analysis}, and \cite{gracy2020asymptotic}.\\
\textit{Proof:} See Appendix.~$\square$ %\axel{Note that with $\beta_{iw}^k$ set to zero for some $i \in [n]$, while maintaining the irreducibility of $B_w^k$, and} 

Proposition~\ref{prop:deltas_chosen_to_heal} represents one strategy to ensure eradication of a virus. By applying this strategy to all viruses, %in line with Proposition~\ref{prop:deltas_chosen_to_heal},
we obtain the following result. %\seb{[COMMENT: I suggest calling the next result as a Theorem, since it makes more sense to highlight this one.]}
\begin{thm} \label{thm:all_deltas_chosen_to_heal}
Consider the SIWS model~\eqref{eq:full} under Assumption~\ref{assum:base}. Suppose, for all~$k \in m$, that $B_w^k$ is irreducible and \eqref{eq:deltas_chosen_to_heal} is fulfilled. %Proposition~\ref{prop:deltas_chosen_to_heal} is fulfilled for all viruses $k \in [m]$. 
%that $B_w^k$ is irreducible, and that the healing rates $\delta_i^k$ are of the form in~\eqref{eq:deltas_chosen_to_heal} for each $i \in [n]$. 
Then, if $\epsilon_i^k > 0$ %in~\eqref{eq:deltas_chosen_to_heal} 
for some $i \in [n]$ and all $k \in [m]$, the healthy state is exponentially stable, with domain of attraction containing $\mathcal{D}$. If not, the healthy state is asymptotically stable, with domain of attraction containing~$\mathcal{D}$.~$\blacksquare$
\end{thm}
\noindent
Theorem~\ref{thm:all_deltas_chosen_to_heal} represents a strategy to eradicate all viruses in a system, which equivalently ensures exponential (resp. asymptotic) convergence to the healthy state. Thus, Theorem~\ref{thm:all_deltas_chosen_to_heal} addresses question~\ref{q6} %fulfilling~\ref{q6}
in Section~\ref{subsec:pb:statement}. %\phil{The following remark provides an epidemiological interpretation of the strategy in Theorem~\ref{thm:all_deltas_chosen_to_heal} [I'm not sure this sentence adds that much value]}.
 
\begin{rem}[Epidemiological interpretation]\label{rem:boost:healing:rates}
The mitigation strategy outlined in Theorem~\ref{thm:all_deltas_chosen_to_heal} could be understood as follows: with respect to each virus, if the healing rate of each subpopulation is sufficiently increased, which could be accomplished %\textcolor{orange}{, for example,} \axel{\textbf{[Could be cut given the later "etc".]}}
by prescribing high dosages of drugs, by administering vaccines, etc.,  then each of the viruses get eradicated exponentially (resp. asymptotically) fast. Observe that this 
% the 
strategy is extreme in the sense that it does not factor in limitations on the availability of resources, and essentially encourages health administration officials to amass (possibly) excessive amounts of resources in order to prevent epidemic outbreaks.~$\blacksquare$
\end{rem}
Note that in the absence of sufficient resources, implementing the aforementioned strategy in practice is impossible. Hence, we are motivated to seek different strategies.
\subsection{Leveraging one virus to eradicate another}\label{subsect:competive:exclusion}
%
%While Theorem~\ref{thm:all_deltas_chosen_to_heal} ensures that all viruses in the system are eradicated, it might not be feasible to manipulate the healing rates in line with~\eqref{eq:deltas_chosen_to_heal}. 
%In real-world scenarios, healing rates are often capped by limiting factors in the form of budgetary or practical constraints \axel{-- for instance, a lack of antidote -- which might prohibit healing rates of the form~\eqref{eq:deltas_chosen_to_heal}}. Then,

%an alternative strategy is
%\seb{[Differently from the strategy in~\eqref{eq:deltas_chosen_to_heal}, we could leverage]} %to leverage
%one virus to eradicate another. \seb{[More specifically]}, consider a bi-virus setting, i.e. $m=2$, where one virus is malign and the other virus is benign. In the case where both viruses are viable, i.e. fulfill Proposition~\ref{prop:necessity}, and the malign virus persists in the population, it turns out that we can leverage the benign virus in order to eradicate the malign virus, as stated in the following theorem.% which makes use of Theorem~\ref{thm:noequi_single-virus}.
%
It turns out that in a bi-virus setting, where one virus is malignant and the other virus is benign, we can leverage the benign virus in order to help eradicate the malignant virus, as stated in the following theorem. 
%\phil{[Same Comment: we probably need to discuss this in the Introduction with some justification via citing the medical literature...]}
\begin{thm} \label{thm:virus-as-vaccine}
Consider the SIWS model~\eqref{eq:full} under Assumption~\ref{assum:base} with $m=2$. Suppose that $B_w^1$ and $B_w^2$ are irreducible matrices; 
% that 
$s(B_w^1 - D_w^1) > 0$; $s(B_w^{2} - D_w^{2}) > 0$; 
% that 
$c^1=c^2$; and $E^2 \subseteq E^1$, where $E^1$ (resp. $E^2$) is the set of directed edges between the population nodes and the shared resource, with respect to virus~$1$ (resp. virus~$2$). %are as defined in Section~\ref{sect:Model}) 
%\phil{[I think we should explain it in words and/or have equation numbers in Section~\ref{sect:Model} because it's far away and hard to find...]} %\sebcancel{[defined where?]} \axel{\textbf{[Early on in Section~\ref{sect:Model}. This is the set of directed edges in a layer.]}} 
% for all $i, j \in [n+1]$; 
If the healing rates for virus~$2$ fulfill
\vspace{-1ex}
\begin{equation} \label{eq:deltas_chosen_to_leverage}
\delta_i^2 > \max_{j\in[n+1]} \left\{\frac{(B_w^2)_{ij}}{((D_w^1)^{-1}B_w^1)_{ij}} \right\}_{(B_w^1)_{ij}>0},
\end{equation}

\vspace{-1ex}

\noindent for all $i \in [n]$, then the only locally asymptotically stable equilibrium in $\mathcal{D}$ is $(\Tilde{y}^1, \textbf{0})$ with $\textbf{0} \ll \Tilde{y}^1 \ll \textbf{1}$.~$\blacksquare$
\end{thm}
\textit{Proof:} See Appendix.~$\square$
%Note that with $\delta_i^2$ of the form given in~\eqref{eq:deltas_chosen_to_leverage} for all $i \in [n]$, since $B_w^1$ and $B_w^2$ are irreducible nonnegative matrices we have $\delta_i^2 > 0$ for all $i \in [n]$. Therefore~\eqref{eq:deltas_chosen_to_leverage} is consistent with Assumption~\ref{assum:base}. Then, it follows from i)~\eqref{eq:deltas_chosen_to_leverage}, ii) $c^1=c^2$, %\seb{[In the theorem statement, you say, before introducing~(89), suppose .. $c_1= c_2$, but here you say~(89) implies $c_1= c_2$...not at all clear ]}
% and iii) $(B_w^2)_{ij} > 0 \implies (B_w^1)_{ij} > 0$, that $(D_w^1)^{-1}B_w^1 > (D_w^2)^{-1}B_w^2$. Hence, since, by assumption, $s(B_w^1 - D_w^1) > 0$ and $s(B_w^2 - D_w^2) > 0$, the conditions for Theorem~\ref{thm:noequi_single-virus} are met. %implying that
% Therefore, the only locally asymptotically stable equilibrium in $\mathcal{D}$ is $(\Tilde{y}^1, \textbf{0})$ with $\textbf{0} \ll \Tilde{y}^1 \ll \textbf{1}$. \hfill $\square$
%If $s(B_w^2 - D_w^2) \leq 0$, then the conditions for Proposition [DOES NOT EXIST YET] are met, implying that the only locally asymptotically stable equilibrium in $\mathcal{D}$ is $(\Tilde{y}^1, \textbf{0})$ with $\textbf{0} \ll \Tilde{y}^1 \ll \textbf{1}$. $\square$}

\noindent
Theorem~\ref{thm:virus-as-vaccine} represents a strategy to eradicate one of the viruses in a bi-virus system, made possible by leveraging the fact that one virus has a stronger set of spread parameters than the other. Thus, Theorem~\ref{thm:virus-as-vaccine} addresses question~\ref{q7} %fulfilling~\ref{q6}
in Section~\ref{subsec:pb:statement}. 
% We discuss an interesting interpretation of the strategy in Theorem~\ref{thm:virus-as-vaccine}, and of the merits of \seb{both the distributed control strategies} \axel{[the distributed control strategies from this section]} in the following remarks. \phil{[this seemed unnecessary, and didn't mirror the rest of the paper so I deleted it]}

\begin{rem}[Virus as vaccine]\label{rem:virus:vaccine}
Since the strategy given in Theorem~\ref{thm:virus-as-vaccine}
% , since it 
ensures local asymptotic convergence to the single-virus endemic equilibrium of the benign virus, %\axel{\textbf{[This sentence could be cut.]}} \seb{[Response: I think that sentence justifies, at least partially, why the interpretation makes sense, so let's keep it.]}
it 
could also be interpreted in the following sense: the benign virus effectively acts as a \emph{vaccine} against the malignant virus. In the context of 
% tackling 
battling %[or eradicating]
epidemic outbreaks, where the goal is to minimize the mortality 
% \axel{(and morbidity)} 
rate, this strategy could potentially provide health administration officials with an effective tool.
~$\blacksquare$
\end{rem}

%\begin{rem} \label{rem:pros and cons:distributed control:strat}
% \seb{
The mitigation strategies detailed in this section can be compared as follows. %Consider a bi-virus epidemic outbreak, with one virus being benign and the other malignant. Suppose that in order to combat such an outbreak, due to resource constraints (e.g. availability of vaccines, drugs, ventilators, etc.), the approach is to eradicate the malignant virus. In such a context, it may be more feasible %prudent \sebcancel{[feasible]}to implement the strategy given in Theorem~\ref{thm:virus-as-vaccine} instead of that in Proposition~\ref{prop:deltas_chosen_to_heal}. }
On the one hand, assuming that the %\seb{approach towards handling a bi-virus epidemic outbreak}
objective of public health  officials  %\phil{[of what?]} %\axel{[of healthcare providers]}
is solely to eradicate one virus 
%~2
in a bi-virus system 
% and factoring in for 
while considering
resource constraints (e.g., availability of vaccines, drugs, ventilators, etc.), %the main advantage of Theorem~\ref{thm:virus-as-vaccine} over Proposition~\ref{prop:deltas_chosen_to_heal} %\seb{[Theorem~\ref{thm:all_deltas_chosen_to_heal}]}
%is that, depending on the constraints involved in boosting $\delta_i^2$, 
it may be more feasible %prudent \sebcancel{[feasible]}
to implement the strategy given in Theorem~\ref{thm:virus-as-vaccine} instead of that in Proposition~\ref{prop:deltas_chosen_to_heal}. 
On the other hand, 
%obtain healing rates $\delta_i^2$ of the form~\eqref{eq:deltas_chosen_to_leverage} instead of~\eqref{eq:deltas_chosen_to_heal}. %Further, given resource constraints, it may not even be possible to achieve~\eqref{eq:deltas_chosen_to_heal}.
%However, such an advantage notwithstanding, 
the strategy in Theorem~\ref{thm:virus-as-vaccine}, 
% differently from
opposed to
that in Theorem~\ref{thm:all_deltas_chosen_to_heal} particularized for a bi-virus setting, requires the persistence of one virus, 
%~1, 
which may be undesirable. 
Moreover, with respect to a given virus, if the healing rate of at least one node is sufficiently boosted, then Proposition~\ref{prop:deltas_chosen_to_heal} guarantees eradication of said virus exponentially fast, whereas Theorem~\ref{thm:virus-as-vaccine} guarantees only asymptotic eradication of a virus. We  %will now
explore the advantages and disadvantages of the strategies outlined in Proposition~\ref{prop:deltas_chosen_to_heal} and Theorem~\ref{thm:virus-as-vaccine} via simulations in Section~\ref{sect:simulations}.%~$\blacksquare$

\section{Simulations}
\label{sect:simulations}
\begin{figure}[h!] 
\centering
    \frame{{\includegraphics[width=0.6\columnwidth]{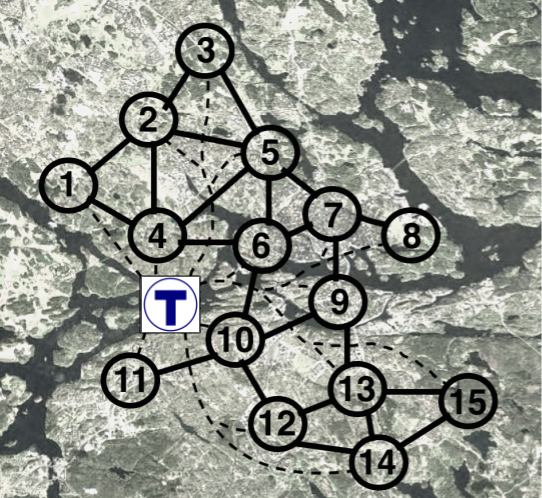}}}
    \caption{A map of~15 city districts in Stockholm. All districts are connected to the metro system.}
    \label{fig:stockholm}
\end{figure}

%\subsection{Simulations to back up our theoretical findings}
%\seb{Comment: Here feel free to reuse some of the simulations from \cite{axel2020multi}.}
In this section, we present a number of simulations to illustrate our theoretical findings, using the city of Stockholm as an example setting. %the setting.
In particular,~15 districts in and around Stockholm are taken to be the population nodes of the network, thus, $n = 15$. All of these districts are connected to the Stockholm metro, which we view as the shared resource of the network; see Figure~\ref{fig:stockholm}.
%\seb{[Note: If not yet done, then we should say that the contact graph for each virus is the same.]}

In all simulated scenarios we consider two competing viruses, namely virus~$1$ and virus~$2$. We denote the average infection ratio of virus~$k$, i.e., $\textstyle \tiny{\frac{1}{n}} \sum_i^n p_i^k(t)$, by $\Bar{p}^k(t)$. The contact network of each virus is the same, i.e., $E^1 = E^2$, and is represented in Figure~\ref{fig:stockholm}. We set $p_i^k(0) = 0.5$ and $z^k(0) = 0.5$ for all $i \in [15]$ and $k \in [2]$, and use the same %initial state
in all simulated scenarios throughout this section. 
%depicted in Figure~\ref{fig:weak_vs_weaker}-\ref{fig:strong_vs_stronger},} %\seb{[Response: you have not told them what Figure~3-6 mean. So this phrasing will have to be modified.]} 
%with $p_i^k(0) = 0.5$ and $z^k(0) = 0.5$ for all $i \in [15]$ and $k \in [2]$. 
%This represents the case where
Observe that the aforementioned choice of initial states represents the case where half of the population in each node is infected by virus~$1$, while the other half is infected by virus~$2$, and the shared resource is contaminated by both viruses. The spread parameters $\beta_{ij}^k$ are taken to be~$1$ if district $i$ is adjacent to
% in walking distance of 
district $j$, or if $i = j$, and $0$ otherwise, for $k \in [2]$. %%for both viruses $k$.
In all scenarios, we also assume that each district is bi-directionally connected to the shared resource, i.e., the Stockholm metro, with $\beta_{iw}^k = 1$ and $c_i^k = 1/15$, for all $i \in [15]$ and $k \in [2]$. As a consequence, $B_w^k$ is irreducible for $k \in [2]$. The following scenarios differ only in terms of the choice of $\delta_i^k$ and $\delta_w^k$. 

% \begin{figure}[h!] 
% \centering
%     \scalebox{0.8}[0.8]{\includegraphics[width=\columnwidth]{Figures/weak_vs_weaker.png}}
%     \caption{Simulation with two viruses (red and blue), both converging exponentially fast to eradication. The average infection ratio of virus~$k$ is denoted by $\Bar{p}^k(t)$.}
%     \label{fig:weak_vs_weaker}
% \end{figure}

%In the simulation depicted in Figure~\ref{fig:weak_vs_weaker}, we chose $\delta_i^1 = 6$, $\delta_w^1 = 6$ and $\delta_i^2 =~10$, $\delta_w^2 =~10$, for all $i \in [15]$. As a consequence, $s(B_w^1-D_w^1) = -0.8$, and $s(B_w^2 - D_w^2) = -4.2$, so the conditions in Theorem~\ref{thm:expo} are satisfied. In line with the result in Theorem~\ref{thm:expo}, we see that the viruses appear to converge exponentially fast to the eradicated state in Figure~\ref{fig:weak_vs_weaker}. Furthermore, virus~$2$ converges faster than virus~$1$, due to $s(B_w^2-D_w^2)$ being more negative than $s(B_w^1-D_w^1)$.

\begin{figure}[h!] 
\centering
    \scalebox{0.8}[0.8]{\includegraphics[width=\columnwidth]{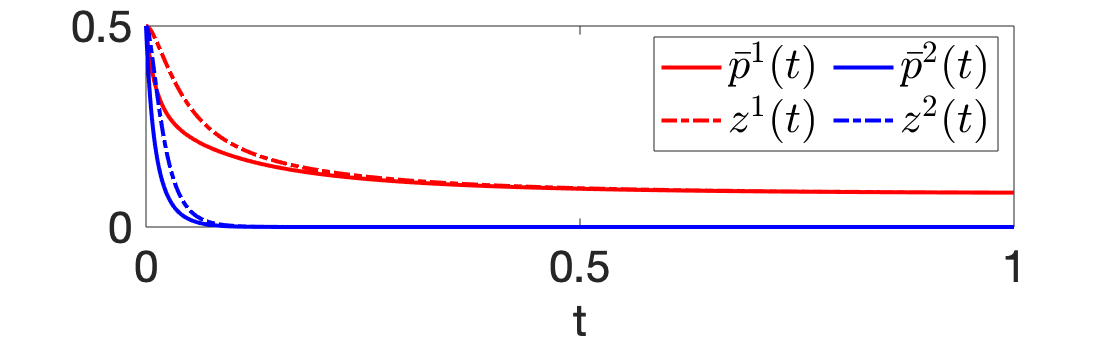}}
    \caption{Simulation with two viruses, one (blue) converging to eradication, while the other (red) converges to its single-virus endemic equilibrium. %The average infection ratio of virus~$k$ is denoted by $\Bar{p}^k(t)$.
    }
    \label{fig:strong_vs_weak}
\end{figure}

In the simulation depicted in Figure~\ref{fig:strong_vs_weak}, we chose $\delta_i^1 = 4.6$, $\delta_w^1 = 4$ and $\delta_i^2 = 10$, $\delta_w^2 = 10$, for all $i \in [15]$. As a consequence, $s(B_w^2 - D_w^2) = -4.2$, so Theorem~\ref{thm:expo} applies to virus~$2$. Consistent with the result in Theorem~\ref{thm:expo}, virus~$2$ becomes eradicated exponentially fast%; see blue line (solid and dashed) in Figure~\ref{fig:strong_vs_weak}
. However, since $s(B_w^1-D_w^1) = 0.3$, virus~$1$ fulfills the conditions for Proposition~\ref{prop:necessity}, and furthermore, once virus~$1$ is eradicated, the system is essentially a single-virus system. Treated as such, virus~$1$ satisfies the conditions in Theorem~\ref{thm:equi}. In line with the result in Theorem~\ref{thm:equi}, %we see that
virus~$1$ converges to a single-virus endemic equilibrium%; see red lines (solid and dashed) in Figure~\ref{fig:strong_vs_weak}
. Insofar as it can be determined by varying $y^1(0)$ in $\mathcal{D}^1$, this single-virus endemic equilibrium appears to be unique and asymptotically stable.  %However, $s(B_w^1-D_w^1) = 0.3$, so virus~$1$ fulfills the conditions for Proposition~\ref{prop:necessity}. Furthermore, at the point where virus~$2$ is eradicated in Figure~\ref{fig:strong_vs_weak}, the system is essentially a single-virus system from thereon out. Treating it as such, virus~$1$ fulfills the conditions for Theorem~\ref{thm:equi}. In line with this, we see that virus~$1$ converges to a single-virus endemic equilibrium in Figure~\ref{fig:strong_vs_weak}, and, insofar as it can be determined by varying the initial condition $y^1(0)$ within $\mathcal{D}^1$, this non-zero equilibrium appears to be unique and asymptotically stable, with domain of attraction containing $\mathcal{D}^1$.

\begin{figure}[h!] 
\centering
    \scalebox{0.8}[0.8]{\includegraphics[width=\columnwidth]{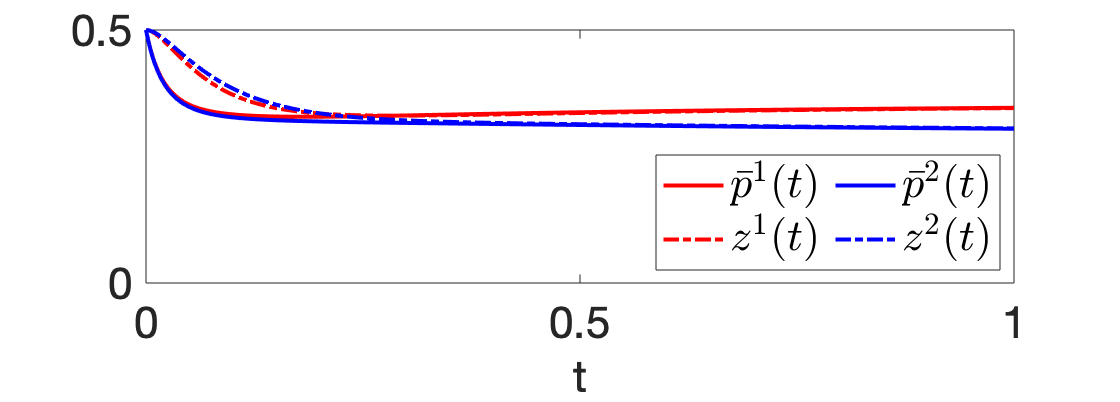}}
    \caption{Simulation with two viruses (red and blue), converging to a coexisting equilibrium. %The average infection ratio of virus~$k$ is denoted by $\Bar{p}^k(t)$.
    }
    \label{fig:coexistence}
\end{figure}

In the simulation depicted in Figure~\ref{fig:coexistence}, we chose $\delta_i^1 = 1.5$ for $i \in [8]$, and $\delta_i^1 = 2$ for 
% the remaining 
$i \in [15]\setminus [8]$, with $\delta_w^1 = 1$. Relating these choices of parameters for virus~$1$ to Figure~\ref{fig:stockholm}, this corresponds to a low healing rate north of the river Mälaren and a high healing rate south of the river. %and a high rate of decay in the shared resource. 
Mirroring this pattern, we chose $\delta_i^2 = 2$ for $i \in [8]$, and $\delta_i^2 = 1.5$ for 
% the remaining 
$i \in [15]\setminus [8]$, with $\delta_w^2 = 1$. With these parameters, it follows that $s(B_w^1-D_w^1) = 2.8$, and $s(B_w^2 - D_w^2) = 2.8$. Hence, both viruses fulfill the conditions for Proposition~\ref{prop:necessity}, providing the existence of exactly two single-virus endemic equilibria, namely $(\Tilde{y}^1, \textbf{0})$ and $(\textbf{0}, \Tilde{y}^2)$. $\Tilde{y}^1$ can be approximated by setting $y^1(0) > \textbf{0}$ and $y^2(0) = \textbf{0}$, and running the simulation for a sufficiently long period of time $T$. Then, assuming that $\Tilde{y}^1 \approx y^1(T)$, and, with an analogous approximation for virus~$2$, $\Tilde{y}^2 \approx y^2(T)$, %we can calculate
we obtain $s((I-X(\Tilde{y}^2))B_w^1-D_w^1) = 0.2$, and $s((I-X(\Tilde{y}^1))B_w^2 - D_w^2) = 0.2$. As a consequence, this pair of viruses fulfills the conditions for Theorem~\ref{thm:joint_eq_exist_shared}. In line with the result in Theorem~\ref{thm:joint_eq_exist_shared}, we see that there exists a coexisting equilibrium. Moreover, our simulations show that the viral infection levels appear to converge to this coexisting equilibrium%; see Figure~\ref{fig:coexistence}
. Additionally, regardless of how the initial condition is varied within $\mathcal{D}$, barring $y^1(0) = \textbf{0}$ or $y^2(0) = \textbf{0}$, we observe that all simulations converge to the same coexisting equilibrium. This observation suggests that the coexisting equilibrium might be unique, as well as asymptotically stable.
%Figure~\ref{fig:coexistence} shows that the viruses appear to converge to some coexisting equilibrium in $\mathcal{D}$, the existence of which is guaranteed by Theorem~\ref{thm:joint_eq_exist_shared}. Beyond just existence, our results indicate that under these conditions, all simulations converge to the same coexisting equilibrium, regardless of how the initial condition is varied within $\mathcal{D}$, barring $y^1(0) = \textbf{0}$ or $y^2(0) = \textbf{0}$. This indicates that the coexisting equilibrium might be unique, as well as asymptotically stable.

\begin{figure}[h!] 
\centering
    \scalebox{0.8}[0.8]{\includegraphics[width=\columnwidth]{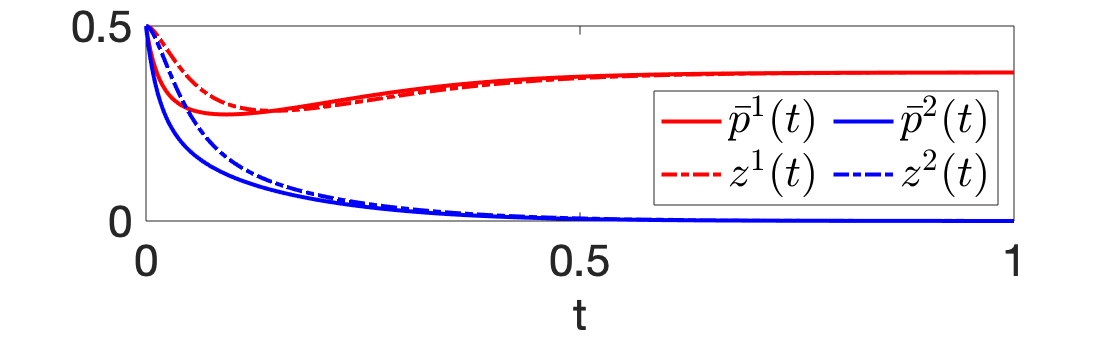}}
    \caption{Simulation with two viruses, with one (blue) converging to eradication, while the other (red) eventually converges to its single-virus endemic equilibrium. %The average infection ratio of virus~$k$ is denoted by $\Bar{p}^k(t)$.
    }
    \label{fig:strong_vs_stronger}
\end{figure}

In the simulation depicted in Figure~\ref{fig:strong_vs_stronger}, we chose $\delta_i^1 = 3$, $\delta_w^1 = 3$ and $\delta_i^2 = 4$, $\delta_w^2 = 4$, for all $i \in [15]$. It follows that $s(B_w^1-D_w^1) = 1.7$, and $s(B_w^2 - D_w^2) = 0.9$. Moreover, we have $(D_w^1)^{-1}B_w^1 > (D_w^2)^{-1}B_w^2$. Therefore, this pair of viruses fulfills the conditions for Theorem~\ref{thm:noequi_single-virus}. In line with the result in Theorem~\ref{thm:noequi_single-virus}, we observe that
virus~$1$ has a competitive edge, allowing it to persist and converge to its single-virus endemic equilibrium. Meanwhile, virus~$2$ is eradicated, despite $s(B_w^2 - D_w^2)$ being positive.

%\subsection{Control simulations}

\begin{figure}[h!] 
\centering
    \scalebox{0.8}[0.8]{\includegraphics[width=\columnwidth]{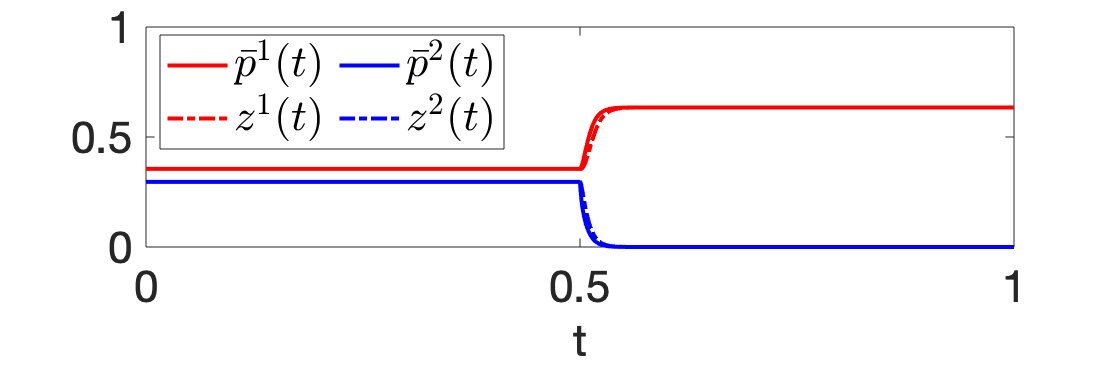}}
    \caption{Simulation with two viruses (red and blue), initialized at a coexisting equilibrium. At $t = 0.5$ the healing rates of virus~$2$ are changed to fulfill Proposition~\ref{prop:deltas_chosen_to_heal}. %The average infection ratio of virus~$k$ is denoted by $\Bar{p}^k(t)$.
    }
    \label{fig:control_heal}
\end{figure}

\begin{figure}[h!] 
\centering
    \scalebox{0.8}[0.8]{\includegraphics[width=\columnwidth]{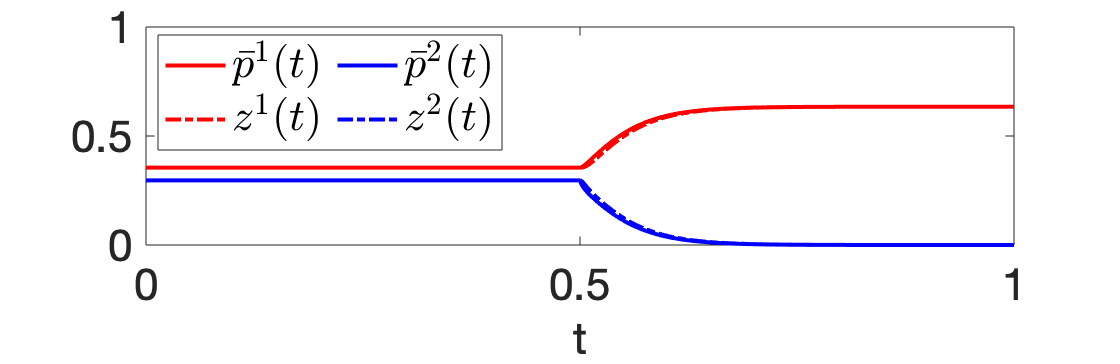}}
    \caption{Simulation with two viruses (red and blue), initialized at a coexisting equilibrium. At $t = 0.5$ the healing rates of virus~$2$ are changed to fulfill Theorem~\ref{thm:virus-as-vaccine}. %The average infection ratio of virus~$k$ is denoted by $\Bar{p}^k(t)$.
    }
    \label{fig:control_exclusion}
\end{figure}

The simulations depicted in Figures~\ref{fig:control_heal} and~\ref{fig:control_exclusion} were both initialized at the coexisting equilibrium from Figure~\ref{fig:coexistence}, with all parameters as in that simulation. %\seb{[Should we not be highlighting Theorem~\ref{thm:all_deltas_chosen_to_heal}?]}
At $t = 0.5$ in the simulation in Figure~\ref{fig:control_heal},
the healing rates of virus~2, i.e., $\delta_i^2$, %are changed
are chosen as in~\eqref{eq:deltas_chosen_to_heal}, 
%to fulfill Proposition~\ref{prop:deltas_chosen_to_heal} for this virus,
with $\epsilon_i^2 = 0$ for all $i \in [15]$. Similarly, at $t = 0.5$ in the simulation in Figure~\ref{fig:control_exclusion}, the healing rates of virus~2, i.e., $\delta_i^2$, are chosen as in~\eqref{eq:deltas_chosen_to_leverage}.  
%are changed to fulfill Theorem~\ref{thm:virus-as-vaccine}. 
More specifically, $\delta_i^2 = 2$ for $i \in [8]$ as before, but $\delta_i^2 = 2.1$ for 
% the remaining 
$i \in [15]\setminus [8]$. %Notably, 
We assume that the cost of a strategy $\ell$ (denoted as $\cost^{\ell}$) is given by the sum of all healing rates (i.e., $\cost^{\ell} = \sum_{i=1}^{15} \delta_i^2$, where $\delta_i^2$ is chosen as in~\eqref{eq:deltas_chosen_to_heal} 
%\axel{\textbf{[The equation references are used here, while the theorem references are used above. Might be confusing.]}}
when $\ell =1$, and as in~\eqref{eq:deltas_chosen_to_leverage} when $\ell =2$). %\phil{$\sum_{i=1}^{15} \delta_i^2$ [should we define some notation for this?]}. 
Our simulations show that $\cost^{1} = 61$, %the first strategy requires $\sum_{i=1}^{15} \delta_i^2 = 59$,
whereas $\cost^{2} = 33$. %the second strategy only requires $\sum_{i=1}^{15} \delta_i^2 = 32$.
However, as seen in Figures~\ref{fig:control_heal} and~\ref{fig:control_exclusion}, the end result of both strategies is the same. That is, virus~1 persists and reaches its single-virus endemic equilibrium, whereas virus~2 is eradicated. Hence, the strategy in Theorem~\ref{thm:virus-as-vaccine} allows us to eradicate virus~2 at a lower cost than the strategy in Proposition~\ref{prop:deltas_chosen_to_heal}, in this case.
%\phil{[Probably worth pointing out that the higher cost strategy does result in faster convergence...]} 
However, the convergence to the single-virus endemic equilibrium of virus~1 is faster with Proposition~\ref{prop:deltas_chosen_to_heal}, as expected from the excessive healing rates.

% \subsection{Exploratory Simulations}
% \seb{This we need to think about. Kalle's email mentions a clustering approach; maybe see the Science paper by Tom Brinton et al. Another option, like Kalle suggests, would be to explore the behavior under time-varying topologies.}

\section{Conclusions}
\label{sect:conclusions}

%\subsection{Summarize what we did}
In this paper, we introduced and analyzed a novel SIWS model of competitive, multi-viral spread across a network of population nodes with a shared resource. We established conditions under which a virus is eradicated exponentially fast (resp. asymptotically). We also provided a condition that ensures a virus can reach and sustain a unique, single-virus endemic equilibrium of infection levels across the network. Taken together, these results allowed us to state a necessary and sufficient condition for convergence to the healthy state.
% \phil{of the population [remove this phrase?]}.
% \phil{Beyond results on independent behavior [not sure what this means...]}
Moreover, we presented sufficient conditions under which two viruses can sustain a coexisting equilibrium, 
% in which
that is, where
neither virus is eradicated. %[The following needs clarification] 
Conversely, we also provided a necessary condition for the existence of such a coexisting equilibrium. To mitigate the spreading process, we proposed two strategies. % that ensure eradication of one or more viruses.
The first strategy involved choosing the healing rate of each subpopulation, with respect to each virus, in a suitable manner so as to ensure that all subpopulations converge to the healthy state. %quarantining some populations nodes from the shared resource, and subsequently fixing the healing parameters of all population nodes.
The second strategy exploited the notion of competitive exclusion to ensure that in a bi-virus setup, with one virus being malignant and the other benign, the malignant virus becomes eradicated.
%
%\axel{\subsection*{Extensions and/or relevant open problems}

%[Stochastic model, more control strategies.]

%In defining~\eqref{eq:orig}, we implicitly allow the number of infected individuals to be treated as a continuous value. This step is not motivated by us here, but it can be derived from first principles by using a stochastic spreading model.}

The present paper concerns time-invariant SIWS models, whereas in order to obtain a better understanding of real-world scenarios (viz. mobile agents, mutating viruses), it is natural to consider time-varying SIWS models.
Secondly, in order to generalize the result on the existence of a coexisting equilibrium (cf. Theorem~\ref{thm:joint_eq_exist_shared}) to more than two viruses, we would need to establish uniqueness of the coexisting equilibrium for the bi-virus case, which is non-trivial.
%\axel{Thirdly}, for the bi-virus case, there is a gap between the sufficient condition and the necessary condition for existence of a coexisting equilibrium; see Theorem~\ref{thm:joint_eq_exist_shared} and Theorem~\ref{thm:noequi_single-virus}, respectively. Bridging this gap by providing a 
%condition that is both necessary and sufficient 
% necessary, and sufficient, condition 
%remains an open problem.
%\axel{Finally}, note that the mitigation strategies proposed in the present paper do not account for resource constraints. Hence, another line of future work could involve developing control strategies that factor in such limitations.

\vspace{-1ex}

\bibliographystyle{IEEEtran}
\bibliography{References-Axel}

\vspace{-1ex}

\vspace{-1ex}

\section*{Appendix}

%\vspace{-1ex}

\subsection*{Proof of Theorem~\ref{thm:expo} }

%\vspace{-1ex} 

Consider a virus~$k$ such that $s(B_w^k - D_w^k) < 0$. By Lemma~\ref{lem:pos} we know that $\mathcal{D}$ is positively invariant with respect to~\eqref{eq:full}, so because $y(0) \in \mathcal{D}$ we have $y(t) \in \mathcal{D}$ %, and hence $y^k(t) \in \mathcal{D}^k$, 
for all $t \geq 0$. Then,~\eqref{eq:yk} with respect to virus~$k$ can be bounded by:
\vspace{-1ex}
\begin{equation} \label{eq:exp_easy_peasy}
 \dot{y}^k(t) \leq (B_w^k - D_w^k)y^k(t).
\end{equation}

\vspace{-1ex}

Since $y^k(t) \geq \textbf{0}$ for all $t \geq 0$, it follows from Grönwall-Bellman's Inequality \cite[pg 651]{khalil2002nonlinear} that the solution of~\eqref{eq:yk} will be bounded above by the solution of the linear system corresponding to~\eqref{eq:exp_easy_peasy} with equality. Since $s(B_w^k - D_w^k) < 0$, we know that $\textbf{0}$ is globally exponentially stable in the linear system. Therefore,  
%we know that $y^k(t)$ converges to $\textbf{0}$ exponentially fast, from any initial condition $y^k(0) \in \mathcal{D}^k$. Thus, 
the eradicated state of virus~$k$ is exponentially stable, with domain of attraction containing $\mathcal{D}^k$.~$\square$

\vspace{-2ex}

\subsection*{Proof of Lemma~\ref{lem:equi_non-zero_nonone} 
% \axel{\textbf{[Cleaned]}}
}
%\vspace{-1ex} 
Consider an equilibrium $y \in \mathcal{D}$ of system~\eqref{eq:full}. 
%First, note that by assumption, $\delta_i^k > 0$,  for all $i \in [n]$ and $k \in [m]$. 
Assume, by way of contradiction, that $\sum_{k=1}^m p_i^k \geq 1$ for some $i \in [n]$. Plugged into~\eqref{eq:split} under Assumption~\ref{assum:base}, we obtain 
%Inserted into~\eqref{eq:split} under Assumption~\ref{assum:base}, this would 
%\seb{[does it follow from~\eqref{eq:split}?]} \axel{\textbf{[Yes, though I suppose it is not trivial. I have changed the wording in this paragraph.]}} 
%imply: 
%
\vspace{-1ex}
\begin{align}
\textstyle \sum_{k=1}^m \dot{p}_i^k(t) %&= \textstyle \sum_{k=1}^m \Big{(} - \delta_i^k p_i^k + \big{(}~1 - \textstyle \sum_{l=1}^m p_i^l \big{)} \nonumber \\
%&\,\,\,\,\,\,\,\,\,\,\,\,\,\,\,\,\,\,\,\,\,\,\,\,\,\,\,\,\,\,\,\,\,\,\,\,\,\,\,\,\,\,\,\,\,\,\,\,\,\,\,\,\,\, \times \big{(} \beta_{iw}^k z^k + \textstyle \sum_{j=1}^{n} \beta_{ij}^k p_j^k \big{)} \Big{)} \nonumber \\
&\leq - \textstyle \sum_{k=1}^m \delta_i^k p_i^k %\label{eq:pder_use_assum_get_money} \\
< 0, \label{eq:pder_use_delta_finish}
\end{align} 

\vspace{-1ex}

\noindent
where %\eqref{eq:pder_use_assum_get_money} 
\eqref{eq:pder_use_delta_finish} follows from i) Assumption~\ref{assum:base}, ii) $\sum_{k=1}^m p_i^k \geq 1$ and iii) that $y \in \mathcal{D}$. %while~\eqref{eq:pder_use_delta_finish} is due to i) $\delta_i^k >0$ for all $i \in [n]$ and for each $k \in [m]$, and ii) the assumption that $\sum_{k=1}^m p_i^k \geq~1$. %\textbf{[I do not know how much to include in the "follows from" part here.]}
Note that~\eqref{eq:pder_use_delta_finish} is a contradiction of the fact that $y$ is an equilibrium, following from the assumption $\sum_{k=1}^m p_i^k \geq 1$ for some $i \in [n]$. Therefore, $\sum_{k=1}^m p^k \ll \textbf{1}$. Now, assume by way of contradiction that $\sum_{k=1}^m z^k \geq 1$. Plugged into~\eqref{eq:split} under Assumption~\ref{assum:base}, we obtain % with $\sum_{k=1}^m p^k \ll \textbf{1}$, this implies that:
\vspace{-1ex}
\begin{align}
\textstyle \sum_{k=1}^m \dot{z}^k %&= \textstyle \sum_{k=1}^m \Big{(} \delta_w^k \big{(} - z^k + \textstyle \sum_{i=1}^{n} c_i^k p_i^k \big{)} \Big{)} \nonumber \\
&< \textstyle %\delta_w^k
\sum_{k=1}^m \delta_w^k \big{(} 1 - z^k \big{)} \label{eq:zder_c_sums_to_one} \\
&\leq 0, \label{eq:zder_use_sum_finish}
\end{align}

\vspace{-1ex}

\noindent
where~\eqref{eq:zder_c_sums_to_one}
is due to the following reason: First, notice that $\textstyle \sum_{i=1}^n c_i^k = 1$ for all $k \in [m]$. Since $\sum_{k=1}^m p^k \ll \textbf{1}$, it follows that %, for each $i \in [n]$, $\textstyle \sum_{k=1}^m p_i^k <1$, which further implies that for any given $i \in [n]$, 
$p_i^k < 1$ for all $i \in [n]$ and $k \in [m]$. Therefore, $\textstyle \sum_{i=1}^n c_i^k p_i^k < 1$. The inequality~\eqref{eq:zder_use_sum_finish} follows from Assumption~1, %the fact that, for all $k \in [m]$, $\delta_w^k >0$
and the assumption that $\sum_{k=1}^m z^k \geq 1$. Note that~\eqref{eq:zder_use_sum_finish} is a contradiction of the fact that $y$ is an equilibrium of system~\eqref{eq:full}. %Hence, it follows that
Since this contradiction follows from the assumption $\sum_{k=1}^m z^k \geq 1$, we must have $\textstyle \sum_{k=1}^m y^k \ll \textbf{1}$, and therefore $y^k \ll \textbf{1}$, for all $k \in [m]$, for any equilibrium $y \in \mathcal{D}$.
%follows from $\textstyle \sum_{i=1}^n c_i^k =~1$ for all $k \in [m]$, and~\eqref{eq:zder_use_sum_finish} follows from $\sum_{k=1}^m z^k \geq~1$. \textbf{[I do not know how much to include in the "follows from" part here.]}~\eqref{eq:zder_use_sum_finish} contradicts that $y$ is an equilibrium. It follows that we must have $\textstyle \sum_{k=1}^m y^k \ll \textbf{1}$, and therefore $y^k \ll \textbf{1}$ for all $k \in [m]$, for any equilibrium $y \in \mathcal{D}$. 
Now, for all $k \in [m]$, $y^k$ is a equilibrium of~\eqref{eq:yk}, %and $D_w^k$ is a positive diagonal matrix and therefore invertible, 
so we have
\vspace{-1ex}
\begin{align}
(- D_w^k + (I - \textstyle \sum_{l=1}^m X(y^l))B_w^k) y^k &= 0, \nonumber\\
\implies (I - \textstyle \sum_{l=1}^m X(y^l)) (D_w^k)^{-1} B_w^k y^k &= y^k. \label{eq:eigenvec_bigger_than_zero}
\end{align}

\vspace{-1ex}

\noindent
Then, since $y^k \ll \textbf{1}$,
%$(I - \textstyle \sum_{l=1}^m X(y^l))$ and $(D_w^k)^{-1}$ are positive diagonal matrices, and $B_w^k$ is an irreducible nonnegative matrix, 
$(I - \textstyle \sum_{l=1}^m X(y^l)) (D_w^k)^{-1} B_w^k$ is an irreducible nonnegative matrix for all $k \in [m]$. Now, for some $k \in [m]$, assume by way of contradiction that $y^k > \textbf{0}$, with $y_i^k = 0$ for all $i \in W$, where $W \subset [n+1]$ is nonempty. Then, by the properties of irreducible nonnegative matrices, $((I - \textstyle \sum_{l=1}^m X(y^l)) (D_w^k)^{-1} B_w^k y^k)_j > 0$ for some $j \in W$. Since $y_j^k = 0$, this contradicts~\eqref{eq:eigenvec_bigger_than_zero}, and therefore we must either have $y^k \gg \textbf{0}$, or $y^k = \textbf{0}$, for each $k \in [m]$.~$\square$

% \subsection*{Proof of Theorem~\ref{thm:expo}}
% Consider a virus~$k$ such that $s(B_w^k - D_w^k) < 0$. By Lemma~\ref{lem:pos} we know that $\mathcal{D}$ is positively invariant with respect to~\eqref{eq:full}, so because $y(0) \in \mathcal{D}$ we have $y(t) \in \mathcal{D}$, and hence $y^k(t) \in \mathcal{D}^k$, for all $t \geq 0$. Then,~\eqref{eq:yk} with respect to virus~$k$ can be bounded by:
% %
% \begin{equation} \label{eq:exp_easy_peasy}
%  \dot{y}^k(t) \leq (B_w^k - D_w^k)y^k(t).
% \end{equation}
% %
% Given that $y^k(t) \in \mathcal{D}^k$ implies $y^k(t) \geq \textbf{0}$ for all $t \geq 0$, it follows from Grönwall-Bellman's Inequality \cite[pg 651]{khalil2002nonlinear} that the solution of~\eqref{eq:yk} will be bounded above by the solution of the linear system~\eqref{eq:exp_easy_peasy}. Since $s(B_w^k - D_w^k) < 0$, we know that $\textbf{0}$ is globally exponentially stable in the linear system. Therefore,  
% %we know that $y^k(t)$ converges to $\textbf{0}$ exponentially fast, from any initial condition $y^k(0) \in \mathcal{D}^k$. Thus, 
% the eradicated state of virus~$k$ is exponentially stable, with domain of attraction containing $\mathcal{D}^k$. \hfill $\square$

%\vspace{-2ex}

\subsection*{Proof of Theorem~\ref{thm:asymp} 
% \axel{\textbf{[Cleaned]}}
}
%\vspace{-1ex}
Consider a virus~$k$ such that $s(B_w^k - D_w^k) \leq 0$. %By Lemma~\ref{lem:pos} we know that $\mathcal{D}$ is positively invariant with respect to~\eqref{eq:full}, so because $y(0) \in \mathcal{D}$ we have $y(t) \in \mathcal{D}$ for all $t \geq 0$. As a consequence, Lemma~\ref{lem:pos} states that $\mathcal{D}^k$ is positively invariant with respect to $\eqref{eq:yk}$, so $y^k(t) \in \mathcal{D}^k$ for all $t \geq 0$. 
Since $y(0) \in \mathcal{D}$, Lemma~\ref{lem:pos} states that we have $y(t) \in \mathcal{D}$ for all $t \geq 0$, and further that $\mathcal{D}^k$ is positively invariant with respect to~\eqref{eq:yk}. Now, note that if $s(B_w^k - D_w^k) < 0$, Theorem~\ref{thm:expo} implies that the eradicated state is exponentially stable for virus~$k$, with domain of attraction containing $\mathcal{D}^k$. %, in turn implying asymptotic stability of the eradicated state with the same domain of attraction. 
Hence, the rest of the proof considers the case when $s(B_w^k - D_w^k) = 0$. 

Since %$B_w^k$ is an irreducible nonnegative matrix, and $D_w^k$ is a diagonal matrix, 
$(B_w^k - D_w^k)$ is an irreducible Metzler matrix, by Lemma~\ref{lem:metz_irreduc} there exists a positive diagonal matrix $Q^k$ such that $\Lambda^k \coloneqq ((B_w^k - D_w^k)^T Q^k + Q^k (B_w^k - D_w^k)) \preccurlyeq 0$. Define the Lyapunov function candidate $V(y^k(t)) = y^k(t)^T Q^k y^k(t)$ with $\mathcal{D}^k$ as the domain. Note that $V(y^k(t)) \succ 0$, and that $V(y^k(t))$ fulfills~\eqref{eq:bounded_contours}. Differentiating $V(y^k(t))$ yields
\vspace{-1ex}
\begin{align} \label{eq:asymp_lyapunov_orig}
\Dot{V}(y^k(t)) = \, &2y^k(t)^T Q^k (B_w^k - D_w^k) y^k(t) \nonumber \\
&- 2y^k(t)^T Q^k \textstyle \sum_{l=1}^m X(y^l(t)) B_w^k y^k(t) \nonumber \\
\leq \, %&2y^k(t)^T Q^k (B_w^k - D_w^k) y^k(t) \nonumber \\
%&-~2y^k(t)^T Q^k X(y^k(t)) B_w^k y^k(t) \nonumber \\
%= \, 
& y^k(t)^T \Lambda^k y^k(t) \nonumber \\
&- 2y^k(t)^T Q^k X(y^k(t)) B_w^k y^k(t).
\end{align}

\vspace{-1ex}

\noindent
We will now show that $\Dot{V}(y^k) < 0$ if $y^k > \textbf{0}$. First, consider the case where %$y^k(t) > \textbf{0}$, with 
$y_i^k = 0$ for some $i \in [n+1]$. Then, noting that %since $Q^k, X(y^k(t))$ and $B_w^k$ are nonnegative matrices, we have 
$y^{kT} Q^k X(y^k) B_w^k y^k \geq 0$, we see that~\eqref{eq:asymp_lyapunov_orig} is bounded by
\vspace{-1ex}
\begin{equation} \label{eq:asymp_lyapunov_y_is_somewhat_zero}
\Dot{V}(y^k) \leq y^{kT} \Lambda^k y^k.
\end{equation}

\vspace{-1ex}

\noindent
Given that %$(B_w^k - D_w^k)$ is an irreducible Metzler matrix, and $Q^k$ is a positive diagonal matrix, $\Lambda^k$ is an irreducible Metzler matrix, %. Because 
$\Lambda^k \preccurlyeq 0$, we have $s(\Lambda^k) \leq 0$. Then, since $\Lambda^k$ is an irreducible Metzler matrix, it follows from Lemma~\ref{lem:perron_frob_metz} that $r \coloneqq s(\Lambda^k)$ is a simple eigenvalue of $\Lambda^k$, with a corresponding eigenvector $\zeta \gg \textbf{0}$. Since we consider the case where $y_i^k = 0$ for some $i \in [n+1]$, $y^k$ can not be parallel to $\zeta$. By the Rayleigh-Ritz Theorem~\cite[Theorem~4.2.2]{horn2012matrix}, $y^{kT} \Lambda^k y^k = r y^{kT} y^k$ only if $y^k$ is parallel to $\zeta$, and $y^{kT} \Lambda^k y^k < r y^{kT} y^k$ otherwise. Since $r \leq 0$, it follows from~\eqref{eq:asymp_lyapunov_y_is_somewhat_zero} that $\Dot{V}(y^k) < 0$ when $y_i^k = 0$ for some $i \in [n+1]$. %\seb{[does this hold even when $r=0$?]} \axel{\textbf{[Yes, see the new clarification.]}} } 
Now, consider the case when $y^k \gg \textbf{0}$. Recall that $\Lambda^k \preccurlyeq 0$, hence,~\eqref{eq:asymp_lyapunov_orig} is bounded by
\vspace{-1.5ex}
\begin{equation} \label{eq:asymp_lyapunov_y_is_not_zero}
\Dot{V}(y^k) \leq - 2y^{kT} Q^k X(y^k) B_w^k y^k.
\end{equation}

\vspace{-1.5ex}

\noindent
Since $B_w^k$ is an irreducible nonnegative matrix, %$X(y^k(t))$ is a nonnegative diagonal matrix with some positive elements, 
$Q^k$ is a positive diagonal matrix and $y^k(t) \gg \textbf{0}$, it follows that $y^{kT} Q^k X(y^k) B_w^k y^k > 0$. Hence,~\eqref{eq:asymp_lyapunov_y_is_not_zero} gives us $\Dot{V}(y^k) < 0$ when $y^k \gg \textbf{0}$. Then we have $\Dot{V}(y^k) < 0$ for all $y^k > \textbf{0}$, and it is clear that $\Dot{V}(\textbf{0}) = 0$. Therefore, $\Dot{V}(y^k) \prec 0$. Finally, since $\mathcal{D}^k$ is positively invariant with respect to~\eqref{eq:yk}, we see that $V(y^k(t))$ meets the conditions for Proposition~\ref{prop:asymp_domatt}. This shows that the eradicated state of virus~$k$ is asymptotically stable, with domain of attraction containing $\mathcal{D}^k$.~$\square$

%\vspace{-2ex}

\subsection*{Proof of Lemma~\ref{lem:pos_never_zero} 
% \axel{\textbf{[Cleaned]}}
}
%\vspace{-1ex}
Lemma~\ref{lem:pos} states that $\mathcal{D}$ is positively invariant with respect to system~\eqref{eq:full}. It remains to be shown that if $y(0) \in \mathcal{D}$, and $y(0) > \textbf{0}$, it follows that $y(t) > \textbf{0}$ for all $t > 0$. With $y(0) > \textbf{0}$ we have $y^k(0) > \textbf{0}$ for some $k \in [m]$, and by Lemma~\ref{lem:pos} we know that $y^k(t) \geq \textbf{0}$ for all $t > 0$. Then,~\eqref{eq:yk} is bounded by
\vspace{-1ex}
\begin{equation} \label{eq:lower_bound_derivative}
\dot{y}^k(t) %&= \big{(} - D_w^k + (I - \textstyle \sum_{l=1}^m X(y^l(t)))B_w^k \big{)} y^k(t) 
\geq -D_w^k y^k(t), 
\end{equation}

\vspace{-1ex}

\noindent
due to i) $y^k(t) \geq \textbf{0}$, and ii) that $(I - \textstyle \sum_{l=1}^m X(y^l(t)))B_w^k$ is a nonnegative matrix. Integrating~\eqref{eq:lower_bound_derivative} yields
\vspace{-1ex}
\begin{equation} \label{eq:lower_bound}
y^k(t) \geq e^{- t D_w^k } y^k(0) > \textbf{0}, 
\end{equation}

\vspace{-1ex}

\noindent
for all $t>0$. The final inequality in~\eqref{eq:lower_bound} follows from the fact that $D_w^k$ is a positive diagonal matrix. Since $y^k(t) > \textbf{0}$, it follows that $y(t) > \textbf{0}$ for all $t > 0$. Therefore, $\mathcal{D} \setminus \{ \textbf{0} \}$ is positively invariant with respect to system~\eqref{eq:full}.~$\square$

\vspace{-2ex}

\subsection*{Proof of Theorem~\ref{thm:equi}}
%\noindent \textit{Proof of Theorem~\ref{thm:equi}:}
% We divide the proof into three parts. First we %show that the conditions in Theorem~\ref{thm:equi} 
% establish the existence of a single-virus endemic equilibrium under the given conditions. Subsequently, we show that this equilibrium is unique. Finally, we show that the equilibrium is asymptotically stable, with domain of attraction containing $\mathcal{D} \setminus \{ \textbf{0} \}$.
% \vspace{1ex}

\noindent \textit{Part~1 -- Proof of existence:}

\noindent Note that if $y>\textbf{0}$, $\diag(D_w^{-1} B_w y)$ is a nonnegative diagonal matrix, and therefore the inverse of $(I + \diag(D_w^{-1} B_w y))$ exists. Define a map $T(y): \mathbb{R}_+^{n+1} \rightarrow \mathbb{R}_+^{n+1}$ such that
% \begin{gather*}
%  X\big{(}y^1(t)\big{)}
%  =
%  \begin{bmatrix}
%  diag(p^1(t)) & 0 \\
%  0 & 0 
%  \end{bmatrix},
% \end{gather*}
% \begin{equation}
%  \dot{y}^1(t) = \Big{(} - D_w^1 + B_w^1 - X\big{(}y^1(t)\big{)}B_w^1 \Big{)} y^1(t).
% \end{equation}
%Let $B_w = B_w^1, D_w = D_w^1, B = B^1, b = b^1, \delta_w = \delta_w^1, z = z^1$ and $y = y^1$.
\vspace{-1ex}
\begin{align*}
T(y) = (&I + \diag(D_w^{-1} B_w y))^{-1} \\
 &\times (D_w^{-1} B_w y + \diag(D_w^{-1} B_w y) [\textbf{0} \, y_{n+1}]^T).
\end{align*}

\vspace{-1.5ex}

\noindent
%Because 
%Since $\diag(D_w^{-1} B_w y)$ is a nonnegative diagonal matrix, the inverse of $(I + \diag(D_w^{-1} B_w y))$ exists, 
Observe that the components of $T(y)$ are
\vspace{-1ex}
\begin{align*}
T_{i}(y) &= \frac{(D_w^{-1} B_w y)_{i}}{1 + (D_w^{-1} B_w y)_{i}}, \text{ for } i \in [n], \\
T_{n+1}(y) &= \frac{(D_w^{-1} B_w y)_{n+1} y_{n+1} + (D_w^{-1} B_w y)_{n+1}}{1 + (D_w^{-1} B_w y)_{n+1}}.
\end{align*}
% 
% \noindent 

\vspace{-1.5ex}

\noindent
Note that the scalar function $s/(1+s)$ is increasing in $s$, and that $D_w^{-1} B_w$ is a nonnegative matrix. Therefore, $v \geq z$ implies $T(v) \geq T(z)$. Now, observe that a fixed point of $T(y)$ fulfills
\vspace{-4ex}
\begin{align} \label{eq:fixed}
 y = (&I + \diag(D_w^{-1} B_w y))^{-1} \nonumber \\
 &\times (D_w^{-1} B_w y + \diag(D_w^{-1} B_w y) [\textbf{0} \, y_{n+1}]^T).
\end{align}

\vspace{-1.5ex}

\noindent
Multiplying~\eqref{eq:fixed} by $(I + \diag(D_w^{-1} B_w y))$ gives us
\vspace{-1ex}
\begin{equation} \label{eq:diagflip}
 D_w^{-1} B_w y + \diag(D_w^{-1} B_w y) [\textbf{0} \, y_{n+1}]^T = (I + \diag(D_w^{-1} B_w y)) y. 
\end{equation}

\vspace{-1ex}

\noindent
Using the identity $\diag(u) v = \diag(v) u$,~\eqref{eq:diagflip} is equivalent to
\vspace{-2ex}
\begin{equation} \label{eq:z_diag_add}
 D_w^{-1} B_w y + \diag([\textbf{0} \, y_{n+1}]^T) D_w^{-1} B_w y = (I + \diag(y) D_w^{-1} B_w) y.
\end{equation}

\vspace{-1ex}

\noindent
For a given $y\in \mathbb{R}^{n+1}_+$, define $X(y)$ to be its diagonalization with the final element $y_{n+1}$ set to zero. 
%\axel{This is kind of shady. I don't think I can just reuse the $X(y)$ function defined in the Model section, since it is defined in the context of the model. This $X(x)$ function is "equivalent" to $X(y)$ whenever $m =~1$.} 
%
As such, subtracting $\diag([\textbf{0} \, y_{n+1}]^T) D_w^{-1} B_w y$ from~\eqref{eq:z_diag_add} yields
%
%\vspace{-1ex}
\begin{equation} \label{eq:X_D_commute}
 D_w^{-1} B_w y = (I + X(y) D_w^{-1} B_w) y.
\end{equation}
%

%\vspace{-1.5ex}

\noindent
Since $X(y)$ and $D_w^{-1}$ are diagonal matrices, they commute. Furthermore, by pre-multiplying~\eqref{eq:X_D_commute} with $D_w$, and suitably rearranging terms, we obtain
% Hence % as such
%~\eqref{eq:X_D_commute} is equivalent to:
% %
% \begin{equation} \label{eq:rearr_Dinv}
%  D_w^{-1} B_w x = (I + D_w^{-1} X(x) B_w) x.
% \end{equation}
% %
% Multiplying~\eqref{eq:rearr_Dinv} by $D_w$, and rearranging gives us:
%
\vspace{-1ex}
\begin{equation} \label{eq:equi_equa}
 (B_w - D_w - X(y) B_w) y = 0. 
\end{equation}

\vspace{-1.5ex}

\noindent
A solution of equation~\eqref{eq:equi_equa} is clearly an equilibrium of~\eqref{eq:yk} with $m=1$. As such, it suffices to show that $T(y)$ has a non-zero fixed point $\Tilde{y}$ in $\mathcal{D}$. %in order to prove the theorem.
We will now show that at least one such fixed point exists.
Since %we have
$s(B_w - D_w) > 0$, by Lemma~\ref{lem:eigspec}, % we know that
$\rho(D_w^{-1} B_w) > 1$. %Further, given that $B_w$ is an irreducible nonnegative matrix and $D_w^{-1}$ is a positive diagonal matrix, 
Note that $D_w^{-1} B_w$ is an irreducible nonnegative matrix. Hence, by item~\ref{item:perfrob_simpleeig} in Lemma~\ref{lem:perron_frob}, %we know that
$\lambda^* = \rho(D_w^{-1} B_w)$ is a simple eigenvalue of $D_w^{-1} B_w$. Furthermore, by item~\ref{item:perfrob_pos_exists} in Lemma~\ref{lem:perron_frob}, we know that the eigenspace of $\lambda^*$ is spanned by a vector $y^* \gg \textbf{0}$. Then, since $\lambda^* > 1$, there exists some constant $\epsilon > 0$ such that, for all $i \in [n+1]$, we have $\epsilon y_i^* \leq (\lambda^* - 1)/\lambda^*$, which implies that $1 \leq \lambda^*/(1 + \lambda^* \epsilon y_i^*)$. %and further 
Hence, $\epsilon y_i^* \leq \lambda^* \epsilon y_i^* / (1 + \lambda^* \epsilon y_i^*)$, which further implies
% 
%\vspace{-1ex}
\begin{equation} \label{eq:i_comp}
 \epsilon y_i^* \leq (D_w^{-1} B_w \epsilon y^*)_i / (1 + (D_w^{-1} B_w \epsilon y^*)_i).
\end{equation}

%\vspace{-1.5ex}

\noindent
% 
% \noindent 
Noting that $(D_w^{-1} B_w \epsilon y^*)_{n+1} \epsilon y_{n+1}^* > 0$, we also have
% 
%\vspace{-1ex}
\begin{equation} \label{eq:n+1_comp}
\epsilon y_{n+1}^* \leq \frac{(D_w^{-1} B_w \epsilon y^*)_{n+1} \epsilon y_{n+1}^* + (D_w^{-1} B_w \epsilon y^*)_{n+1}}{1 + (D_w^{-1} B_w \epsilon y^*)_{n+1}}.
\end{equation}

%\vspace{-1ex}

\noindent
Due to the inequalities~\eqref{eq:i_comp} and~\eqref{eq:n+1_comp}, we have $T(\epsilon y^*) \geq \epsilon y^*$. Since $y \geq z$ implies %that
$T(y) \geq T(z)$, it follows that for any $y \geq \epsilon y^*$ we have $T(y) \geq \epsilon y^*$. %Now, studying 
Consider $T(\textbf{1})$ %we see that,
for $i\in [n]$,
\vspace{-1ex}
\begin{equation} \label{eq:i_one_ineq}
T_{i}(\textbf{1}) = (D_w^{-1} B_w \textbf{1})_{i} / (1 + (D_w^{-1} B_w \textbf{1})_{i}) \leq 1.
\end{equation}

%\vspace{-1.5ex}

\noindent
For $i = n+1$, %note that:
% 
% \begin{gather*}
%  D_w^{-1} B_w = 
%  \begin{bmatrix}
%  D^{-1} & 0 \\
%  0 & \delta_w^{-1} 
%  \end{bmatrix}
%  \begin{bmatrix}
%  B & b \\
%  \delta_w c & 0 
%  \end{bmatrix}
%  =
%  \begin{bmatrix}
%  D^{-1} B & D^{-1} b \\
%  c & 0 
%  \end{bmatrix}
%  ,
% \end{gather*}
% % 
% $$\implies 
%$$T_{n+1}(x) = \frac{([c \,\, 0] x) \, x_{n+1} + [c \,\, 0] x}{1 + [c \,\, 0] x}.$$
% 
% \noindent 
%Since $c \,\, \textbf{1} =~1$,
we have
% 
%\vspace{-1ex}
\begin{equation} \label{eq:n+1_one_ineq}
T_{n+1}(\textbf{1}) = 2[c \,\, 0] \textbf{1} / (1 + [c \,\, 0] \textbf{1}) \leq 1.
\end{equation}

%\vspace{-1ex}

\noindent
% 
% \noindent 
Due to~\eqref{eq:i_one_ineq} and~\eqref{eq:n+1_one_ineq}, we have $T(\textbf{1}) \leq \textbf{1}$. Since $v \geq w$ implies $T(v) \geq T(w)$, it follows that $T(y) \leq \textbf{1}$ if $y \leq \textbf{1}$. By Brouwer's fixed-point theorem \cite[Theorem~9.3]{starr_2011}, there is at least one fixed point of $T(y)$ in the domain $\{ y : \epsilon y^* \leq y \leq \textbf{1}\}$. Since a fixed point of $T(y)$ is equivalent to an equilibrium of~\eqref{eq:full}, by Lemma~\ref{lem:equi_non-zero_nonone}, any fixed point must fulfill $y \ll \textbf{1}$.

In conclusion, the map $T(y)$ has at least one fixed point in the domain $\{ y : \epsilon y^* \leq y \ll \textbf{1}\}$, and therefore system~\eqref{eq:full} has at least one equilibrium $\Tilde{y} \in \mathcal{D}$ such that $\textbf{0} \ll \Tilde{y} \ll \textbf{1}$. 

\vspace{.75ex}

\noindent \textit{Part~2 -- Proof of uniqueness}

\noindent We will now prove that the single-virus endemic equilibrium is unique. Suppose that there are two single-virus endemic equilibria, $\Tilde{y}$ and $\Tilde{\mathbf{y}}$. By Lemma~\ref{lem:equi_non-zero_nonone} we have $\Tilde{y} \gg \textbf{0}$ and $\Tilde{\mathbf{y}} \gg \textbf{0}$. Let $\kappa = \max_{i \in [n+1]} \Tilde{y}_i / \Tilde{\mathbf{y}}_i$. First we show that $\kappa$ is given by
\vspace{-1ex}
\begin{equation} \label{eq:unique_single-virus_kappadef}
\kappa = \max_{i \in [n]} \Tilde{y}_i/\Tilde{\mathbf{y}}_i.
\end{equation}

\vspace{-1.5ex}

\noindent
% First we will show that $\kappa$ is given by:
% %
% \begin{equation} \label{eq:unique_single-virus_kappadef}
% \kappa = \max_{i \in [n]} \frac{\Tilde{y}_i}{\Tilde{\mathbf{y}}_i}.
% \end{equation}
% %
To do this, assume by way of contradiction that $\kappa = \Tilde{y}_{n+1}/\Tilde{\mathbf{y}}_{n+1}$, and $\kappa > \Tilde{y}_i/\Tilde{\mathbf{y}}_i$, for all $i \in [n]$. Note that since both $\Tilde{y}$ and $\Tilde{\mathbf{y}}$ are equilibria of system~\eqref{eq:yk}, we have
\vspace{-1.5ex}
\begin{equation} \label{eq:shared_resource_weighted_average}
\begin{split}
\Tilde{y}_{n+1} &= [c \,\, 0] \Tilde{y} = \textstyle \sum_i^{n} c_i \Tilde{y}_i,\\
\Tilde{\mathbf{y}}_{n+1} &= [c \,\, 0] \Tilde{\mathbf{y}} = \textstyle \sum_i^{n} c_i \Tilde{\mathbf{y}}_i.
\end{split}
\end{equation}

\vspace{-1.5ex}

\noindent
Since we assume that $\kappa > \Tilde{y}_i/\Tilde{\mathbf{y}}_i$, we have $\kappa \Tilde{\mathbf{y}}_i > \Tilde{y}_i$, for all $i \in [n]$. Then,~\eqref{eq:shared_resource_weighted_average} gives us
\vspace{-1ex}
\begin{equation*}
\Tilde{y}_{n+1} = \textstyle \sum_i^{n} c_i \Tilde{y}_i %\nonumber \\
< \textstyle \kappa \sum_i^{n} c_i \Tilde{\mathbf{y}}_i %\label{eq:singleequi_unique_notn+1_use_kappaassum}\\
= \kappa \Tilde{\mathbf{y}}_{n+1}, %\nonumber
\end{equation*}

\vspace{-1.5ex}

\noindent
%This would imply that
%where~\eqref{eq:singleequi_unique_notn+1_use_kappaassum} follows from $\kappa \Tilde{\mathbf{y}}_i > \Tilde{y}_i$. 
Hence, $\kappa > \Tilde{y}_{n+1}/\Tilde{\mathbf{y}}_{n+1}$, which contradicts the assumption that $\kappa = \Tilde{y}_{n+1}/\Tilde{\mathbf{y}}_{n+1}$. 
% Hence, $\kappa \leq \Tilde{y}_i/\Tilde{\mathbf{y}}_i$ for at least some %\seb{[should this not be for all $i \in [n]$?]}
% $i \in [n]$, meaning that 
Therefore, $\kappa$ must be given by equation~\eqref{eq:unique_single-virus_kappadef}.
Now, by~\eqref{eq:unique_single-virus_kappadef} we know that $\Tilde{y} \leq \kappa \Tilde{\mathbf{y}}$. For some $j \in [n]$ we have $\Tilde{y}_j = \kappa \Tilde{\mathbf{y}}_j$. Assume, by way of contradiction, that $\kappa > 1$. Then, using the fact that an equilibrium of~\eqref{eq:yk} also constitutes a fixed point of $T(y)$ (see part~1 of this proof), we have
% can show that:
%
\vspace{-1ex}
\begin{align}
\Tilde{y}_j &= (D_w^{-1} B_w \Tilde{y})_{j} / (1 + (D_w^{-1} B_w \Tilde{y})_{j}) \nonumber \\
&\leq (D_w^{-1} B_w \kappa \Tilde{\mathbf{y}})_{j} / (1 + (D_w^{-1} B_w \kappa \Tilde{\mathbf{y}})_{j}) \label{eq:singleequi_unique_kappaybigger} \\
&< \kappa (D_w^{-1} B_w \Tilde{\mathbf{y}})_{j} / (1 + (D_w^{-1} B_w \Tilde{\mathbf{y}})_{j}) \label{eq:singleequi_unique_kappabiggerthanone} \\
&= \kappa \Tilde{\mathbf{y}}_j \label{eq:singleequi_unique_alsoequilibrium}  \\
&= \Tilde{y}_j, \label{eq:singleequi_unique_j_hasequality}
\end{align}
%

%\vspace{-1.5ex}

\noindent
where~\eqref{eq:singleequi_unique_kappaybigger} follows from $\Tilde{y} \leq \kappa \Tilde{\mathbf{y}}$ and that $T(v) \geq T(w)$ whenever $v \geq w$,~\eqref{eq:singleequi_unique_kappabiggerthanone} follows from the assumption $\kappa > 1$, and~\eqref{eq:singleequi_unique_alsoequilibrium} follows from the fact that $\Tilde{\mathbf{y}}$ is an equilibrium of~\eqref{eq:yk}. Note that~\eqref{eq:singleequi_unique_j_hasequality} is a contradiction, following from our assumption that $\kappa >1$. Hence, $\kappa \leq 1$, meaning that $\Tilde{y} \leq \Tilde{\mathbf{y}}$. Switching the roles of $\Tilde{y}$ and $\Tilde{\mathbf{y}}$, we see that $\Tilde{\mathbf{y}} \leq \Tilde{y}$. Therefore, $\Tilde{y} = \Tilde{\mathbf{y}}$, and thus the equilibrium is unique. 

\vspace{1.5ex}

\noindent \textit{Part 3 -- Proof of asymptotic stability}\\
%\axel{\textit{Part 3 of the proof of Theorem~\ref{thm:equi}:} \seb{[TBD: Use the previous heading. The current one is not symmetric with the headings for parts~1 and~2.]}} 
Recall that $\Tilde{y}$ %\seb{[TBD: I just noticed: in part~1, you refer to the endemic equilibrium as $y^{\star}$, whereas hereafter it seems like you are using $\tilde{y}$. I think it would make sense to go back to Part~1, and change $y^{\star}$ to $\tilde{y}$, so as not to confuse the reader.]}
%\seb{[at this point you have already established that $\Tilde{y}$, as defined previously, is the unique equilibrium. So you are not \emph{letting} $\Tilde{y}$ to be the unique non-zero equilibrium, but rather \emph{recalling} it is so. Makes sense?]} 
is the unique single-virus endemic equilibrium of system~\eqref{eq:yk} in $\mathcal{D}$,
%Employing Proposition~\ref{prop:asymp_domatt}, we intend to show that $\Tilde{y}$ is asymptotically stable, with domain of attraction containing $\mathcal{D} \setminus \{ \textbf{0} \}$. \seb{[I am not sure what this paragraph does. I think everything in this paragraph up to the next sentence could be skipped.]}} 
where $\textbf{0} \ll \Tilde{y} \ll \textbf{1}$, and
\vspace{-1ex}
\begin{align} \label{eq:equi_necess_single-virus}
(- D_w + (I - X(\Tilde{y}))B_w) \Tilde{y} = \textbf{0}.
\end{align}
%

%\vspace{-1.5ex}

\noindent
%\seb{[Does not~\eqref{eq:equi_necess_single-virus}  follow from the fact that $\tilde{y}$ is an equilibrium? If so, then why do we need Lemma~\ref{lem:equi_non-zero_nonone} here?]}
For $y(t) \in \mathcal{D}$, let $\Delta y(t) = y(t) - \Tilde{y}$. %and $X(\Delta y(t)) = X(y(t)) - X(\Tilde{y})$. %\seb{[$\Delta y(t)$ is defined for some arbitrary $y(t)$, is it not?]} \axel{\textbf{[In a sense. In the following equations $\Delta y(t)$ is meant to be a dynamic variable, changing with $y(t)$.]}}. 
From \eqref{eq:yk} it follows that
%
%\vspace{-1ex}
\begin{align}
\Delta \dot{y}(t) &= (- D_w + (I - X(\Delta y) - X(\Tilde{y}))B_w) (\Delta y + \Tilde{y}) \nonumber \\
%&= (- D_w + (I - X(\Delta y) - X(\Tilde{y}))B_w) \Delta y \nonumber \\ 
%& \,\,\,\,\,\,\,\,\, - X(\Delta y)B_w\Tilde{y} \label{eq:use_equi_is_zero}\\
&= (- D_w + (I - X(\Tilde{y}))B_w) \Delta y - X(\Delta y)B_w y \label{eq:use_equi_is_zero} %\nonumber 
\\
&= (- D_w + (I - X(\Tilde{y}))B_w - X(B_w y)) \Delta y, \label{eq:use_X_diag_relation}
\end{align}

%\vspace{-1.5ex}

\noindent
where~\eqref{eq:use_equi_is_zero} follows from~\eqref{eq:equi_necess_single-virus}, and~\eqref{eq:use_X_diag_relation} follows from %the relation 
$X(u) v = X(v) u$. Now, since $D_w$ is %a positive diagonal matrix and therefore 
invertible,~\eqref{eq:equi_necess_single-virus} is equivalent to
\vspace{-1ex}
\begin{equation} \label{eq:equi_proper_single-virus}
(I - X(\Tilde{y})) D_w^{-1} B_w \Tilde{y} = \Tilde{y}.
\end{equation}
%
%\seb{Very minor thing: do we need to reuse~\eqref{eq:equi_proper_single-virus}? it does not look like, so maybe the numbering could be dropped\\}

%\vspace{-1.5ex}

\noindent
Since $B_w$ is an irreducible nonnegative matrix, and $D_w$ is a positive diagonal matrix, $\Tilde{y} \ll \textbf{1}$ ensures that $(I - X(\Tilde{y}))D_w^{-1} B_w$ is an irreducible nonnegative matrix, and in turn that $(- D_w + (I - X(\Tilde{y}))B_w)$ is an irreducible Metzler matrix. Therefore, since $\Tilde{y} \gg \textbf{0}$, item~\ref{item:perfrob_pos_necess} in Lemma~\ref{lem:perron_frob} applied to~\eqref{eq:equi_proper_single-virus} gives us $\rho((I - X(\Tilde{y})) D_w^{-1} B_w) = 1$,
%\seb{[Lemma~\ref{lem:perron_frob}, as currently constituted, tells us about the algebraic multiplicity of the largest eigenvalue; nothing about its value. But if you invoke equation~\ref{eq:equi_proper_single-virus}, then it's more clearer ]} 
which by Lemma~\ref{lem:eigspec} is equivalent to $s(- D_w + (I - X(\Tilde{y}))B_w) = 0$. Then, Lemma~\ref{lem:metz_irreduc} guarantees the existence of a positive diagonal matrix $Q$ such that $\Psi \coloneqq (- D_w + (I - X(\Tilde{y}))B_w)^T Q + Q (- D_w + (I - X(\Tilde{y}))B_w) \preccurlyeq 0$. 
%\axel{\textbf{[I am not sure that we even need Q in this proof.]}} \seb{[Response: We need it for showing positive definiteness, in line with a requirement in Prop.~\ref{prop:asymp_domatt}]} \axel{\textbf{[Q is used for this purpose right now, however I think that~\eqref{eq:equi_necess_single-virus} can be used without Q to get the requisite definiteness on its own.]}} 
Define the Lyapunov function candidate $V(\Delta y(t)) = \Delta y(t)^T Q \Delta y(t)$, with $y(t) \in \mathcal{D} \setminus \{ \textbf{0}\}$ as the domain. Note that $V(\Delta y(t)) \succ 0$ and that $V(\Delta y(t))$ fulfills~\eqref{eq:bounded_contours}. Differentiating $V(\Delta y(t))$ with respect to $t$ yields
\vspace{-1ex}
\begin{align}
\dot{V}(\Delta y(t)) %&=~2 \Delta y(t)^T Q \Delta \dot{y}(t) \nonumber \\
&= 2 \Delta y^T Q (- D_w + (I - X(\Tilde{y}))B_w) \Delta y \nonumber \\
& \,\,\,\,\, - 2  \Delta y^T Q X(B_w y) \Delta y \label{eq:using_delta_from_above} \\
&= \Delta y^T \Psi \Delta y - 2  \Delta y^T Q X(B_w y) \Delta y \label{eq:lyapunov_diff_singlevirusendemic}
\end{align}

\vspace{-1.5ex}

\noindent
where~\eqref{eq:using_delta_from_above} makes use of~\eqref{eq:use_X_diag_relation}. We want to show that $\dot{V}(\Delta y(t)) < 0$ for all $y(t) \in \mathcal{D} \setminus \{ \textbf{0}\}$ such that $\Delta y(t) = y(t) - \Tilde{y} \neq \textbf{0}$. First, consider all $y(t) \in \mathcal{D} \setminus \{ \textbf{0}\}$ such that $y(t) \gg \textbf{0}$. Combining~\eqref{eq:lyapunov_diff_singlevirusendemic} and the fact that $\Psi \preccurlyeq 0$ gives us
\vspace{-1ex}
\begin{align}
\dot{V}(\Delta y(t)) &\leq - 2 \Delta y(t)^T Q X(B_w y(t)) \Delta y(t). \label{eq:using_negative_semidefiniteness}
\end{align}

\vspace{-1.5ex}

\noindent
Note that, given $y(t) \gg \textbf{0}$, $Q X(B_w y(t))$ is a positive diagonal matrix, and thus $Q X(B_w y(t)) \succ 0$. Therefore,~\eqref{eq:using_negative_semidefiniteness} gives us $\dot{V}(\Delta y(t)) < 0$ for all $y(t) \in \mathcal{D} \setminus \{ \textbf{0}\}$ such that $y(t) \gg \textbf{0}$, $\Delta y(t) = y(t) - \Tilde{y} \neq \textbf{0}$. Now, consider all $y(t) \in \mathcal{D} \setminus \{ \textbf{0}\}$ such that $y(t) > \textbf{0}$, $y_i(t) = 0$ for some $i \in [n+1]$. Note that, given $y(t) > \textbf{0}$, $Q X(B_w y(t))$ is a nonnegative diagonal matrix. Then~\eqref{eq:lyapunov_diff_singlevirusendemic} can be bounded by
\vspace{-1ex}
\begin{align}
\dot{V}(\Delta y(t)) &\leq \Delta y(t)^T \Psi \Delta y(t). \label{eq:using_nonparallelism}
\end{align}

\vspace{-1.5ex}

\noindent
Given that $(- D_w + (I - X(\Tilde{y}))B_w)$ is an irreducible Metzler matrix and $Q$ is a positive diagonal matrix, $\Psi$ is an irreducible Metzler matrix. %\seb{[I think there may be a problem here. As far as I know, in general the product of two irreducible matrices is not necessarily irreducible.]}.
Employing~\eqref{eq:equi_necess_single-virus}, we see that
\vspace{-1ex}
\begin{equation} \label{eq:cool_trick}
\Tilde{y}^T \Psi \Tilde{y} = 0. 
\end{equation}

\vspace{-1.5ex}

\noindent
Item~\ref{item:perfrob_simpleeig} in Lemma~\ref{lem:perron_frob_metz} stipulates that $r \coloneqq s(\Psi)$ is a simple eigenvalue of $\Psi$. Due to $\Psi \preccurlyeq 0$ and the Rayleigh-Ritz Theorem \cite[Theorem~4.2.2]{horn2012matrix}, it follows from~\eqref{eq:cool_trick} 
%demonstrates \seb{[from~\eqref{eq:cool_trick} it follows]} 
that $r=0$, and that $\Tilde{y}$ spans the eigenspace of $r$. As such, due to $\Tilde{y} \gg \textbf{0}$, and $y(t) > \textbf{0}$ with $y_i(t) = 0$ for some $i \in [n+1]$, $y(t)$ can not be parallel to $\Tilde{y}$. As a consequence, $\Delta y(t)$ can not be parallel to $\Tilde{y}$. By the Rayleigh-Ritz Theorem \cite[Theorem~4.2.2]{horn2012matrix}, $x^T \Psi x = r x^T x$ only if $x$ is parallel to $\Tilde{y}$, and $x^T \Psi x < r x^T x$ otherwise. Therefore, $r = 0$ together with~\eqref{eq:using_nonparallelism} gives us $\dot{V}(\Delta y(t)) < 0$ for all $y(t) \in \mathcal{D} \setminus \{ \textbf{0}\}$ such that $y(t) \gg \textbf{0}$ and $y(t) - \Tilde{y} \neq \textbf{0}$. 

Thus, we have $\dot{V}(\Delta y(t)) < 0$ for all $y(t) \in \mathcal{D} \setminus \{ \textbf{0}\}$ such that $\Delta y(t) = y(t) - \Tilde{y} \neq \textbf{0}$, and it is clear that $\dot{V}(\textbf{0}) = 0$. Therefore, $\dot{V}(\Delta y(t)) \prec 0$ for $y(t) \in \mathcal{D} \setminus \{ \textbf{0}\}$.

Finally, from Lemma~\ref{lem:pos_never_zero} we have that $\mathcal{D} \setminus \{ \textbf{0}\}$ is a positively invariant set with respect to~\eqref{eq:full}. Thus, we see that $V(\Delta y(t))$ meets the conditions for Proposition~\ref{prop:asymp_domatt} with respect to the shifted coordinates $\Delta y(t) = y(t) - \Tilde{y}$, for all $y(t) \in \mathcal{D} \setminus \{ \textbf{0}\}$. This shows that the unique single-virus endemic equilibrium $\Tilde{y}$ is asymptotically stable, with domain of attraction containing $\mathcal{D} \setminus \{ \textbf{0}\}$. Thus, with parts~1,~2 and~3 in place, the proof of Theorem~\ref{thm:equi} is concluded.~$\square$

%\vspace{-2ex}

\subsection*{Proof of Proposition~\ref{prop:necessity}}

%\vspace{-1ex}

Suppose that, for some $k \in [m]$, $B_w^k$ is irreducible and $s(B_w^k - D_w^k) > 0$, and that $y^l = \textbf{0}$ for all $l \in [m]$, $l \neq k$. Then the dynamics of virus~$k$ can be written as
\vspace{-1ex}
\begin{align} \label{eq:yk_new}
 \dot{y}^k(t) =\big{(} - D_w^k + (I - X(y^k(t)))B_w^k \big{)}y^k(t).
\end{align}
%

%\vspace{-1ex}

\noindent
Note that~\eqref{eq:yk_new} corresponds to the dynamics of the single-virus case. Therefore, since $B_w^k$ is irreducible and $s(B_w^k - D_w^k) > 0$, it follows from the first and second parts of the proof of Theorem~\ref{thm:equi} that there exists a unique single-virus endemic equilibrium of the form $(\textbf{0}, \dots, \Tilde{y}^k, \dots, \textbf{0})$, with $\textbf{0} \ll \Tilde{y}^k \ll \textbf{1}$ in $\mathcal{D}$. This holds for each $k \in [m]$ such that $B_w^k$ is irreducible and $s(B_w^k - D_w^k) > 0$, by repeating the arguments above. \hfill $\square$

%\vspace{-2ex}

\subsection*{Proof of Theorem~\ref{thm:joint_eq_exist_shared}}
%\vspace{-1ex}
Recall that for $k\in [2]$, Assumption~\ref{assum:base} implies that $D_w^k$ is a positive diagonal matrix, and therefore invertible. Furthermore, note that $(I + \diag((D_w^1)^{-1} B_w^1 y^1))$ and $(I + \diag((D_w^2)^{-1} B_w^2 y^2))$ are positive diagonal matrices whenever $y^1 \geq 0$ and $y^2 \geq 0$, and are then also invertible. For $y \in \mathbb{R}^{n+1}_+$, define $X(y)$ to be $\diag (y)$ with $y_{n+1}$ set to zero. Define the maps $T^1(y^1, y^2): [0,1]^{n+1} \times [0,1]^{n+1} \rightarrow [0,1]^{n+1}$, and $T^2(y^1, y^2): [0, 1]^{n+1} \times [0,1]^{n+1} \rightarrow [0,1]^{n+1}$, such that
\vspace{-3ex}

\small
\begin{align*}
&T^1(y^1, y^2) = (I + \diag((D_w^1)^{-1} B_w^1 y^1))^{-1} \\
&\times ((I - X(y^2))(D_w^1)^{-1} B_w^1 y^1 + \diag((D_w^1)^{-1} B_w^1 y^1) [\textbf{0} \, y_{n+1}^1]^T), \nonumber 
\end{align*}
\small
\begin{align*}
&T^2(y^1, y^2) = (I + \diag((D_w^2)^{-1} B_w^2 y^2))^{-1} \\
&\times ((I - X(y^1))(D_w^2)^{-1} B_w^2 y^2 + \diag((D_w^2)^{-1} B_w^2 y^2) [\textbf{0} \, y_{n+1}^2]^T). 
\end{align*}

\normalsize

\vspace{-1ex}

\noindent
For $i \in [n]$, the $i^{\text{th}}$ components of the maps are
\vspace{-1ex}
\begin{align*}
T_i^1(y^1, y^2) &= \frac{(1 - y_i^2)((D_w^1)^{-1} B_w^1 y^1)_i}{1 + ((D_w^1)^{-1} B_w^1 y^1)_i}, 
%\text{ for all } i \in [n], 
\nonumber \\
T_i^2(y^1, y^2) &= \frac{(1 - y_i^1)((D_w^2)^{-1} B_w^2 y^2)_i}{1 + ((D_w^2)^{-1} B_w^2 y^2)_i}
%, \text{ for all } i \in [n]
.
\end{align*}

\vspace{-1ex}

\noindent Furthermore, the $(n+1)^{\text{th}}$ components of the maps are
\vspace{-2ex}

\small
\begin{align*}
T_{n+1}^1(y^1, y^2) &= \frac{((D_w^1)^{-1} B_w^1 y^1)_{n+1} + (D_w^1)^{-1} B_w^1 y^1)_{n+1} y_{n+1}^1}{1 + ((D_w^1)^{-1} B_w^1 y^1)_{n+1}}, 
%\text{ for all } i \in [n], 
\nonumber \\
T_{n+1}^2(y^1, y^2) &= \frac{((D_w^2)^{-1} B_w^2 y^2)_{n+1} + ((D_w^2)^{-1} B_w^2 y^2)_{n+1} y_{n+1}^2}{1 + ((D_w^2)^{-1} B_w^2 y^2)_{n+1}}
%, \text{ for all } i \in [n]
.
\end{align*}

\normalsize

\vspace{-1ex}

\noindent
Note that the scalar function $s/(1+s)$ is increasing in $s$, and for $k \in [2]$, the matrix $(D_w^k)^{-1} B_w^k$ is nonnegative, therefore, $T_i^k$ is an increasing function in $y_j^k$ for all $i, j \in [n+1]$.
%its $k^{\rm{th}}$ argument. 
Moreover, $T_i^1$ is a decreasing function in $y_i^2$ and $T_i^2$ is a decreasing function in $y_i^1$, for all $i \in [n]$. 
%\seb{[this only partly explains how you get the inequlaities in (38); the reader is unlikely to immediately grasp why, for instance, $T^{1}$ is decreasing in the second argument]} \axel{\textbf{[True. I am not sure how to explain the inequalities concisely, without causing more confusion. The shared resource component is just so bloated.]}}. \seb{[Response: For the $n+1^{th}$ component, it seems like the $k^{th}$ map depends only on the $k^{th}$ argument. For all other components, each map has a  scaling factor in the numerator (which is in terms of the non $k^{th}$ argument). The latter part is what we should highlight......makes sense?]}
Hence, for any $y^1, y^2 \in [0,1]^{n+1}$, if $v \geq z$ it follows that
%
%\vspace{-1ex}
\begin{equation} \label{eq:inequality_bonanza_shared}
\begin{split} 
T^1(v, y^2) \geq T^1(z, y^2)&, \,\, T^1(y^1, v) \leq T^1(y^1, z), \\
T^2(v, y^2) \leq T^2(z, y^2)&, \,\, T^2(y^1, v) \geq T^2(y^1, z).
\end{split}
\end{equation}

%\vspace{-1.5ex}

\noindent
%\phil{[I think we should maybe change the order of the arguments so the superscript comes first. This will simplify the process when we want to generalize to $m$ viruses, and make the directions of the inequalities above be the same (obviously)... now that I read the next part, this may be a bit more difficult to rearrange... thoughts?] \sebcancel{[Response: Yeah, the rearrangement does not seem so immediate. I was under the erroneous impression that extending it for arbitrary $m$ (using the same techniques as in $m=2$) would be straightforward.]}}
%
The inequalities in~\eqref{eq:inequality_bonanza_shared} state that $T^k(y^1, y^2)$ is increasing in its $k^{\rm{th}}$ argument and decreasing in its other argument, for $k \in [2]$. Let $y = (y^1, y^2)$, and let $T(y): [0,1]^{2(n+1)} \rightarrow [0,1]^{2(n+1)}$ be the map $T(y) = (T^1(y), T^2(y))$. A fixed point of $T(y)$ fulfills
\vspace{-6ex}

\small
\begin{align} \label{eq:double_eq_fixed_shared}
&y^1 = (I + \diag((D_w^1)^{-1} B_w^1 y^1))^{-1} \nonumber \\
&\times ((I - X(y^2))(D_w^1)^{-1} B_w^1 y^1 + \diag((D_w^1)^{-1} B_w^1 y^1) [\textbf{0} \, y_{n+1}^1]^T), \nonumber \\
&y^2 = (I + \diag((D_w^2)^{-1} B_w^2 y^2))^{-1} \nonumber \\
&\times ((I - X(y^1))(D_w^2)^{-1} B_w^2 y^2 + \diag((D_w^2)^{-1} B_w^2 y^2) [\textbf{0} \, y_{n+1}^2]^T). 
    % (I + \diag((D_w^1)^{-1} B_w^1 y^1))^{-1} ((&I - X(y^2))(D_w^1)^{-1} B_w^1 y^1 \nonumber \\
    % & + \diag((D_w^1)^{-1} B_w^1 y^1) [\textbf{0} \, y_{n+1}^1]^T) &= y^1, \nonumber \\
    % (I + \diag((D_w^2)^{-1} B_w^2 y^2))^{-1} ((&I - X(y^1))(D_w^2)^{-1} B_w^2 y^2 \nonumber \\
    % & + \diag((D_w^2)^{-1} B_w^2 y^2) [\textbf{0} \, y_{n+1}^2]^T) &= y^2.
\end{align}

\normalsize

%\vspace{-1.5ex}

\noindent
Pre-multiplying the first line (resp. second line) of~\eqref{eq:double_eq_fixed_shared} by $(I + \diag((D_w^1)^{-1} B_w^1 y^1))$ (resp. $(I + \diag((D_w^2)^{-1} B_w^2 y^2))$) gives us
%
%\vspace{-1.5ex}
\begin{align} 
    &(I + \diag((D_w^1)^{-1} B_w^1 y^1)) y^1 = \nonumber \\
    &(I - X(y^2))(D_w^1)^{-1} B_w^1 y^1 + \diag((D_w^1)^{-1} B_w^1 y^1) [\textbf{0} \, y_{n+1}^1]^T, \nonumber \\
    &(I + \diag((D_w^2)^{-1} B_w^2 y^2)) y^2 = \label{eq:double_eq_rearr_shared}\\
    &(I - X(y^1))(D_w^2)^{-1} B_w^2 y^2 + \diag((D_w^2)^{-1} B_w^2 y^2) [\textbf{0} \, y_{n+1}^2]^T.  \nonumber
\end{align}
%

%\vspace{-1.5ex}

\noindent
Rearranging~\eqref{eq:double_eq_rearr_shared}, making use of the identity $\diag(u) v = \diag(v) u$, and subtracting $\diag([\textbf{0} \, y_{n+1}^1]^T) (D_w^1)^{-1} B_w^1 y^1$ (resp. $\diag([\textbf{0} \, y_{n+1}^2]^T) (D_w^2)^{-1} B_w^2 y^2$) from the first (resp. second line) yields
\vspace{-1ex}
\begin{equation} \label{eq:double_eq_Dinv_shared}
\begin{split}  
(I - X(y^1) - X(y^2))(D_w^1)^{-1} B_w^1 y^1 &= y^1, \\
(I - X(y^1) - X(y^2))(D_w^2)^{-1} B_w^2 y^2 &= y^2.
\end{split}
\end{equation}
%

%\vspace{-1.5ex}

\noindent
Making use of the fact that diagonal matrices commute, pre-multiplying the first line (resp. second line) of~\eqref{eq:double_eq_Dinv_shared} by $D_w^1$ (resp. $D_w^2$), and rearranging terms gives us
\vspace{-1ex}
\begin{equation} \label{eq:double_eq_orig_shared}
\begin{split}
(-D_w^1 + (I - X(y^1) - X(y^2))B_w^1) y^1 &= \textbf{0}, \\
(-D_w^2 + (I - X(y^1) - X(y^2))B_w^2) y^2 &= \textbf{0}.
\end{split}
\end{equation}

\vspace{-1ex}

\noindent
Comparing~\eqref{eq:double_eq_orig_shared} with~\eqref{eq:yk}, it follows that a fixed point of $T(y)$ constitutes an equilibrium of system~\eqref{eq:full} and vice versa.
%\seb{[do we need the vice-versa part? I don't think we are using it anywhere.]} \axel{\textbf{[It is used to invoke Lemma~\ref{lem:equi_non-zero_nonone} near the end of this proof.]}} 
It suffices to show that $T(y)$ has a fixed point $\hat{y} = (\hat{y}^1, \hat{y}^2) \gg \textbf{0}$, such that $\hat{y}^1 + \hat{y}^2 \leq \textbf{1}$. 

Recall that $(\Tilde{y}^1, \textbf{0})$ and $(\textbf{0}, \Tilde{y}^2)$ are single-virus endemic equilibria of system~\eqref{eq:full}. Consider $T^1(\Tilde{y}^1, y^2)$. 
%\seb{[in this proof, this is the first time we are using $\Tilde{y}^1$. So we need to say the following:  By assumption,  $s(B_w^1 - D_w^1) > 0$ and $s(B_w^2 - D_w^2) > 0$. Consequently, by Proposition~\ref{prop:necessity} there exist two non-zero equilibria, namely $(\Tilde{y}^1, \textbf{0})$ and $(\textbf{0}, \Tilde{y}^2)$, such that $\textbf{0} \ll \Tilde{y}^1 \ll \textbf{1}$ and $\textbf{0} \ll \Tilde{y}^2 \ll \textbf{1}$]} \axel{\textbf{[Maybe we could say something like "Recall that $(\Tilde{y}^1, \textbf{0})$ and $(\textbf{0}, \Tilde{y}^2)$ are non-zero equilibria of system~\eqref{eq:full}.". We have already motivated the existence of and defined these equilibria in the theorem statement.]}}.
By assumption, $(\Tilde{y}^1, \textbf{0})$ is an equilibrium of~\eqref{eq:full}, therefore $T^1(\Tilde{y}^1, \textbf{0}) = \Tilde{y}^1$. By the inequalities in~\eqref{eq:inequality_bonanza_shared} we have $T^1(\Tilde{y}^1, y^2) \leq \Tilde{y}^1$, and thus $T^1(y^1, y^2) \leq \Tilde{y}^1$, for all $y^1 \leq \Tilde{y}^1$. Analogously, it can be shown that we have $T^2(y^1, y^2) \leq \Tilde{y}^2$, for all $y^2 \leq \Tilde{y}^2$. Thus,
%
%\vspace{-1.5ex}
\begin{equation}\label{eq:Tleq_shared}
    T(y^1, y^2) \leq (\Tilde{y}^1, \Tilde{y}^2), %\text{for all } (y^1, y^2) \leq (\Tilde{y}^1, \Tilde{y}^2).
\end{equation}
%

%\vspace{-1.5ex}

\noindent
whenever $(y^1, y^2) \leq (\Tilde{y}^1, \Tilde{y}^2)$. Now, by assumption, $s(- D_w^1 + (I - X(\Tilde{y}^2))B_w^1) > 0$, and, since $D_w^1$ and $(I - X(\Tilde{y}^2))$ are positive diagonal matrices and $B_w^1$ is an irreducible nonnegative matrix, $(- D_w^1 + (I - X(\Tilde{y}^2))B_w^1)$ is an irreducible Metzler matrix. Then, by Lemma~\ref{lem:eigspec} and the fact that diagonal matrices commute, we have $\rho((I - X(\Tilde{y}^2))(D_w^1)^{-1} B_w^1) > 1$. Further, since $((I - X(\Tilde{y}^2))(D_w^1)^{-1} B_w^1)$ is an irreducible nonnegative matrix, by item~\ref{item:perfrob_simpleeig} in Lemma~\ref{lem:perron_frob} we know that $\lambda^1 = \rho((I - X(\Tilde{y}^2))(D_w^1)^{-1} B_w^1)$ is a simple eigenvalue of this matrix. Furthermore, by item~\ref{item:perfrob_pos_exists} in Lemma~\ref{lem:perron_frob}, we know that the eigenspace of $\lambda^1$ is spanned by a vector $\Bar{\Bar{y}}^1 \gg \textbf{0}$. Analogously we get $\lambda^2 = \rho((I - X(\Tilde{y}^1))(D_w^2)^{-1} B_w^2)>1$, and the corresponding eigenvector $\Bar{\Bar{y}}^2 \gg \textbf{0}$. 

\noindent With the eigenvectors $\Bar{\Bar{y}}^1, \Bar{\Bar{y}}^2$ in place, we see that since $(D_w^1)^{-1} B_w^1$ and $(D_w^2)^{-1} B_w^2$ are irreducible nonnegative matrices, we have $((D_w^1)^{-1} B_w^1 \Bar{\Bar{y}}^1)_i > 0$, and $((D_w^2)^{-1} B_w^2 \Bar{\Bar{y}}^2)_i > 0$, for all $i \in [n+1]$. Further, given that $\Bar{\Bar{y}}^1 \gg \textbf{0}, \Bar{\Bar{y}}^2 \gg \textbf{0}, \Tilde{y}^1 \gg \textbf{0}$, and $\Tilde{y}^2 \gg \textbf{0}$, we have $\Tilde{y}_i^1 / \Bar{\Bar{y}}_i^1 > 0$, and $\Tilde{y}_i^2 / \Bar{\Bar{y}}_i^2 > 0$, for all $i \in [n+1]$. Moreover, note that $\lambda^1 - 1 > 0$ and $\lambda^2 - 1 > 0$. %\seb{[here you need to also argue that the first term within the paranthesis in the next line is positive, neither of the terms are zero, and therefore you have $\epsilon^1 > 0$.......makes sense? ]} \axel{\textbf{[How about now?]}}
Hence, there exist $\epsilon^1 > 0$ and $\epsilon^2 > 0$ such that 
\vspace{-1.5ex}
\begin{equation} \label{eq:joint_eq_exist_epsilonineq}
\begin{split}
\epsilon^1 &< \min \left\{\frac{\lambda^1 - 1}{\max_{i \in [n+1]}((D_w^1)^{-1} B_w^1 \Bar{\Bar{y}}^1)_i}, \, \min_{i \in [n+1]} \frac{\Tilde{y}_i^1}{\Bar{\Bar{y}}_i^1}\right\}, \\
\epsilon^2 &< \min \left\{\frac{\lambda^2 - 1}{\max_{i \in [n+1]}((D_w^2)^{-1} B_w^2 \Bar{\Bar{y}}^2)_i}, \, \min_{i \in [n+1]} \frac{\Tilde{y}_i^2}{\Bar{\Bar{y}}_i^2}\right\}. 
\end{split}
\end{equation}
%
% \begin{align*}
% \epsilon^1 &< \min \{\frac{\lambda^1 -~1}{\max_{i \in [n+1]}((D_w^1)^{-1} B_w^1 \Bar{\Bar{y}}^1)_i}, \, \min_{i \in [n+1]} \frac{\Tilde{y}_i^1}{\Bar{\Bar{y}}_i^1}\}, \\ \epsilon^2 &< \min \{\frac{\lambda^2 -~1}{\max_{i \in [n+1]}((D_w^2)^{-1} B_w^2 \Bar{\Bar{y}}^2)_i}, \, \min_{i \in [n+1]} \frac{\Tilde{y}_i^2}{\Bar{\Bar{y}}_i^2}\}.
% \end{align*}
%

\vspace{-1ex}

\noindent
From~\eqref{eq:joint_eq_exist_epsilonineq} follows that
%
%\vspace{-1ex}
\begin{equation} \label{eq:joint_eq_exist_epsilonineq_employed}
 \begin{split} 
 1 + \textstyle\max_{i \in [n]} ((D_w^1)^{-1} B_w^1 \epsilon^1 \Bar{\Bar{y}}^1)_i &< \lambda^1, \\ 
 1 + \textstyle\max_{i \in [n]} ((D_w^2)^{-1} B_w^2 \epsilon^2 \Bar{\Bar{y}}^2)_i &< \lambda^2.
 \end{split}
\end{equation}
%

%\vspace{-1ex}

%
\noindent Employing~\eqref{eq:joint_eq_exist_epsilonineq_employed} it follows that, for all $i \in [n]$, we have
\begin{align*}
T_i^1(\epsilon^1 \Bar{\Bar{y}}^1, \Tilde{y}^2) &= \frac{((I - X(\Tilde{y}^2))(D_w^1)^{-1} B_w^1 \epsilon^1 \Bar{\Bar{y}}^1)_i}{1 + ((D_w^1)^{-1} B_w^1 \epsilon^1 \Bar{\Bar{y}}^1)_i} \\
&= \frac{\lambda^1 \epsilon^1 \Bar{\Bar{y}}_i^1}{1 + ((D_w^1)^{-1} B_w^1 \epsilon^1 \Bar{\Bar{y}}^1)_i} > \epsilon^1 \Bar{\Bar{y}}_i^1, 
\\
% \end{align*}
% \vspace{-2ex}
% \begin{align*}
T_i^2(\Tilde{y}^1, \epsilon^2 \Bar{\Bar{y}}^2) &= \frac{((I - X(\Tilde{y}^1))(D_w^2)^{-1} B_w^2 \epsilon^2 \Bar{\Bar{y}}^2)_i}{1 + ((D_w^2)^{-1} B_w^2 \epsilon^2 \Bar{\Bar{y}}^2)_i} \\
&= \frac{\lambda^2 \epsilon^2 \Bar{\Bar{y}}_i^2}{1 + ((D_w^2)^{-1} B_w^2 \epsilon^2 \Bar{\Bar{y}}^2)_i} > \epsilon^2 \Bar{\Bar{y}}_i^2.
\end{align*}
For $T_{n+1}^1$ and $T_{n+1}^2$, note that $((D_w^1)^{-1} B_w^1 \epsilon^1 \Bar{\Bar{y}}^1)_{n+1} \Bar{\Bar{y}}_{n+1}^1 > 0$ and $((D_w^2)^{-1} B_w^2 \epsilon^2 \Bar{\Bar{y}}^2)_{n+1} \Bar{\Bar{y}}_{n+1}^2 > 0$, respectively. Therefore
\vspace{-1ex}
\small
\begin{align*}
&T_{n+1}^1(\epsilon^1 \Bar{\Bar{y}}^1, \Tilde{y}^2) \\
&= 
% \frac{((I - X(\Tilde{y}^2))(D_w^1)^{-1} B_w^1 \epsilon^1 \Bar{\Bar{y}}^1)_{n+1} + ((D_w^1)^{-1} B_w^1 \epsilon^1 \Bar{\Bar{y}}^1)_{n+1} \Bar{\Bar{y}}_{n+1}^1}{1 + ((D_w^1)^{-1} B_w^1 \epsilon^1 \Bar{\Bar{y}}^1)_{n+1}} 
\frac{((I - X(\Tilde{y}^2))(D_w^1)^{-1} B_w^1 \epsilon^1 \Bar{\Bar{y}}^1)_{n+1} }{1 + ((D_w^1)^{-1} B_w^1 \epsilon^1 \Bar{\Bar{y}}^1)_{n+1}} 
% \\
% &
%\, \, \, \, \, \, \, \, \,
+
\frac{ ((D_w^1)^{-1} B_w^1 \epsilon^1 \Bar{\Bar{y}}^1)_{n+1} \Bar{\Bar{y}}_{n+1}^1}{1 + ((D_w^1)^{-1} B_w^1 \epsilon^1 \Bar{\Bar{y}}^1)_{n+1}}\\ 
&\geq \frac{\lambda^1 \epsilon^1 \Bar{\Bar{y}}_{n+1}^1}{1 + ((D_w^1)^{-1} B_w^1 \epsilon^1 \Bar{\Bar{y}}^1)_{n+1}} > \epsilon^1 \Bar{\Bar{y}}_{n+1}^1,
\end{align*}
\vspace{-3.25ex}
\begin{align*}
&T_{n+1}^2(\Tilde{y}^1, \epsilon^2 \Bar{\Bar{y}}^2) %&
\\
&= 
% \frac{((I - X(\Tilde{y}^1))(D_w^2)^{-1} B_w^2 \epsilon^2 \Bar{\Bar{y}}^2)_{n+1} + ((D_w^2)^{-1} B_w^2 \epsilon^2 \Bar{\Bar{y}}^2)_{n+1} \Bar{\Bar{y}}_{n+1}^2}{1 + ((D_w^2)^{-1} B_w^2 \epsilon^2 \Bar{\Bar{y}}^2)_{n+1}} 
\frac{((I - X(\Tilde{y}^1))(D_w^2)^{-1} B_w^2 \epsilon^2 \Bar{\Bar{y}}^2)_{n+1}}{1 + ((D_w^2)^{-1} B_w^2 \epsilon^2 \Bar{\Bar{y}}^2)_{n+1}} 
% \\
% &\, \, \, \, \, \, \, \, \,
+
\frac{((D_w^2)^{-1} B_w^2 \epsilon^2 \Bar{\Bar{y}}^2)_{n+1} \Bar{\Bar{y}}_{n+1}^2}{1 + ((D_w^2)^{-1} B_w^2 \epsilon^2 \Bar{\Bar{y}}^2)_{n+1}} 
\\
&\geq \frac{\lambda^2 \epsilon^2 \Bar{\Bar{y}}_{n+1}^2}{1 + ((D_w^2)^{-1} B_w^2 \epsilon^2 \Bar{\Bar{y}}^2)_{n+1}} > \epsilon^2 \Bar{\Bar{y}}_{n+1}^2.
\end{align*}

\normalsize

\vspace{-1.25ex}

\noindent
Given that~\eqref{eq:joint_eq_exist_epsilonineq} implies $\epsilon^1 \Bar{\Bar{y}}^1 < \Tilde{y}^1$ and $\epsilon^2 \Bar{\Bar{y}}^2 < \Tilde{y}^2$, by the inequalities in~\eqref{eq:inequality_bonanza_shared} we have $T^1(\epsilon^1 \Bar{\Bar{y}}^1, y^2) > \epsilon^1 \Bar{\Bar{y}}^1$ 
%
%\begin{equation} \label{eq:joint_eq_exist_first_inner_limit_1}
%T^1(\epsilon^1 \Bar{\Bar{y}}^1, y^2) > \epsilon^1 \Bar{\Bar{y}}^1, 
%\,\, \text{for all } y^2 \text{ such that } \epsilon^2 \Bar{\Bar{y}}^2 \leq y^2 \leq \Tilde{y}^2, 
%\end{equation}
%
%for all $y^2$ such that 
whenever $\epsilon^2 \Bar{\Bar{y}}^2 \leq y^2 \leq \Tilde{y}^2$, and $T^2(y^1, \epsilon^2 \Bar{\Bar{y}}^2) > \epsilon^2 \Bar{\Bar{y}}^2$ 
%
%\begin{equation} \label{eq:joint_eq_exist_first_inner_limit_2}
%T^2(y^1, \epsilon^2 \Bar{\Bar{y}}^2) > \epsilon^2 \Bar{\Bar{y}}^2, 
%\,\, \text{for all } y^1 \text{ such that } \epsilon^1 \Bar{\Bar{y}}^1 \leq y^1 \leq \Tilde{y}^1,
%\end{equation}
%
%for all $y^1$ such that 
whenever $\epsilon^1 \Bar{\Bar{y}}^1 \leq y^1 \leq \Tilde{y}^1$.
Further application of the inequalities in~\eqref{eq:inequality_bonanza_shared} %to~\eqref{eq:joint_eq_exist_first_inner_limit_1} and~\eqref{eq:joint_eq_exist_first_inner_limit_2} 
yields
\vspace{-1.5ex}
\begin{equation} \label{eq:joint_eq_exist_second_inner_limit}
%\begin{split}
T^1(y^1, y^2) > \epsilon^1 \Bar{\Bar{y}}^1, \,\,\, %\\ %\,\, \text{for all } (y^1, y^2) \text{ such that } \\ 
%&\, \, \, \, \, \, \, \, \, (\epsilon^1 \Bar{\Bar{y}}^1, \epsilon^2 \Bar{\Bar{y}}^2) \leq (y^1, y^2) \leq (\Tilde{y}^1, \Tilde{y}^2),
%
%for all $(y^1, y^2)$ such that $(\epsilon^1 \Bar{\Bar{y}}^1, \epsilon^2 \Bar{\Bar{y}}^2) \leq (y^1, y^2) \leq (\Tilde{y}^1, \Tilde{y}^2)$, and
%
T^2(y^1, y^2) > \epsilon^2 \Bar{\Bar{y}}^2, %\,\, \text{for all }  (y^1, y^2) \text{ such that } \\ 
%& \, \, \, \, \, \, \, \, \, (\epsilon^1 \Bar{\Bar{y}}^1, \epsilon^2 \Bar{\Bar{y}}^2) \leq (y^1, y^2) \leq (\Tilde{y}^1, \Tilde{y}^2).
%\end{split}
\end{equation}

\vspace{-1.25ex}

\noindent
%
%for all $(y^1, y^2)$ such that 
whenever $(\epsilon^1 \Bar{\Bar{y}}^1, \epsilon^2 \Bar{\Bar{y}}^2) \leq (y^1, y^2) \leq (\Tilde{y}^1, \Tilde{y}^2)$.
Then, \eqref{eq:Tleq_shared} and~\eqref{eq:joint_eq_exist_second_inner_limit} show that $(\epsilon^1 \Bar{\Bar{y}}^1, \epsilon^2 \Bar{\Bar{y}}^2) \leq T(y^1, y^2) \leq (\Tilde{y}^1, \Tilde{y}^2)$ %for all $(y^1, y^2)$ such that 
whenever $(\epsilon^1 \Bar{\Bar{y}}^1, \epsilon^2 \Bar{\Bar{y}}^2) \leq (y^1, y^2) \leq (\Tilde{y}^1, \Tilde{y}^2)$. By Brouwer's fixed point theorem \cite[Theorem~9.3]{starr_2011}, there exists at least one fixed point of $T(y)$ in the domain $\{y = (y^1, y^2) : (\epsilon^1 \Bar{\Bar{y}}^1, \epsilon^2 \Bar{\Bar{y}}^2) \leq (y^1, y^2) \leq (\Tilde{y}^1, \Tilde{y}^2)\}$. Recall that a fixed point of $T(y)$ is equivalent to an equilibrium of~\eqref{eq:full},
%\seb{[maybe we should add a line immediately after defining $T(y)$ that a fixed point of $T(y)$ is an equilirbium of system~(5)]}
hence, by Lemma~\ref{lem:equi_non-zero_nonone}, any fixed point of $T(y)$ must fulfill $y^1 + y^2 \leq \textbf{1}$.
%\noindent \seb{[have you not said this already in the previous paragraph?]} \axel{\textbf{[The difference is that the set in which the fixed point exists is now further restricted by the sum inequality.]}} 
In conclusion, 
%the map $T(y)$ has at least one fixed point in the domain $\{y = (y^1, y^2) : (\epsilon^1 \Bar{\Bar{y}}^1, \epsilon^2 \Bar{\Bar{y}}^2) \leq (y^1, y^2) \leq (\Tilde{y}^1, \Tilde{y}^2), y^1 + y^2 \leq \textbf{1}\}$, and therefore 
system~\eqref{eq:full} has at least one coexisting equilibrium $(\hat{y}^1, \hat{y}^2) \gg \textbf{0}$ in $\mathcal{D}$, such that $\hat{y}^1 + \hat{y}^2 \leq \textbf{1}$.~$\square$
%

%\vspace{-2ex}

\subsection*{Proof of Theorem~\ref{thm:noequi_single-virus}}
%\vspace{-1ex}
\noindent In order to prove Theorem~\ref{thm:noequi_single-virus}, we require the following lemma.
%
%\vspace{-3ex}
\begin{lem} \label{lem:ineq_singlevirus_equi}
Consider system~\eqref{eq:full} under Assumption~\ref{assum:base} with $m=2$. Suppose that $B_w^1$ and $B_w^2$ are irreducible matrices, that $s(B_w^1 - D_w^1) > 0$ and $s(B_w^2 - D_w^2) > 0$, and that $(D_w^1)^{-1}B_w^1 > (D_w^2)^{-1}B_w^2$. If $(\Tilde{y}^1, \textbf{0})$ and $(\textbf{0}, \Tilde{y}^2)$ are single-virus endemic equilibria of~\eqref{eq:full}, then $\Tilde{y}^1 > \Tilde{y}^2$.~$\blacksquare$
\end{lem}
%\vspace{-1ex}
%
%\seb{[Maybe the proof could be presented better; splitting into two cases perhaps, Case~1: $\tilde{y}^2 > \tilde{y}^1$ and case~2: $\tilde{y}^1 = \tilde{y}^2$]}
\textit{Proof:} Given that $\Tilde{y}^1$ and $\Tilde{y}^2$ are equilibria of~\eqref{eq:yk}, and observing that diagonal matrices commute, we obtain
%
%\vspace{-1ex}
\begin{equation}\label{eq:lem_ineq_singlevirus_equiequa}
\begin{split}
(I - X(\Tilde{y}^1))(D_w^1)^{-1}B_w^1 \Tilde{y}^1 &= \Tilde{y}^1,\\
(I - X(\Tilde{y}^2))(D_w^2)^{-1}B_w^2 \Tilde{y}^2 &= \Tilde{y}^2. 
\end{split}
\end{equation}
%

%\vspace{-1ex}

\noindent
By Lemma~\ref{lem:equi_non-zero_nonone} we have $\textbf{0} \ll \Tilde{y}^1 \ll \textbf{1}$ and $\textbf{0} \ll \Tilde{y}^2 \ll \textbf{1}$. Let $\kappa = \max_{i \in [n+1]} \Tilde{y}_i^2 / \Tilde{y}_i^1$, and thus $\Tilde{y}^2 \leq \kappa \Tilde{y}^1$. Since $(D_w^1)^{-1}B_w^1 > (D_w^2)^{-1}B_w^2$, and $\textstyle \sum_{i=1}^n c_i^k = 1$ for all $k \in [2]$, we have $c^1 = c^2$. Then, by analogous arguments to those in part~2 of the proof of Theorem~\ref{thm:equi}, %particularily~\eqref{eq:singleequi_unique_notn+1_use_kappaassum}, 
we know that $\kappa = \max_{i \in [n]} \Tilde{y}_i^2 / \Tilde{y}_i^1$. Let $j$ be the index in $[n]$ such that $\Tilde{y}_j^2 = \kappa \Tilde{y}_j^1$. Assume, by way of contradiction, that $\kappa > 1$, implying $\Tilde{y}_j^2 > \Tilde{y}_j^1$. Note that since $(D_w^1)^{-1}B_w^1 > (D_w^2)^{-1}B_w^2$, with both matrices being irreducible and nonnegative, it follows that $(D_w^1)^{-1}B_w^1 y > (D_w^2)^{-1}B_w^2 y$ for any $y \gg \textbf{0}$. Further, note that $\Tilde{y}^1 \ll \textbf{1}$, ensuring $(1-\Tilde{y}_j^1) > 0$. Then,~\eqref{eq:lem_ineq_singlevirus_equiequa} gives us
%
%\vspace{-1ex}
\begin{align}
\Tilde{y}_j^2 &= (1 - \Tilde{y}_j^2)((D_w^2)^{-1}B_w^2 \Tilde{y}^2)_j \nonumber \\
&\leq (1 - \Tilde{y}_j^2)((D_w^1)^{-1}B_w^1 \Tilde{y}^2)_j \label{eq:lem_ineq_singlevirus_seq_use_reproddiff}\\
&< (1 - \Tilde{y}_j^2)((D_w^1)^{-1}B_w^1 \kappa \Tilde{y}^1)_j \label{eq:lem_ineq_singlevirus_seq_kappalarger} \\
%&= \frac{1 - \Tilde{y}_j^2}{1 - \Tilde{y}_j^1} ((I - X(\Tilde{y}^1))(D_w^1)^{-1}B_w^1 \kappa \Tilde{y}^1)_j \nonumber \\
&= 
% \frac
{(1 - \Tilde{y}_j^2)}{(1 - \Tilde{y}_j^1)}^{-1} \kappa \Tilde{y}_j^1 \label{eq:lem_ineq_singlevirus_seq_eigenvec} \\
&= 
% \frac
{(1 - \Tilde{y}_j^2)}{(1 - \Tilde{y}_j^1)}^{-1} \Tilde{y}_j^2 \label{eq:lem_ineq_singlevirus_seq_argmin} \\
&< \Tilde{y}_j^2, \label{eq:lem_ineq_singlevirus_seq_end}
\end{align}
%

%\vspace{-1ex}

\noindent
where~\eqref{eq:lem_ineq_singlevirus_seq_use_reproddiff} follows from $(D_w^1)^{-1}B_w^1 > (D_w^2)^{-1}B_w^2$,~\eqref{eq:lem_ineq_singlevirus_seq_kappalarger} follows from $\Tilde{y}^2 < \kappa \Tilde{y}^1$,~\eqref{eq:lem_ineq_singlevirus_seq_eigenvec} follows from~\eqref{eq:lem_ineq_singlevirus_equiequa},~\eqref{eq:lem_ineq_singlevirus_seq_argmin} follows from $\Tilde{y}_j^2 = \kappa \Tilde{y}_j^1$, and~\eqref{eq:lem_ineq_singlevirus_seq_end} follows from $\Tilde{y}_j^2 > \Tilde{y}_j^1$. 
Note that~\eqref{eq:lem_ineq_singlevirus_seq_end} is a contradiction, following from our assumption that $\kappa > 1$. Therefore, $\kappa \leq 1$, and hence, $\Tilde{y}^1 \geq \Tilde{y}^2$.
%
%\begin{equation} \label{eq:lem_ineq_singlevirus_almost_right}
%\Tilde{y}^1 \geq \Tilde{y}^2. 
%\end{equation}
%
Assume, by way of contradiction, that $\Tilde{y}^1 = \Tilde{y}^2 = \Tilde{y}$. Then $(I - X(\Tilde{y}))(D_w^1)^{-1}B_w^1 > (I - X(\Tilde{y}))(D_w^2)^{-1}B_w^2$, and it follows from~\eqref{eq:lem_ineq_singlevirus_equiequa} that
% 
%\vspace{-1.5ex}
\begin{align}
\Tilde{y} &= (I - X(\Tilde{y}))(D_w^1)^{-1}B_w^1 \Tilde{y} \nonumber \\
&> (I - X(\Tilde{y}))(D_w^2)^{-1}B_w^2 \Tilde{y} = \Tilde{y}. \label{eq:lem_ineq_singlevirus_equal_contradict}
\end{align}
%

%\vspace{-1ex}

\noindent
It is clear that~\eqref{eq:lem_ineq_singlevirus_equal_contradict} is a contradiction, following from our assumption that $\Tilde{y}^1 = \Tilde{y}^2$. %Hence, in conjunction with~\eqref{eq:lem_ineq_singlevirus_almost_right}, 
Therefore we have $\Tilde{y}^1 > \Tilde{y}^2$.~$\square$\\

\textit{Proof of Theorem~\ref{thm:noequi_single-virus}:} \\%Without loss of generality, \seb{[I wouldn't phrase it this way. I would rather structure the proof as follows: case~1: suppose $(D_w^1)^{-1}B_w^1 > (D_w^2)^{-1}B_w^2$, and have whatever you have now. case~2: $(D_w^2)^{-1}B_w^2 > (D_w^1)^{-1}B_w^1$ and say "the proof is analogous to case~1, and therefore omitted" ]} let $(D_w^1)^{-1}B_w^1 > (D_w^2)^{-1}B_w^2$. 
Recall that the healthy state is an equilibrium of~\eqref{eq:full}. Since $s(B_w^1 - D_w^1) > 0$ and $s(B_w^2 - D_w^2) > 0$, by Proposition~\ref{prop:necessity} there are exactly two single-virus endemic equilibria of system~\eqref{eq:full}, namely $(\Tilde{y}^1, \textbf{0})$ and $(\textbf{0}, \Tilde{y}^2)$, such that $\textbf{0} \ll \Tilde{y}^1 \ll \textbf{1}$ and $\textbf{0} \ll \Tilde{y}^2 \ll \textbf{1}$. We will now show that with $(D_w^1)^{-1}B_w^1 > (D_w^2)^{-1}B_w^2$, there are no equilibria in $\mathcal{D}$ other than the healthy state, $(\Tilde{y}^1, \textbf{0})$ and $(\textbf{0}, \Tilde{y}^2)$. First, note that by Lemma~\ref{lem:equi_non-zero_nonone}, any additional equilibrium in $\mathcal{D}$ must be of the form $\hat{y} = (\hat{y}^1, \hat{y}^2)$, such that $\hat{y}^1 \gg \textbf{0}$, $\hat{y}^2 \gg \textbf{0}$, and $\hat{y}^1 + \hat{y}^2 \ll \textbf{1}$. Assume, by way of contradiction, that such an equilibrium $\hat{y}$ exists. %\seb{[is there something more here? coz this sentence is a bit ambiguous]} %\axel{\textbf{[How about now?]}}
Since $\hat{y}$ is an equilibrium of~\eqref{eq:full} it follows that
% 
%\vspace{-1ex}
\begin{equation} \label{eq:prop_noequi_single-virus_doubleequi}
\begin{split}
(I - X(\hat{y}^1) - X(\hat{y}^2))(D_w^1)^{-1}B_w^1 \hat{y}^1 &= \hat{y}^1, \\
(I - X(\hat{y}^1) - X(\hat{y}^2))(D_w^2)^{-1}B_w^2 \hat{y}^2 &= \hat{y}^2.
\end{split}
\end{equation}
%

%\vspace{-1.25ex}

\noindent
Given that $\Tilde{y}^1$ and $\Tilde{y}^2$ are equilibria of~\eqref{eq:yk}, and observing that diagonal matrices commute, we obtain
% 
%\vspace{-1ex}
\begin{equation} \label{eq:prop_noequi_single-virus_singleequis}
\begin{split}
(I - X(\Tilde{y}^1))(D_w^1)^{-1}B_w^1 \Tilde{y}^1 &= \Tilde{y}^1, \\
(I - X(\Tilde{y}^2))(D_w^2)^{-1}B_w^2 \Tilde{y}^2 &= \Tilde{y}^2. 
\end{split}
\end{equation}
%

%\vspace{-1.5ex}

\noindent
Since $\Tilde{y}^1 \gg \textbf{0}$, we can define
% 
%\vspace{-1.5ex}
\begin{equation}\label{eq:kappa}
    \kappa = \max_{i \in [n+1]} (\hat{y}^1 + \hat{y}^2)_i / \Tilde{y}_i^1.
\end{equation}

%\vspace{-1ex}

\noindent
%
%By Lemma~\ref{lem:max_not_shared} we know that: 
%
%\begin{equation}\label{eq:kappa_actual}
%    \kappa = \max_{i \in [n]} \frac{(\hat{y}^1 + \hat{y}^2)_i}{\Tilde{y}_i^1}.
%\end{equation}
%
Thus, $\hat{y}^1 + \hat{y}^2 \leq \kappa \Tilde{y}^1$. Let $j\in [n+1]$ be the index such that $(\hat{y}^1 + \hat{y}^2)_j = \kappa \Tilde{y}_j^1$. Assume, by way of contradiction, that $\kappa \geq 1$, implying $(\hat{y}^1 + \hat{y}^2)_j \geq \Tilde{y}_j^1$. Note that since $(D_w^1)^{-1}B_w^1 > (D_w^2)^{-1}B_w^2$, with both matrices being irreducible and nonnegative, it follows that $(D_w^1)^{-1}B_w^1 y > (D_w^2)^{-1}B_w^2 y$ for any $y \gg \textbf{0}$.
%\phil{[should we cite a lemma for this?]} \axel{[I think it is trivial enough to be skipped.]} \seb{[Response: Yeah, I don't think a lemma is needed here.]}
Further, note that $\Tilde{y}^1 \ll \textbf{1}$, ensuring $(1-\Tilde{y}_j^1) > 0$. 
% \axel{\textbf{[It is used to establish the first inequality in the sequence below.]}}. 
We will now show that $\kappa \geq 1$ leads to contradiction, irrespective of whether $j \in [n]$ or $j=n+1$. First, consider the case where $j \in [n]$. Then,~\eqref{eq:prop_noequi_single-virus_doubleequi} gives us %\phil{for $j$ that is the arg$\min$ of~\eqref{eq:kappa_actual}}:
\vspace{-1ex}
\begin{align}
(\hat{y}^1 + \hat{y}^2)_j &= (1 - \hat{y}_j^1 - \hat{y}_j^2) ((D_w^1)^{-1} B_w^1 \hat{y}^1 + (D_w^2)^{-1} B_w^2 \hat{y}^2)_j \nonumber\\
&< (1 - \hat{y}_j^1 - \hat{y}_j^2) ((D_w^1)^{-1} B_w^1 (\hat{y}^1 + \hat{y}^2))_j \label{eq:strict:due:assumption} \\
&\leq (1 - \hat{y}_j^1 - \hat{y}_j^2) ((D_w^1)^{-1} B_w^1 \kappa \Tilde{y}^1)_j \label{eq:kappa_bigger_employ} \\
%&= \frac{(1 - \hat{y}_j^1 - \hat{y}_j^2)}{1 - \Tilde{y}_j^1} ((I-X(\Tilde{y}^1))(D_w^1)^{-1} B_w^1 \kappa \Tilde{y}^1)_j \nonumber\\
&= 
% \frac
{(1 - \hat{y}_j^1 - \hat{y}_j^2)}{(1 - \Tilde{y}_j^1)}^{-1} \kappa \Tilde{y}_j^1  \label{eq:seq_tilde}
\\
&= 
% \frac
{(1 - \hat{y}_j^1 - \hat{y}_j^2)}{(1 - \Tilde{y}_j^1)}^{-1} (\hat{y}^1 + \hat{y}^2)_j   \label{eq:argmin}
% \phil{\text{[this equality only holds when $j$ is the argmin of~\eqref{eq:kappa}, so I added the note before the align environment]}} 
\\
&\leq (\hat{y}^1 + \hat{y}^2)_j,\label{eq:seq_end}
\end{align}

\vspace{-1ex}

\noindent
where~\eqref{eq:strict:due:assumption} follows from the assumption that $(D_w^1)^{-1}B_w^1 > (D_w^2)^{-1}B_w^2$,~\eqref{eq:kappa_bigger_employ} follows from $\hat{y}^1 + \hat{y}^2 \leq \kappa \Tilde{y}^1$,~\eqref{eq:seq_tilde} follows from~\eqref{eq:prop_noequi_single-virus_singleequis},~\eqref{eq:argmin} holds since $(\hat{y}^1 + \hat{y}^2)_j = \kappa \Tilde{y}_j^1$, and~\eqref{eq:seq_end} holds due to $(\hat{y}^1 + \hat{y}^2)_j \geq \Tilde{y}_j^1$. Note that~\eqref{eq:seq_end} is a contradiction, following from the assumption that $\kappa \geq 1$. Now, consider $j = n+1$. Then, from~\eqref{eq:prop_noequi_single-virus_doubleequi} we have 
%
%\vspace{-1.5ex}
\begin{align}
(\hat{y}^1 + \hat{y}^2)_{n+1} &= ((D_w^1)^{-1} B_w^1 \hat{y}^1 + (D_w^2)^{-1} B_w^2 \hat{y}^2)_{n+1} \label{eq:X_does_not_matter_nplusone} \\
&< ((D_w^1)^{-1} B_w^1 (\hat{y}^1 + \hat{y}^2))_{n+1} \label{eq:strict:due:assumption_nplusone} \\
&\leq ((D_w^1)^{-1} B_w^1 \kappa \Tilde{y}^1)_{n+1} \label{eq:kappa_bigger_employ_nplusone} \\
&= ((I-X(\Tilde{y}^1))(D_w^1)^{-1} B_w^1 \kappa \Tilde{y}^1)_{n+1} \label{eq:kappa_bigger_employ_nplusone_sebin} \\
&= \kappa \Tilde{y}_{n+1}^1  \label{eq:seq_tilde_nplusone} \\
&\leq (\hat{y}^1 + \hat{y}^2)_{n+1}, \label{eq:argmin_nplusone}
\end{align}
%

%\vspace{-1.5ex}

\noindent
where~\eqref{eq:X_does_not_matter_nplusone} follows from the observation that the matrix $X(y)$ has a $0$ in the $(n+1)^{\rm{th}}$ position along its diagonal for any $y \in \mathcal{D}^k$,
%thus ensuring that $[I-X(y)]_{n+1, n+1} =1$] for any $y \in \mathcal{D}$,
\eqref{eq:strict:due:assumption_nplusone} follows from the assumption that $(D_w^1)^{-1}B_w^1 > (D_w^2)^{-1}B_w^2$,~\eqref{eq:kappa_bigger_employ_nplusone} follows from $\hat{y}^1 + \hat{y}^2 \leq \kappa \Tilde{y}^1$,~\eqref{eq:kappa_bigger_employ_nplusone_sebin} holds for the same reason as~\eqref{eq:X_does_not_matter_nplusone},~\eqref{eq:seq_tilde_nplusone} follows from~\eqref{eq:prop_noequi_single-virus_singleequis}, and~\eqref{eq:argmin_nplusone} holds since $(\hat{y}^1 + \hat{y}^2)_{n+1} = \kappa \Tilde{y}_{n+1}^1$. Note that~\eqref{eq:argmin_nplusone} is a contradiction, following from the assumption that $\kappa \geq 1$. Since the assumption that $\kappa \geq 1$ leads to contradiction for all $j \in [n+1]$, we have $\kappa < 1$, implying that $\hat{y}^1 + \hat{y}^2 \ll \Tilde{y}^1$.

Now, note that since $\Tilde{y}^1 \ll \textbf{1}$, and $\hat{y}^1 + \hat{y}^2 \ll \textbf{1}$, we know that $(I - X(\hat{y}^1) - X(\hat{y}^2))(D_w^1)^{-1}B_w^1$ and $(I - X(\Tilde{y}^1))(D_w^1)^{-1}B_w^1$ are irreducible nonnegative matrices. Then, since $\Tilde{y}^1 \gg \textbf{0}$, it follows from~\eqref{eq:prop_noequi_single-virus_singleequis} and item~\ref{item:perfrob_pos_necess} in Lemma~\ref{lem:perron_frob} that $\rho((I - X(\Tilde{y}^1))(D_w^1)^{-1}B_w^1) = 1$. Likewise, $\hat{y}^1 \gg \textbf{0}$,~\eqref{eq:prop_noequi_single-virus_doubleequi} and item~\ref{item:perfrob_pos_necess} in Lemma~\ref{lem:perron_frob} give us $\rho ((I - X(\hat{y}^1) - X(\hat{y}^2))(D_w^1)^{-1}B_w^1) = 1$. Following from $\hat{y}^1 + \hat{y}^2 \ll \Tilde{y}^1$, we have
%
%\vspace{-1ex}
\begin{equation} \label{eq:prop_noequi_important_matrix_ineq}
\small (I - X(\hat{y}^1) - X(\hat{y}^2))(D_w^1)^{-1}B_w^1 < (I - X(\Tilde{y}^1))(D_w^1)^{-1}B_w^1.
\end{equation}
%

%\vspace{-1.5ex}

\noindent
Applying item~\ref{item:perfrob_pos_necess} in Lemma~\ref{lem:perron_frob} to~\eqref{eq:prop_noequi_single-virus_doubleequi},~\eqref{eq:prop_noequi_single-virus_singleequis}, and~\eqref{eq:prop_noequi_important_matrix_ineq} yields
%
%\vspace{-1ex}
\begin{align}
1 &= \rho((I - X(\hat{y}^1) - X(\hat{y}^2))(D_w^1)^{-1}B_w^1) \nonumber \\
&< \rho((I - X(\Tilde{y}^1))(D_w^1)^{-1}B_w^1) = 1. \label{eq:prop_noequi_important_specrad_ineq}
\end{align}
%

%\vspace{-1.5ex}

\noindent
Clearly,~\eqref{eq:prop_noequi_important_specrad_ineq} is a contradiction following from the assumption that $\hat{y}$ exists, and hence, $\hat{y}$ does not exist. Therefore, the only equilibria in $\mathcal{D}$ are the healthy state, $(\Tilde{y}^1, \textbf{0})$ and $(\textbf{0}, \Tilde{y}^2)$. 

It remains to be shown that the healthy state and $(\textbf{0}, \Tilde{y}^2)$ are unstable, and that $(\Tilde{y}^1, \textbf{0})$ is locally exponentially stable. Since, by assumption, $s(B_w^1-D_w^1) > 0$ and $s(B_w^2-D_w^2) > 0$, instability of the healthy state follows directly. Moreover, since, by assumption, $(D_w^1)^{-1}B_w^1 > (D_w^2)^{-1}B_w^2$, from Lemma~\ref{lem:ineq_singlevirus_equi}, it follows that $\Tilde{y}^1 > \Tilde{y}^2$. 
% Instability of the healthy state follows directly from Proposition~\ref{prop:violated_condition_asymp} and $s(B_w^1-D_w^1) > 0$. Now, note that Lemma~\ref{lem:ineq_singlevirus_equi} applies, giving us $\Tilde{y}^1 > \Tilde{y}^2$. 
Recall that $\rho(I - X(\Tilde{y}^1)(D_w^1)^{-1}B_w^1) = 1$. Then, since $\Tilde{y}^1 > \Tilde{y}^2$, item~\ref{item:perfrob_matrix_ineq} in Lemma~\ref{lem:perron_frob} implies that $\rho(I - X(\Tilde{y}^2)(D_w^1)^{-1}B_w^1) > \rho(I - X(\Tilde{y}^1)(D_w^1)^{-1}B_w^1)$, which by Lemma~\ref{lem:eigspec} is equivalent to $s((I - X(\Tilde{y}^1))B_w^1 - D_w^1) > 0$. Observe that the Jacobian of~\eqref{eq:full} evaluated at $(\textbf{0}, \Tilde{y}^2)$ is
%
%\vspace{-1ex}
\begin{gather*} 
    J(\textbf{0}, \Tilde{y}^2) 
    =
    \tiny
    \begin{bmatrix}
    (I - X(\Tilde{y}^2))B_w^1 - D_w^1 & 0\\
    - \diag(B_w^2 \Tilde{y}^2) & (I - X(\Tilde{y}^2))B_w^2 - D_w^2 - X(B_w^2 \Tilde{y}^2) %\\
    \end{bmatrix}.
\end{gather*}

\vspace{-1ex}

\noindent
Since $s((I - X(\Tilde{y}^2))B_w^1 - D_w^1) > 0$, by the properties of block-triangular matrices, $s(J(\textbf{0}, \Tilde{y}^2)) > 0$. Hence, $(\textbf{0}, \Tilde{y}^2)$ is unstable. Now, note that $(D_w^1 + X(B_w^1 \Tilde{y}^1)) > D_w^1$, in turn implying $(D_w^1 + X(B_w^1 \Tilde{y}^1))^{-1} < (D_w^1)^{-1}$, which, since $(I - X(\Tilde{y}^1))$ is a positive diagonal matrix, further implies that $(I - X(\Tilde{y}^1))(D_w^1 + X(B_w^1 \Tilde{y}^1))^{-1}B_w^1 < (I - X(\Tilde{y}^1))(D_w^1)^{-1} B_w^1$. Hence, applying item~\ref{item:perfrob_matrix_ineq} in Lemma~\ref{lem:perron_frob} yields $\rho((I - X(\Tilde{y}^1))(D_w^1 + X(B_w^1 \Tilde{y}^1))^{-1}B_w^1) < \rho(I - X(\Tilde{y}^1)(D_w^1)^{-1}B_w^1)$, which by Lemma~\ref{lem:eigspec} is equivalent to $s((I - X(\Tilde{y}^1))B_w^1 - D_w^1 - X(B_w^1 \Tilde{y}^1)) < 0$. Furthermore, note that since $\Tilde{y}
^2 \gg \textbf{0}$, it follows from~\eqref{eq:prop_noequi_single-virus_singleequis} and item~\ref{item:perfrob_pos_necess} in Lemma~\ref{lem:perron_frob} that $\rho(I - X(\Tilde{y}^2)(D_w^2)^{-1}B_w^2) = 1$. Then, since $\Tilde{y}^1 > \Tilde{y}^2$, item~\ref{item:perfrob_matrix_ineq} in Lemma~\ref{lem:perron_frob} implies that $\rho(I - X(\Tilde{y}^1)(D_w^2)^{-1}B_w^2) < \rho(I - X(\Tilde{y}^2)(D_w^2)^{-1}B_w^2)$, which by Lemma~\ref{lem:eigspec} is equivalent to $s((I - X(\Tilde{y}^1))B_w^2 - D_w^2) < 0$. Observe that the Jacobian of~\eqref{eq:full} evaluated at $(\Tilde{y}^1, \textbf{0})$ is
\vspace{-1ex}
\begin{gather*} 
    J(\Tilde{y}^1, \textbf{0})
    =
    \tiny
    \begin{bmatrix}
    (I - X(\Tilde{y}^1))B_w^1 - D_w^1 - X(B_w^1 \Tilde{y}^1)& - X(B_w^1 \Tilde{y}^1)\\
    0 & (I - X(\Tilde{y}^1))B_w^2 - D_w^2  %\\
    \end{bmatrix}.
\end{gather*}

\vspace{-1ex}

\noindent
Since $s((I - X(\Tilde{y}^1))B_w^1 - D_w^1 - X(B_w^1 \Tilde{y}^1)) < 0$ and $s((I - X(\Tilde{y}^1))B_w^2 - D_w^2) < 0$, by the properties of block-triangular matrices, we have $s(J(\Tilde{y}^1, \textbf{0})) < 0$. Hence, $(\Tilde{y}^1, \textbf{0})$ is locally exponentially stable.~$\square$

 \subsection*{Proof of Proposition~\ref{prop:deltas_chosen_to_heal}}
 With $\delta_i^k$ chosen according to~\eqref{eq:deltas_chosen_to_heal}, the fact that $B_w^k$ is an irreducible nonnegative matrix
 ensures that $\delta_i^k>0$ for all $i \in [n]$. Hence $D_w^k$ is invertible, and we have

 \vspace{-2ex}
 \begin{gather*}
     (D_w^k)^{-1}B_w^k = 
     \begin{bmatrix}
         (D^k)^{-1}B^k & (D^k)^{-1}b^k \\
         c^k & 0
     \end{bmatrix}
     .
 \end{gather*}
 Note that since $B_w^k$ is an irreducible nonnegative matrix, and $D_w^k$ is a positive diagonal matrix, $(D_w^k)^{-1}B_w^k$ is an irreducible nonnegative matrix. 
 Consider the submatrix
 \vspace{-2.25ex}
 \begin{gather*}
     \begin{bmatrix}
         (D^k)^{-1}B^k \,\,\, (D^k)^{-1}b^k
     \end{bmatrix}
     =
     \tiny
     \begin{bmatrix}
         \beta_{11}^k/\delta_1^k & \dots & \beta_{1w}^k/\delta_1^k \\
         \vdots & \ddots & \vdots \\
         \beta_{n1}^k/\delta_n^k & \dots & \beta_{nw}^k/\delta_n^k
     \end{bmatrix}
     .
 \end{gather*}
 First, consider the case where $\epsilon_i^k = 0$ for all $i \in [n]$. Then, given~\eqref{eq:deltas_chosen_to_heal}, each row of $[(D^k)^{-1}B^k \,\,\, (D^k)^{-1}b^k]$ sums to $1$. Further, by definition $\textstyle \sum_{j=1}^n c_i^k =~1$. Therefore, $(D_w^k)^{-1}B_w^k \textbf{1} = \textbf{1}$, which by item~\ref{item:perfrob_pos_necess} in Lemma~\ref{lem:perron_frob} implies that $\rho((D_w^k)^{-1}B_w^k) =~1$, in turn implying that $s(B_w^k - D_w^k) = 0$ by Lemma~\ref{lem:eigspec}. Hence, the conditions for Theorem~\ref{thm:asymp} are fulfilled by virus~$k$, meaning that its eradicated state is asymptotically stable, with domain of attraction containing $\mathcal{D}^k$.

 Next, consider the case where $\epsilon_i^k > 0$ for some $i \in [n]$. Due to the irreducibility of $B_w^k$, each of its rows has at least one positive element. Then, compared to the previous case where $\epsilon_i^k = 0$ for all $i \in [n]$, this case involves increasing some $\epsilon_i^k$, which implies decreasing at least some elements in the $i^{\rm{th}}$ row of $(D_w^k)^{-1}B_w^k$. Recalling that previously we had $\rho((D_w^k)^{-1}B_w^k)=1$, invoking item~\ref{item:perfrob_matrix_ineq} in Lemma~\ref{lem:perron_frob} now gives us $\rho((D_w^k)^{-1}B_w^k) <~1$. Then, by Lemma~\ref{lem:eigspec} it follows that $s(B_w^k - D_w^k) < 0$. Hence, the conditions for Theorem~\ref{thm:expo} are fulfilled by virus~$k$, causing its eradicated state to be exponentially stable, with domain of attraction containing~$\mathcal{D}^k$.~$\square$
%\vspace{-2ex}
\subsection*{Proof of Theorem~\ref{thm:virus-as-vaccine}}
%\vspace{-1ex}
Note that with $\delta_i^2$ given by~\eqref{eq:deltas_chosen_to_leverage} for all $i \in [n]$, since $B_w^1$ and $B_w^2$ are irreducible nonnegative matrices we have $\delta_i^2 > 0$ for all $i \in [n]$. Therefore~\eqref{eq:deltas_chosen_to_leverage} is consistent with Assumption~\ref{assum:base}. Then, it follows from i)~\eqref{eq:deltas_chosen_to_leverage}, ii) $c^1=c^2$, %\seb{[In the theorem statement, you say, before introducing~(89), suppose .. $c_1= c_2$, but here you say~(89) implies $c_1= c_2$...not at all clear ]}
and iii) $E^2 \subseteq E^1$, that $(D_w^1)^{-1}B_w^1 > (D_w^2)^{-1}B_w^2$. Hence, since, by assumption, $s(B_w^1 - D_w^1) > 0$ and $s(B_w^2 - D_w^2) > 0$, the conditions for Theorem~\ref{thm:noequi_single-virus} are met. %implying that
Therefore, the only locally asymptotically stable equilibrium in $\mathcal{D}$ is $(\Tilde{y}^1, \textbf{0})$ with $\textbf{0} \ll \Tilde{y}^1 \ll \textbf{1}$.~$\square$

\vspace{-10mm}
\begin{IEEEbiography}[{\includegraphics[width=1in,height=1.5in,
%trim = 150 200 150 20,
clip,keepaspectratio]{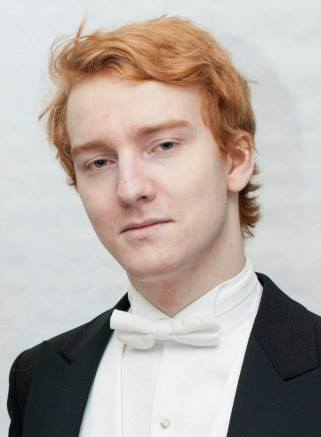}}]
{Axel Janson} is an M.S. student at the Department of Mathematics in the School of Engineering Sciences at KTH Royal Institute of Technology, Stockholm, Sweden. He obtained his B.S. degree in Engineering from KTH Royal Institute of Technology in 2020. His research interests include the analysis and control of networked dynamical systems. %He is an aspiring ventriloquist and one day hopes to take over the world, leaning on this and his many other talents. 
%Biography text here.
\end{IEEEbiography}\vspace{-15mm}

\begin{IEEEbiography}[{\includegraphics[width=1in,height=1.5in,
%trim = 150 200 150 20,
 trim = 450 200 250 0, 
clip,keepaspectratio]{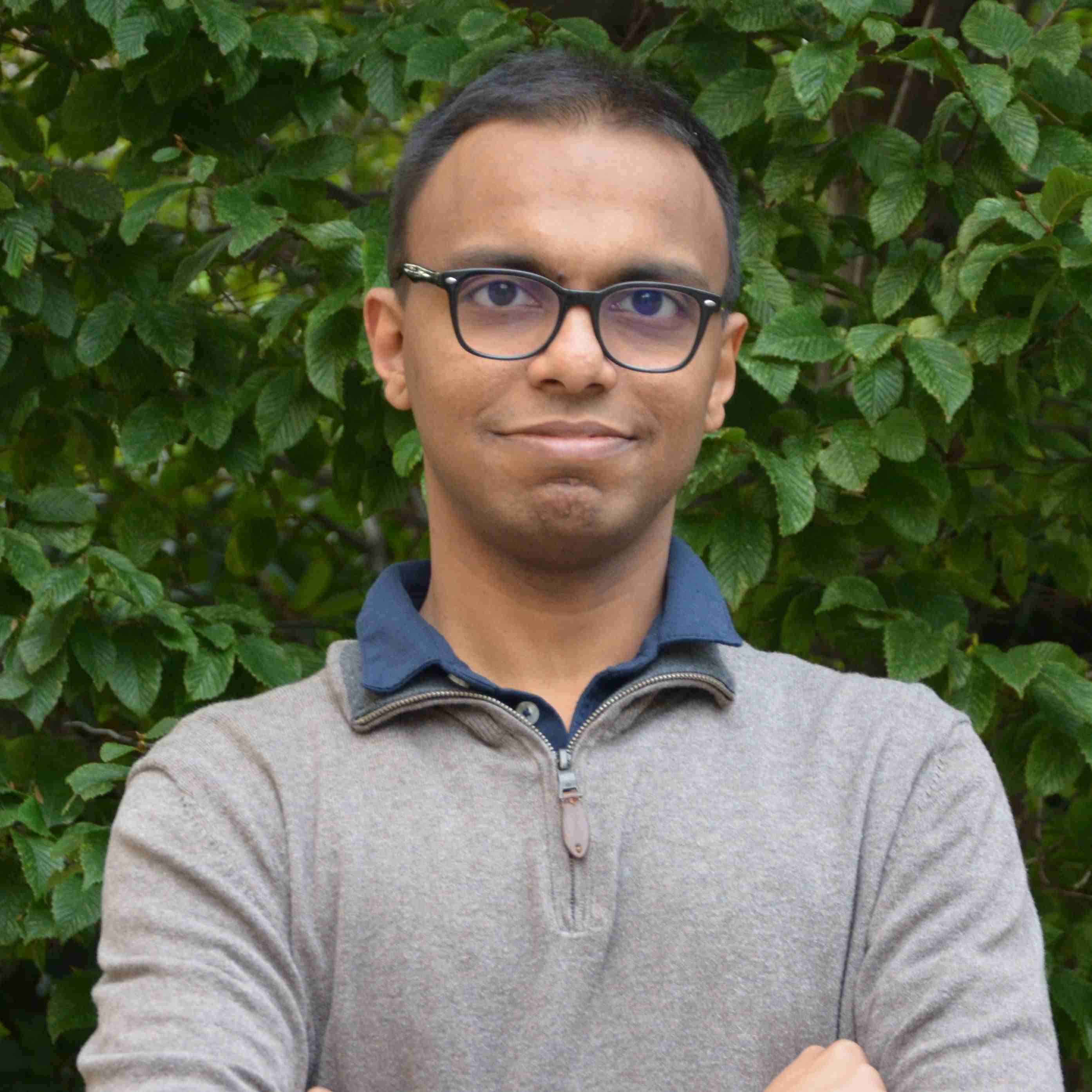}}]
{Sebin Gracy} is a Post-Doctoral Researcher in the Division of Decision and Control Systems in the School of Electrical Engineering and Computer Science at KTH Royal Institute of Technology. He obtained his Ph.D. degree at Universit\'e Grenoble-Alpes in November, 2018. Prior to that, he obtained his M.S. and B.E. degrees in Electrical Engineering from the University of Colorado at Boulder and the University of Mumbai, in December, 2013 and June 2010, respectively. %His research interests are in the realm of networked control systems.
%Biography text here.
\end{IEEEbiography}\vspace{-20mm}
\begin{IEEEbiography}[{\includegraphics[width=1in,height=1.25in,
trim = 350 300 300 0,
% trim = 200 0 200 0,
clip,keepaspectratio]{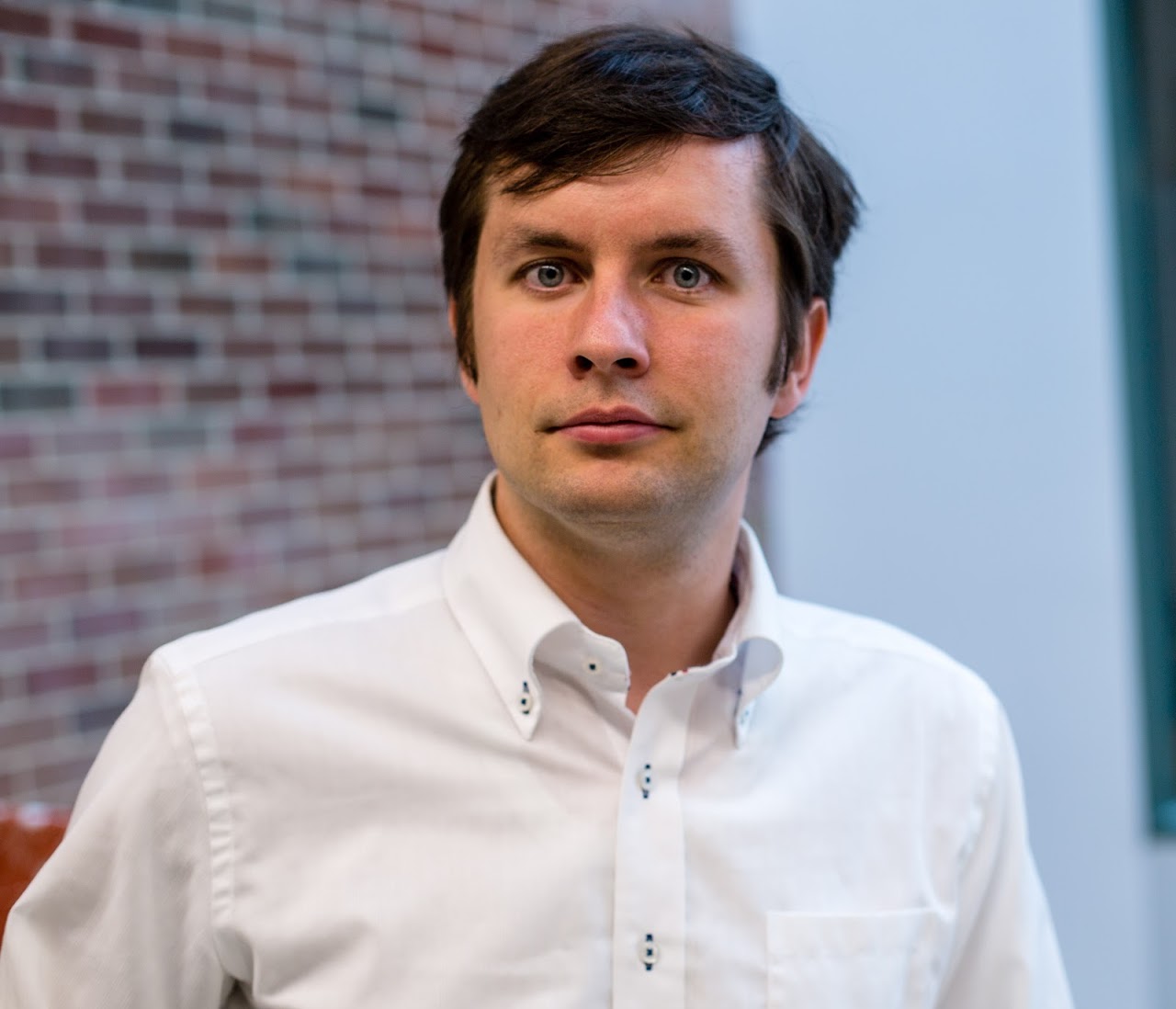}}]%{Philip E. Par\'{e}}
{Philip E. Par\'{e}} is an Assistant Professor in the School of Electrical and Computer Engineering at Purdue University. He received his Ph.D. in Electrical and Computer Engineering (ECE) from the University of Illinois at Urbana-Champaign (UIUC) in 2018, after which he went to KTH Royal Institute of Technology in Stockholm, Sweden to be a Post-Doctoral Scholar. He received his B.S. in Mathematics with University Honors and his M.S. in Computer Science from Brigham Young University in 2012 and 2014, respectively. 
% Philip was the recipient of the 2017-2018 Robert T. Chien Memorial Award for excellence in research from the UIUC ECE Department and named a 2017-2018 UIUC College of Engineering Mavis Future Faculty Fellow. 
His research focuses on networked control systems, namely modeling, analysis, and control of virus spread over networks.
\end{IEEEbiography}%\vspace{-5mm}

\begin{IEEEbiography}[{\includegraphics[width=1in,height=1.5in,
trim = 0 30 0 0,
clip,keepaspectratio]{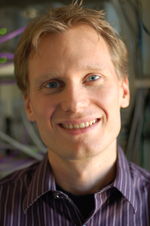}}]
{Henrik Sandberg} is Professor at the Division of Decision and Control Systems, KTH Royal Institute of Technology, Stockholm, Sweden. He received the M.Sc. degree in engineering physics and the Ph.D. degree in automatic control from Lund University, Lund, Sweden, in 1999 and 2004, respectively. His current research interests include security of cyber-physical systems, power systems, model reduction, and fundamental limitations in control. Dr. Sandberg was a recipient of the Best Student Paper Award from the IEEE Conference on Decision and Control in 2004, an Ingvar Carlsson Award from the Swedish Foundation for Strategic Research in 2007, and a Consolidator Grant from the Swedish Research Council in 2016. He has served on the editorial boards of IEEE Transactions on Automatic Control and the IFAC Journal Automatica.

%is Professor at the Division of Decision and Control Systems, KTH Royal Institute of Technology, Stockholm, Sweden. He received the M.Sc. degree in engineering physics and the Ph.D. degree in automatic control from Lund University, Lund, Sweden, in 1999 and 2004, respectively. From 2005 to 2007, he was a Post-Doctoral Scholar at the California Institute of Technology, Pasadena, USA. In 2013, he was a visiting scholar at the Laboratory for Information and Decision Systems (LIDS) at MIT, Cambridge, USA. He has also held visiting appointments at the Australian National University and the University of Melbourne, Australia. His current research interests include security of cyber-physical systems, power systems, model reduction, and fundamental limitations in control. Dr. Sandberg was a recipient of the Best Student Paper Award from the IEEE Conference on Decision and Control in 2004, an Ingvar Carlsson Award from the Swedish Foundation for Strategic Research in 2007, and a Consolidator Grant from the Swedish Research Council in 2016. He has served on the editorial boards of IEEE Transactions on Automatic Control and the IFAC Journal Automatica.

\end{IEEEbiography}\vspace{-30mm}

\begin{IEEEbiography}[{\includegraphics[width=1in,height=1.5in,trim = 0 0 0 70,clip,keepaspectratio]{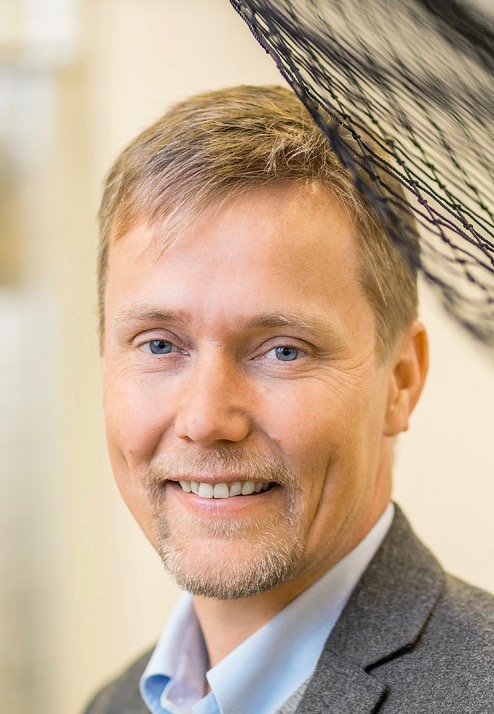}}]
{Karl H. Johansson} is Professor with the School of Electrical Engineering and Computer Science at KTH Royal Institute of Technology in Sweden and Director of Digital Futures. He received M.Sc. and Ph.D. degrees from Lund University. He has held visiting positions at UC Berkeley, Caltech, NTU, HKUST Institute of Advanced Studies, and NTNU. His research interests are in networked control systems and cyber-physical systems with applications in transportation, energy, and automation networks. He is a member of the Swedish Research Council's Scientific Council for Natural Sciences and Engineering Sciences. He has served on the IEEE Control Systems Society Board of Governors, the IFAC Executive Board, and is currently Vice-President of the European Control Association. He has received several best paper awards and other distinctions from IEEE, IFAC, and ACM. He has been awarded Distinguished Professor with the Swedish Research Council and Wallenberg Scholar with the Knut and Alice Wallenberg Foundation. He has received the Future Research Leader Award from the Swedish Foundation for Strategic Research and the triennial Young Author Prize from IFAC. He is Fellow of the IEEE and the Royal Swedish Academy of Engineering Sciences, and he is IEEE Control Systems Society Distinguished Lecturer.
\end{IEEEbiography}
\end{document}